\documentclass{revtex4}%
\usepackage {amsmath, amssymb, mathscinet, color, xfrac}
\makeatletter
\newtheorem{thm}{Theorem}[section]
\newtheorem{cor}[thm]{Corollary}
\newtheorem{lem}[thm]{Lemma}
\newtheorem{defn}[thm]{Definition}
\newtheorem{prop}[thm]{Proposition}

\newtheorem{preremark}[thm]{Remark}
\newenvironment{remark}%
  {\begin{preremark}\upshape}{\end{preremark}}
\newtheorem{preexample}[thm]{Example}
\newenvironment{example}%
  {\begin{preexample}\upshape}{\end{preexample}}
\newtheorem{notation}[thm]{Notation}
\numberwithin{equation}{section}
\newtheorem{fact}[thm]{Fact}
\numberwithin{equation}{section}

\numberwithin{thm}{section}
\newtheorem{lem/defn}[thm]{Lemma/Definition}
\newtheorem{preex/defn}[thm]{Example/Definition}
\newenvironment{ex/defn}%
  {\begin{preex/defn}\upshape}{\end{preex/defn}}

\newcommand{\ten}{\otimes}
\newcommand{\del}{\bigtriangleup}

\newcommand{\abs}[1]{\lvert#1\rvert}

\newcommand{\second}{\prime \prime }

\DeclareMathOperator{\End}{End}

\DeclareMathOperator{\Res}{Res}

\begin{document}

\title[Twisted vertex algebras]{Twisted vertex algebras, bicharacter construction  and boson-fermion correspondences}
\author{Iana I. Anguelova}

\address{Department of Mathematics,  College of Charleston,
Charleston SC 29424 }
\email{anguelovai@cofc.edu}


\begin{abstract}
 The boson-fermion correspondences are an important phenomena on the
intersection of several areas in mathematical physics: representation theory, vertex algebras
and conformal field theory, integrable systems, number theory, cohomology. Two such correspondences are well known: the types A and B (and their super extensions). As a main result of this paper  we present a new boson-fermion correspondence, of type D-A. Further, we define a new concept of twisted vertex algebra of order $N$, which  generalizes  super vertex algebra.  We develop the bicharacter construction  which we use for constructing classes of examples of twisted vertex algebras, as well as for deriving formulas for the operator product expansions (OPEs), analytic continuations and normal ordered products. By using the underlying Hopf algebra structure  we prove general bicharacter formulas for the vacuum expectation values for three important groups of examples.  We  show that  the correspondences of type B, C and D-A  are isomorphisms of twisted vertex algebras.
\end{abstract}

\maketitle

\tableofcontents

\section{Introduction}
\label{sec:intro}

\subsection{Motivation}

In 2 dimensions  the bosons and fermions  are related by the boson-fermion correspondences.
The best known  boson-fermion correspondence
is that of type A (see e.g. \cite{Frenkel-BF}, \cite{DJKM-KP}, \cite{KacRaina}).
There are many properties and applications of this correspondence, often called the charged
free fermion-boson correspondence,  and an  exposition
of some of them is given in \cite{KacRaina}, \cite{Miwa-book}.  \cite{DJKM-KP} and \cite{Frenkel-BF}
discovered its connection to the theory of integrable systems (e.g. the KP and KdV
hierarchies), to the theory of symmetric polynomials and the representation theory of the infinite dimensional
Lie algebra  $a_{\infty}$  (whence the name "type A" derives),
as well as to  $\hat{sl}_n$  and other affine Lie algebras.
As this boson-fermion correspondence has so many applications and
connections to various mathematical areas (including number theory and
geometry, as well as random matrix theory and random processes), a deeper understanding of the types of mathematical
structures that were being equated by it was sought. A partial answer early on
was given by Igor Frenkel in \cite{Frenkel-BF}, but a complete  answer had to wait for the development
of the theory of vertex algebras. Vertex operators were introduced in
string theory and now play an important role in many areas such as quantum field theory,
integrable models, representation theory, random matrix theory, statistical physics, and many others. The
theory of super vertex algebras axiomatizes the properties of some, simplest, "algebras" of
vertex operators (see for instance  \cite{BorcVA}, \cite{FLM}, \cite{FHL}, \cite{Kac}, \cite{LiLep}, \cite{FZvi}). Thus, as an application  of the theory of super vertex algebras the
question "what is the boson-fermion correspondence of type A?" can be answered precisely as follows:
the boson-fermion correspondence of type A is an isomorphism of two super vertex
algebras (\cite{Kac}).

 There are other examples of boson-fermion correspondences, for
instance the boson-fermion correspondence of type B (e.g. \cite{DJKM-4}, \cite{YouBKP}, \cite{OrlovVanDeLeur}), the super extensions of the boson-fermion correspondences
of type A and B (\cite{KacLeur}, \cite{Kac}), and others (\cite{DJKM6}, \cite{OrlovLeur}). The first of those, the type B, was introduced in \cite{DJKM-4}, and was discovered in  connection to the theory
of integrable systems, e.g.  the BKP hierarchy (\cite{DJKM-4}), and to the representation theory of the $b_{\infty}$ algebra (whence the name "type B" derives). Connections to the theory of symmetric polynomials and the symmetric group are shown in \cite{YouBKP}.  There is also
ongoing work on other boson-fermion and boson-boson correspondences, e.g.
the CKP correspondence (see \cite{DJKM6}, \cite{OrlovLeur}).

Our motivation thus is twofold: first, we considered the questions "Are all boson-fermion correspondences isomorphisms of super vertex algebras? And if not, what are the boson-fermion correspondences, i.e., what mathematical structures do they equate?".  And second, we considered the question "are there other boson-fermion correspondences?".

The first  part of the first question is readily answered by the fact that in all other cases except the type A, the boson-fermion correspondences cannot  be isomorphisms between
two super vertex algebras, since the associated
operator product expansions  have singularities at both $z = w$ and $z = -w$. Hence, in order to answer the  question, "what mathematical structures do these correspondences equate",  we introduce the concept of a twisted vertex algebra of order $N$ (where $N$ is a positive integer). It   generalizes the concept of a super vertex algebra in the sense that a super vertex algebra can be considered a twisted vertex algebra of order 1. The boson-fermion
correspondence of type B is then an  example of an isomorphism
 of twisted vertex algebras of order 2. There are  other examples of isomorphisms of twisted vertex algebras in the literature: the correspondence of type CKP, which was introduced in \cite{DJKM6} and developed further in \cite{OrlovLeur},
 while strictly speaking not a boson-fermion correspondence since the generating operator product expansions are  bosonic on both sides, is another such example, as is the super extension of the boson-fermion correspondence of type B.

 As one of the main results of this paper, and as an answer to the second question posed,  we introduce the new example of the boson-fermion correspondence of type D-A.
 This boson-fermion correspondence of type D-A completes the bosonisation of the 4 double-infinite rank affine Kac-Moody Lie algebras $a_{\infty}$, $b_{\infty}$, $c_{\infty}$, $d_{\infty}$. Although the super vertex algebra of the neutral fermion of type D is well known (see \cite{Kac}, \cite{WangKac}, \cite{Wang}) as it gives the basic representation of $d_{\infty}$, this  neutral fermion super vertex algebra does not itself have a bosonic equivalent,  but as we show in this paper its  twisted double cover does.  We further show that it is another example of an isomorphism of twisted vertex algebras. Hence all the correspondences of type B, C and D-A are isomorphisms of twisted vertex algebras.

\subsection{Overview of the paper}
After detailing the notation and some background that we will use throughout
the paper, we first briefly describe two examples of super vertex algebras. We assume the reader is familiar
with the definition and properties of super vertex algebras, hence we directly list the
relevant example of the boson-fermion correspondence of type A. We list only one property
which is an "imprint"of the boson-fermion correspondence of type A: the Cauchy
determinant identity which is a direct corollary of  the equality between the vacuum
expectation values of the two sides of the correspondence. (As discussed above, there are
many  properties and applications of any boson-fermion correspondence, which can,
and do, occupy many papers).
Next we proceed with the definition of a twisted vertex algebra (Section \ref{sec:dca}). Twisted vertex algebras generalize super vertex algebras in two main directions. First,  they have finitely many points of locality at $z=\epsilon ^i w$, where $\epsilon $ is a primitive $N$th root of unity (whereas super vertex algebras are only local at $z=w$). As such, twisted vertex algebras are similar to the   $\Gamma$-vertex algebras of \cite{Li2}, but are more general as they were designed to incorporate normal ordered products and operator product expansion coefficients as descendant fields (for a more detailed discussion on the comparison see \cite{ACJ}). The second  main difference between twisted vertex algebras and super vertex algebras (and $\Gamma$-vertex algebras of \cite{Li2}) is that in most of the examples of twisted vertex algebras the space of fields strictly contains the space of states. In fact a projection map from the space of fields to the space of states is part of the definition (see Definition \ref{defn:twistedVA}). In this aspect twisted vertex algebras more closely resemble the deformed chiral algebras defined in \cite{FR}. We do not discuss many properties of twisted vertex algebras in this paper due to length, instead we refer the interested readers to the follow-up paper \cite{ACJ}. In Section \ref{sec:dca} we continue with the description
of the examples of two pairs of twisted vertex algebras which constitute two examples
of boson-fermion correspondences: the boson-fermion correspondence of type B (reformulated from \cite{DJKM-4} and \cite{YouBKP} in the language of vertex algebras) and the
boson-fermion correspondence of type D-A, which is new to this paper. We list only one property out of many  for each correspondence,
a representative  identity which  is a direct corollary of the equality between the vacuum expectation values of the two sides of the correspondence.
For the correspondence of type B, this equality is actually the Schur Pfaffian
identity (Lemma \ref{lem:VacuumExpEqB}), and for the type D-A it is a different Pfaffian identity (Lemma \ref{lem:VacuumExpEqD}).
 As the definition of a twisted vertex algebra  is  new and technical, its necessity can only be justified by some very meaningful examples of twisted vertex algebras---which indeed are the boson-fermion correspondences.
  Although there are  many other examples of twisted vertex algebras, as we will see in theorem \ref{thm:Main}, the boson-fermion correspondences are important enough phenomena to justify this new definition of a twisted vertex algebra.
  Since the proofs of the
statements in the narrative of the examples are lengthy enough to impede the overview, we
give the proofs in the latter part of the paper, after we have introduced the requisite
technical set of tools, namely the bicharacter construction.  The bicharacter construction  was first introduced in the context of vertex algebras in \cite{BorcQVA}, the super bicharacters were introduced in \cite{A2}. The bicharacter construction is used in this paper in two ways.  First, to give a general construction and description of variety of examples of twisted vertex algebras. We prove  formulas for the analytic continuations, operator product expansions, vacuum expectation values using the bicharacters. (Such explicit formulas, for example the analytic continuations formulas, are very hard if not impossible to obtain otherwise  even for specific examples.)  Secondly, we use the  bicharacter construction extensively  to prove the properties of the new  boson-fermion correspondence of type D-A.

 There is a major  difference between the bicharacter description of a vertex algebra and the operator-based description typically used: In the operator-based description  the examples  are presented in terms of generating fields (vertex operators)   and their  OPEs (or commutation relations). With the bicharacter construction one starts instead with a (supercommutative supercocommutative) Hopf algebra $M$ and its free Leibnitz module (for the definition and examples of   free Leibnitz module see  sections \ref{subsection:super-bichLeibnitz} and \ref{subsection:freeLeModEx}).  The commutation relations then result from the choice of a bicharacter $r$--- different bicharacter $r$ will dictate different commutation relations. Moreover, for each Hopf algebra $M$ there are many choices of a  bicharacter $r$, hence each such pair $(M, r)$ will give rise to a different twisted vertex algebra, even if the underlying spaces of states are the same as Hopf algebras. It is in fact  the field-state  correspondence $Y$ that changes  with each choice of a bicharacter, and hence the fields in the vertex algebra. This is the case for instance for  the fermionic sides of both the B and D-A boson-fermion correspondences: they have identical spaces of states as Hopf algebras (which is not altogether surprising as they are both neutral fermions), but the collections of fields describing them differ substantially. Therefore in the bicharacter construction examples are grouped based on  the underlying Hopf algebra $M$---one starts by keeping $M$ the same, but changing the bicharacter $r$. We want to stress that there is a  variety of examples even after we fix the  algebra $M$.  This description based on the Hopf algebra $M$ is further used to prove  general formulas for the vacuum expectation values based on the Hopf algebra $M$  (sections \ref{subsection:Pfaffian}, \ref{subsection:Determinant}  and \ref{subsection:Product}). This is the explanation why the  vacuum expectation values for both the B and the D fermion are Pfaffians, although, of course,  different ones.  In  section \ref{section:examplesexpl}  the particular  examples of the fermionic and bosonic examples are detailed and their properties are proved, including the statements and assertions from sections \ref{subsection:examplesB} and \ref{subsection:examplesD}.

\section{Notation and background}

\subsection{Notation}

In this section we list notations we will continuously use throughout the paper.
Throughout we assume $N$ is a positive integer. We work over the field of complex numbers $\mathbb{C}$, with the category of super vector spaces, i.e., $\mathbb  {Z}_{2}$ graded vector spaces. The flip map $\tau $  is defined by
\begin{equation}
\label{eq:flip}
 \tau (a\ten b) =(-1)^{\tilde{a} \cdot \tilde{b}} (b\ten a)
\end{equation}
for any homogeneous elements $a, b$ in the super vector space, where $\tilde{a}$, $\tilde{b}$ denote correspondingly  the parity of $a$, $b$.

A superbialgebra $H$ is a an associative algebra with a $\mathbb  {Z}_{2}$ grading and a compatible coalgebra
structure, in the sense that the coproduct denoted by $\del$ and the counit denoted by $\eta$ are algebra maps.
A Hopf superalgebra is a superbialgebra with an antipode $S$.
For a superbialgebra $H$  we will write $\del (a)=\sum \ a^{\prime}\ten a^{\second}$ for the coproduct of $a\in H$ (Sweedler's notation); i.e., we will usually omit  the indexing in $\del (a)=\sum_k \ a^{\prime}_k\ten a^{\second}_k$, especially when it is clear from the context.
Recall in a super  Hopf algebra the product on
 $H\ten H$ is defined by
\begin{equation}
\label{eq:tespr}
(a\ten b)(c\ten d)= (-1)^{\tilde{b} \cdot \tilde{c} } (ac\ten bd)
\end{equation}
for any $a, b, c, d$ homogeneous elements in $H$.
A supercocommutative bialgebra is a superbialgebra with
$\tau (\del(a))=\del(a)$.
If $H$ is a Hopf super algebra, an element  $a\in H$ satisfying  $\del
(a)=a\ten 1+1\ten a$, $\eta (a)=0$, and $S(a)=-a$ is called ``primitive"
A  element $g\in H$  satisfying $\del(g)=g\ten g, \ \eta (g)=1$, $S(g)=g^{-1}$ is called ``grouplike".
\begin{notation}
For any $a\in A$, where $A$ is a   commutative associative $\mathbb{C}$ algebra denote $a^{(n)}:=\frac{a^n}{n!}$.
\end{notation}
\begin{defn}\begin{bf}(The Hopf algebra $H_D=\mathbb{C} [D]$)\end{bf}\\
The Hopf algebra $H_D=\mathbb{C} [D]$ is  the polynomial algebra with a primitive generator $D$.
We have
\begin{equation}
\label{eq:D^n}
\del D^{(n)}=\sum _{k+l=n} D^{(k)}\ten D^{(l)}.
\end{equation}
\end{defn}
\begin{defn}\begin{bf}(The Hopf algebra $H^N_{T_{\epsilon}}$)\end{bf}\\
Let $\epsilon$ be  a primitive root of unity of order $N$.
The Hopf algebra $H^N_{T_{\epsilon}}$ is the Hopf algebra with  a primitive generator $D$ and a grouplike generator $T_{\epsilon}$ subject to the  relations:
\begin{equation}
DT_{\epsilon }=\epsilon T_{\epsilon } D, \quad  (T_{\epsilon })^N=1.
\end{equation}
$H^N_{T_{\epsilon}}$ contains $H_D$ as a Hopf subalgebra. Both $H_D$ and $H^N_{T_{\epsilon}}$ are  entirely even.
\end{defn}
\begin{notation}\begin{bf}(The function spaces $\mathbf{F}^N(z, w)$, $\mathbf{F}^N(z, w)^{+, w}$)\end{bf}\\
Let $\epsilon$ be  a primitive root of unity of order $N$.
Denote by $\mathbf{F}^N(z, w)$ the space of rational functions in the  formal variables $z, w$ with only poles at $z=0, w=0, \ z= \epsilon^i w$, $i=1, \dots , N$:
\begin{equation}
\mathbf{F}^N(z, w)=\mathbb C[z,w][z^{-1},w^{-1},(z-w)^{-1},(z-\epsilon w)^{-1},\cdots, (z-\epsilon^{N-1}w)^{-1}].
\end{equation}
Also, denote $\mathbf{F}^N(z, w)^{+, w}$ the space of rational functions in the formal variables $z, w$ with only poles at $z=0,  \ z= \epsilon^i w$, $i=1, \dots , N$.
Note that we do not allow a pole at $w=0$ in $\mathbf{F}^N(z, w)^{+, w}$, i.e., if $f(z, w)\in \mathbf{F}^N(z, w)$, then $f(z, 0)$ is well defined.
 More generally, let $\mathbf{F}^N(z_1, z_2, \dots , z_l)$ is the space of rational functions in variables $z_1, z_2, \dots, z_l$ with only poles at $z_m=0$, $m=1,\dots, l$, or at $z_i= \epsilon^{k} z_j$, and $i\neq j=1,\dots, l$, $k=1, \dots , N$.
 Lastly $\mathbf{F}^N(z_1, z_2, \dots , z_l)^{+, z_l}$ is the space of rational functions in variables $z_1, z_2, \dots, z_l$ with only poles at $z_m=0$, $m=1,\dots, l-1$, or at $z_i= \epsilon^{k} z_j$, $i\neq j=1,\dots, l$, $k=1, \dots , N$.
\end{notation}
\begin{fact}
$\mathbf{F}^N(z, w)$ is an  $H^N_{T_{\epsilon}}\ten H^N_{T_{\epsilon}}$ (and consequently an $H_D\ten H_D$)  Hopf module by
\begin{align}
&D_z f(z, w)=\partial_z f(z, w), \quad (T_{\epsilon}) _{z} f(z, w)=f(\epsilon z, w) \\
&D_w f(z, w)=\partial_w f(z, w), \quad (T_{\epsilon}) _{w} f(z, w)=f(z, \epsilon  w)
\end{align}
We will denote the action of elements $h\ten 1 \in H^N_{T_{\epsilon}}\ten H^N_{T_{\epsilon}}$  on $\mathbf{F}^N(z, w)$ by $h_z\cdot$, correspondingly $h_w\cdot$ will denote the action of the elements $1\ten h \in H^N_{T_{\epsilon}}\ten H^N_{T_{\epsilon}}$.
\end{fact}
\begin{notation}
For a rational function $f(z,w)$ we denote by $i_{z,w}f(z,w)$
the expansion of $f(z,w)$ in the region $\abs{z}\gg \abs{w}$, and correspondingly for
$i_{w,z}f(z,w)$. Similarly, we will denote by \mbox{$i_{z_1,z_2,\dots,z_n}$} the expansion in the region \mbox{$|z_1|\gg \dots \gg|z_n|$}.
\end{notation}

\subsection{Super vertex algebras and the boson-fermion correspondence of type A}
\label{sec:type A}

The following definitions are well known (see for instance \cite{FLM}, \cite{FHL},  \cite{Kac}, \cite{LiLep} and others).
\begin{defn}
\begin{bf} (Field)\end{bf}\label{defn:field-fin}
 A field $a(z)$ on a vector space $V$ is a series of the form
\begin{equation*}
a(z)=\sum_{n\in \mathbf{Z}}a_{(n)}z^{-n-1}, \ \ \ a_{(n)}\in
\End(V), \ \ \text{such that }\ a_{(n)}v=0 \ \ \forall \ v\in V, \ n\gg 0.
\end{equation*}
\end{defn}
\begin{remark} \label{remark:fieldindexing}
The coefficients $a_{(n)}, \ n\in \mathbb{Z}$, are often called modes. The indexing above is  typically used in super vertex algebras: if a field  $a(z)$ with this indexing is a vertex operator in a super  vertex algebra, then the modes $a_{(n)}$ with $n\geq 0$  annihilate the vacuum vector (hence are called annihilation operators), and the modes $a_{(n)}$ with $n< 0$ are the creation operators.
Denote
\begin{equation}
a(z)_-:=\sum_{n\geq 0}a_nz^{-n-1},\quad a(z)_+:=\sum_{n<0}a_nz^{-n-1}.
\end{equation}
\end{remark}
\begin{defn} {\bf (Normal ordered product)}
Let $a(z), b(z)$ be fields on a vector space $V$.  Define the normal ordered product of these fields by
\begin{equation}
\label{eqn:normordDef}
:a(z)b(w):=a(z)_+b(w)+(-1)^{\tilde{a}\tilde{b}}b(w)a_-(z).
\end{equation}
\end{defn}
\begin{remark}
Let  $a(z), b(z) \in End(V)[[ z^{\pm 1}]]$ be fields on a vector space $V$. Then
\mbox{$:a(z)b(\lambda z):$} and $:a(\lambda z)b( z):$ for any $\lambda \in \mathbb{C}^*$ are well defined elements of  $End(V)[[ z^{\pm 1}]]$, and are also fields on $V$ .
\end{remark}
\begin{remark}
The space of fields on a vector space $V$ is an $H^N_{T_{\epsilon}}$ module via  the same action $h_z\cdot$ used on  the space $\mathbf{F}^N(z, w)$.
\end{remark}
The definition of a super vertex algebra is well known, we refer the reader for example to  \cite{FLM}, \cite{FHL}, \cite{Kac}, \cite{LiLep}, as well as for  \ notations, details and theorems. Super vertex algebras have two  important properties which we would like to carry over to
the case of twisted vertex algebras. These are the properties of analytic continuation and completeness with respect to  operator product expansions (OPEs). In fact our definition of a twisted vertex algebra is based on enforcing these two properties. Recall we have for the OPE of two local (at $z=w$) fields  (see e.g. \cite{Kac})
\begin{equation}
\label{eqn:OPEnorm}
a(z)b(w)=\sum _{j=0}^{N-1} i_{z,w}\frac{c^j(w)}{(z-w)^{j+1}} +:a(z)b(w):.
\end{equation}
 Moreover, if the two fields $a(z), b(w)$ are vertex operators in a super vertex algebra we have $Res_{(z-w)}a(z)b(w)(z-w)^j=c^j(w)=(a_{(j)}b) (w)$, i.e., the coefficients of the OPEs are vertex operators  in the \textbf{same} super vertex algebra (i.e., attached to elements of the vertex algebra). Since the commutation relations are determined only by the singular part of the OPE, we abbreviate the OPE above as:
\begin{equation}
\label{eqn:OPE}
a(z)b(w)\sim \sum _{j=0}^{N-1}\frac{c^j(w)}{(z-w)^{j+1}}.
\end{equation}
Also, an analytic continuation property for a super vertex algebra holds: for any $a_i \in V, i=1,\dots ,k$,  there exist a  rational vector valued   function
\begin{equation*}
\label{eq:defX}
X_{z_1, z_2, \dots , z_k} : V^{\ten k}
\to W[[z_1, z_2, \dots , z_k]]\ten \mathbf{F}^1_{\epsilon}(z_1, z_2, \dots , z_k)^{+, z_k},
\end{equation*}
    such that
\begin{equation*}
Y(a_1, z_1)Y(a_2, z_2)\dots Y (a_k, z_k)1=i_{z_1,z_2,\dots,z_k} X_{z_1, z_2, \dots , z_k}(a_1\ten a_2\ten \dots \ten a_k).
\end{equation*}
For many examples, super vertex algebras are generated by a much smaller number of generating  fields (see e.g. \cite{Kac}),  with imposing the condition that the resulting space  of fields of the vertex algebra has to be closed under certain operations: for any vertex operator  $a(z)$ the field $Da(z) =\partial_z a(z)$  has to also be a vertex operator in the vertex algebra. Also, the  OPEs coefficients  ($c^j(w)$ from  \eqref{eqn:OPE})  and  normal ordered  products $:a(z)b(z):$ of  any two vertex operators $a(z)$ and $b(w)$ have to be  vertex operators in  the vertex algebra.
 Note that the identity operator on $V$ is always a  vertex operator in the vertex algebra corresponding to the vacuum vector $|0\rangle  \in V$.

Next we briefly recall the boson-fermion correspondence of type A, which is an isomorphism of super vertex algebras.
The fermion side of the boson-fermion correspondence of type A is a super vertex algebra generated by  two nontrivial odd fields---two charged fermions: the fields $\phi (z)$ and $\psi (z)$ with only nontrivial operator product expansion (OPE):
\begin{equation}
\phi (z)\psi (w)\sim \frac{1}{z-w}\sim \psi(z)\phi(w), \quad \phi (z)\phi (w)\sim 0 \sim \psi(z)\psi(w),
\end{equation}
where the $1$ above denotes the identity map $Id$.
The modes $\phi_n$ and $\psi_n$, $n\in \mathbb{Z}$, of the fields  $\phi (z)$ and $\psi (z)$, which we index as follows:
\begin{equation}
\phi (z) =\sum _{n\in \mathbf{Z}} \phi_n z^{n}, \quad \psi (z) =\sum _{n\in \mathbf{Z}} \psi_n z^{n},
\end{equation}
form a Clifford algebra $\mathit{Cl_A}$  with relations
\begin{equation}
[\phi_m,\psi_n]_{\dag}=\delta _{m+n, -1}1, \quad [\phi_m,\phi_n]_{\dag}=[\psi_m,\psi_n]_{\dag}=0.
\end{equation}
The indexing of the generating fields vary depending on the point of view; our indexing here corresponds to $\phi_n =\hat{v}_{n+1}, \quad \psi_n=\check{v}^*_{-n}$ of \cite{KacRaina}. Here our choice corresponds to our bicharacter description of this example. This indexing and the properties of the vertex algebra dictate that
the underlying  space of states of this super vertex algebra---the fermionic Fock space, denoted by $\mathit{F_A}$ --- is the highest weight   representation of $\mathit{Cl_A}$ generated by the vacuum  vector $|0\rangle $,  so that $\phi_n|0\rangle=\psi_n|0\rangle=0 \ \text{for} \  n<0$.
We denote  by  $\mathit{F_A}$ both the space of states and the resulting vertex algebra generated by the fields $\phi (z)$ and $\psi (z)$; it is often called  the charged free fermion vertex algebra.

We can calculate vacuum expectation values if we have a  symmetric bilinear form $\langle
\ \mid\ \rangle: V\ten V \to \mathbb{C}$ on the space of states of the vertex algebra $V$. There is a very important concept  of an invariant bilinear form on a vertex algebra, for details see for example \cite{FHL}, \cite{Li-bilinear}.  We will not recall the full definition  here, but we will require  that  the bilinear form is such that the vacuum vector $|0\rangle$ spans an  orthogonal subspace on its own (for a precise statement  in the context of the bicharacter construction see  sections \ref{subsection:Pfaffian} and \ref{subsection:Determinant}). We also require that the bilinear form is  normalized on the vacuum vector:
\begin{equation}
\langle \langle 0| \mid |0\rangle \rangle =1.
\end{equation}
By abuse of notation we will just write $\langle  0 \mid 0 \rangle$ instead of $\langle \langle 0| \mid |0\rangle \rangle$.
\begin{lem}
The following  determinant formula for the vacuum expectation values on the fermionic side $\mathit{F_A}$ holds:
\begin{equation}
\label{eqn:FermVEV-A}
\langle  0 |\phi (z_1)\phi (z_2)\dots \phi (z_n)\psi (w_1)\psi (w_2)\dots \psi (w_n) | 0 \rangle =
(-1)^{n(n-1)/2}i_{z, w} det \Big(\frac{1}{z_i -w_j}\Big)_{i, j =1}^n.
\end{equation}
Here $i_{z; w}$  stands for the expansion $i_{z_1,z_2,\dots,z_n, w_1, \dots, w_n}$.
\end{lem}
This formula is well known, new  proof as a corollary of  the bicharacter description will be given in the section \ref{section:freefermionA}.

The boson-fermion correspondence of type A is  determined once we write the images of  generating fields $\phi (z)$ and $\psi (z)$ under the correspondence. In order to do that, an \textbf{essential} ingredient is the so-called Heisenberg field $h(z)$ given by
\begin{equation}
\label{eqn:normal-order-h-A}
h(z)=:\phi (z)\psi (z):
\end{equation}
It follows that the Heisenberg field $h(z)=\sum _{n\in \mathbb{Z}} h_n z^{-n-1}$ has OPEs with itself given by:
\begin{equation}
\label{eqn:HeisOPEsA}
h(z)h(w)\sim \frac{1}{(z-w)^2}, \quad \text{in \ modes:} \ [h_m,h_n]=m\delta _{m+n,0}1.
\end{equation}
i.e., its  modes $h_n, \ n\in \mathbb{Z}$, generate a Heisenberg algebra $\mathcal{H}_{\mathbb{Z}}$ (which is the reason for the name Heisenberg field).
It is well known that any irreducible highest weight module of this Heisenberg algebra is isomorphic as Heisenberg module  to the polynomial algebra $\mathit{B_m}\cong \mathbb{C}[x_1, x_2, \dots , x_n, \dots ]$ with countably many variables  via the action:
\begin{equation}
h_n\mapsto \partial _{x_{n}}, \quad h_{-n} \mapsto nx_n\cdot, \quad \text{for any} \ \ n\in \mathbb{N}, \quad h_0\mapsto m\cdot.
\end{equation}
The fermionic Fock space decomposes (via the charge decomposition, for details see for example \cite{KacRaina}) as $\mathit{F_A}=\oplus _{i\in \mathbb{Z}} B_i$, which we can write as
\begin{equation}
\mathit{F_A}=\oplus _{i\in \mathbb{Z}} B_i \cong \mathbb{C}[e^{\alpha}, e^{-\alpha}]\ten \mathbb{C}[x_1, x_2, \dots , x_n, \dots ],
\end{equation}
where by $\mathbb{C}[e^{\alpha}, e^{-\alpha}]$ we mean the Laurent polynomials with one variable   $e^{\alpha}$ (the reason for this notation is that the resulting vertex algebra on the right hand side is a lattice vertex algebra.) The isomorphism is as  Heisenberg modules, where $e^{n\alpha}$ identifies the highest weight vector for the irreducible Heisenberg module $B_n$. We denote the  vector space on the right-hand-side of this $\mathcal{H}_{\mathbb{Z}}$-module isomorphism by $B_A: =\mathbb{C}[e^{\alpha}, e^{-\alpha}]\ten \mathbb{C}[x_1, x_2, \dots , x_n, \dots ]$.  $B_A$ is the underlying vector space of the bosonic side of the boson-fermion correspondence of type A.

Now we can write the images of  generating fields $\phi (z)$ and $\psi (z)$ under the correspondence:
\begin{equation}
\label{eqn:imageA}
\phi (z)\mapsto e^{\alpha}_A(z), \quad  \psi (z)\mapsto e^{-\alpha}_A(z),
\end{equation}
where the generating fields $e^{\alpha}_A(z)$, $e^{-\alpha}_A(z)$ for the bosonic part of the correspondence are given by
\begin{align}
\label{eqn:ExponA-1}
e^{\alpha}_A(z)&=\exp (\sum _{n\ge 1}\frac{h_{-n}}{n} z^n)\exp (-\sum _{n\ge 1}\frac{h_{n}}{n} z^{-n})e^{\alpha}z^{\partial_{\alpha}}, \\
\label{eqn:ExponA-2}
e^{-\alpha}_A(z)& =\exp (-\sum _{n\ge 1}\frac{h_{-n}}{n} z^n)\exp (\sum _{n\ge 1}\frac{h_{n}}{n} z^{-n})e^{-\alpha}z^{-\partial_{\alpha}},
\end{align}
the operators $e^{\alpha}$, $e^{-\alpha}$, $z^{\partial_{\alpha}}$ and $z^{-\partial_{\alpha}}$ act in an obvious way on the space $B_A$.

The resulting super vertex algebra generated by the fields $e^{\alpha}_A(z)$ and $e^{-\alpha}_A(z)$ acting on the underlying vector space $B_A$ we denote also by $B_A$.
\begin{lem}
The following product formula for the vacuum expectation values on the bosonic side $B_A$ holds:
\begin{equation}
\label{eqn:BosVEV-A}
\langle  0 |e^{\alpha}_A (z_1)e^{\alpha}_A (z_2)\dots e^{\alpha}_A (z_n)e^{-\alpha}_A (w_1)e^{-\alpha}_A (w_2)\dots e^{-\alpha}_A (w_n) | 0 \rangle =i_{z, w}\frac{\prod_{i<j}^n ((z_i-z_j)(w_i-w_j))}{\prod_{i, j=1}^n (z_i-w_j)}
\end{equation}
Here $i_{z, w}$  stands for the expansion $i_{z_1,z_2,\dots,z_n, w_1, \dots, w_n}$.
\end{lem}
The proof will be given in the section \ref{section:freebosonA}.
\begin{thm}(\cite{Kac})
The boson-fermion correspondence of type A is the isomorphism between the charged free fermions super vertex algebra $\mathit{F_A}$ and the bosonic super vertex algebra $B_A$.
\end{thm}
\begin{cor} \label{cor:VacuumExpEqA} The Cauchy's determinant identity follows from the equality of the vacuum expectation values ($AC$ stands for Analytic Continuation):
\begin{align*}
(-1)^{n(n-1)/2} det \Big(\frac{1}{z_i -w_j}\Big)_{i, j =1}^n &=AC \langle  0 |\phi (z_1)\dots \phi (z_n)\psi (w_1)\dots \psi (w_n) | 0 \rangle \\
&=AC\langle  0 |e^{\alpha}_A(z_1)\dots e^{\alpha}_A (z_n)e^{-\alpha}_A (w_1)\dots e^{-\alpha}_A (w_n) | 0 \rangle =\frac{\prod_{i<j} ((z_i-z_j)(w_i-w_j))}{\prod_{i, j=1}^{n} (z_i-w_j)}.
\end{align*}
\end{cor}
It is important to note that although this determinant identity is obviously well known and has been proved in a variety of ways since Cauchy, it is also a  direct corollary of   the boson-fermion correspondence of type A.

\section{Twisted vertex algebras: definition, overview, examples}
\label{sec:dca}

\subsection{Twisted vertex algebras: definition and overview}
\begin{defn}\label{defn:twistedVA} \begin{bf}(Twisted vertex algebra of order $N$)\end{bf}
A twisted vertex algebra  of order $N$ is a collection $(V, W, \pi, Y)$  of  the following data :
\begin{itemize}
\item the space of fields $V$: a vector super space  and  an $H^N_{T_{\epsilon}}$ module graded as an  $H_D$-module;
\item the space of states $W$: a vector super subspace of $V$;
\item a linear surjective projection map $\pi: V\to W$, such that $\pi\arrowvert _W=Id_W$
\item a field-state correspondence $Y$: a linear map  from $V$ to the space of fields on $W$;
\item a vacuum vector: a vector $|0\rangle \in W\subset V$.
\end{itemize}
This data should satisfy the following  set of axioms:
\begin{itemize}
\item vacuum axiom: \ \ $Y(|0\rangle, z)=Id_W$;
\item modified creation axiom: \ \ $Y(a, z)|0 \rangle \arrowvert _{z=0}=\pi(a)$, for any $a\in V$;
\item transfer of action:\ \  $Y(ha,z)=h_z\cdot Y(a, z)$ for any $h\in H^N_{T_{\epsilon}}$;
\item analytic continuation: For any $a, b, c\in V$ exists  $X_{z, w, 0}(a\ten b\ten c)\in W[[z, w]]\ten \mathbf{F}^N(z, w)$ such that
\begin{equation}
Y(a, z)Y(b, w)\pi(c)=i_{z, w} X_{z, w,  0}(a\ten b\ten  c)
\end{equation}
\item symmetry: $X_{z, w, 0}(a\ten b\ten c)=X_{w, z, 0}(\tau (a\ten b)\ten c)$;
\item Completeness with respect to Operator Product Expansions (OPE's): For each $i\in 0, 1, \dots, N-1$, \ $k\in \mathbb{Z}$, and any $a, b, c\in V$,  where $a, b$ are homogeneous  with respect to the grading by $D$, there exist $l^s_{k, i}\in \mathbb{Z}, \ |l^s_{k, i}|\leq  (N-1)(k+1)$, such that
\begin{equation}
Res_{z=\epsilon ^i w}X_{z, w, 0}(a\ten b\ten c)(z-\epsilon ^i w)^k =\sum_s^{\text{finite}} w^{l^s_{k,i}}  Y(v^s_{k, i}, w)\pi(c)
\end{equation}
for some \ homogeneous \ elements  $v^s_{k, i}\in V, \ \ l^s_{k,i}\in \mathbb{Z}$.
\end{itemize}
\end{defn}
\begin{remark} If $V$ is an (ordinary) super vertex algebra, then the data $(V, V, \pi=Id_V, Y)$ is a twisted vertex algebra of order 1. Moreover, it can be proved that a twisted vertex algebra $(V, V, \pi=Id_V, Y)$ of order 1 is a super vertex algebra, due to the restrictions $|l^s_{k, i}|\leq (N-1)(k+1)$ in the shifts $l^s_{k, i}$.
\end{remark}
The definition of a twisted vertex algebra is very similar to the definition of a deformed chiral algebra given by E. Frenkel and Reshetikhin in \cite{FR}.
\begin{remark} (\textbf{OPE shifts})
\label{remark:shift}
The axiom  requiring completeness with respect to the OPEs is  weaker  than the OPE property of super  vertex algebras. We can express this weaker axiom as follows (for more details see \cite{ACJ}):  the OPE coefficient, the residue \mbox{$Res_{z=\epsilon ^i w}X_{z, w, 0}(a\ten b\ten c)(z-\epsilon ^i w)^k$}, gives the values of a field  $v(w)$ which may not be a vertex operator associated to an element of  $V$. But each mode $v_{(n)}$ is a finite sum of the  modes $v^s_{k, i}(w)_{(n-l^s_{k,i})}$ of the corresponding  fields $v^s_{k, i}(w)$. Each of these  modes $v^s_{k, i}(w)_{(n-l^s_{k,i})}$ may require a different shift $l^s_{k, i}$ in its index.  As was mentioned in the previous section, in a super vertex algebra we have a stronger property, requiring  that the OPE coefficients automatically, without need of a shift,   be vertex operators  in the same vertex algebra (i.e., attached to an element of $V$).  This stronger property cannot hold in the interesting examples, which forced the modification of the OPE completeness axiom (see remark \ref{remark:shiftinB} below).
\end{remark}
\begin{defn}(\textbf{Isomorphism of twisted vertex algebras})
\label{defn:isomTVA}
Two twisted vertex algebras $(V, W, \pi, Y)$ and $(\widetilde{V}, \widetilde{W}, \tilde{\pi}_{f}, \widetilde{Y})$ are said to be isomorphic via a linear bijective map $\Phi: V\to \widetilde{V}$ if $\Phi (|0\rangle_{W}) =|0\rangle_{\widetilde{W}}$ and the following holds: for any
 $v\in V$, $\tilde{v}\in \widetilde{V}$ and  $w\in W$,  $\tilde{w}\in \widetilde{W}$ we have
 \begin{equation*}
\Phi(v)=\sum _{\text{finite}}c_{k}\tilde{v}_{k}, \ \ c_{k}\in \mathbb{C}, \ \ \tilde{v}_{k}\in \widetilde{V},\quad
\Phi^{-1}(\tilde{v})=\sum _{\text{finite}}d_{m}v_{m}, \ \ d_{m}\in \mathbb{C}, \ \ v_{m}\in V;
\end{equation*}
so that
\begin{align}
& \Phi(Y(v, z)w) =\sum_{\text{finite}} z^{l_k}c_{k} \widetilde{Y} (\tilde{v}_{k}, z)\tilde{\pi}_{f}\circ \Phi(w), \ \ \ l_k \in \mathbb{Z}; \\
& \Phi^{-1}(\widetilde{Y}(\tilde{v}, z)\tilde{w}) =\sum_{\text{finite}} z^{l_m}d_{m} Y (v_{m}, z)\pi\circ \Phi ^{-1}(\tilde{w}), \ \ \ l_m \in \mathbb{Z}.
\end{align}
\end{defn}
\begin{remark}
This definition is much more complicated than in the super vertex algebra case due to the allowance for the shifts in the OPEs. This can cause each of the summands in the linear sum $\Phi(v)$ to appear with a different shift in the sum of the corresponding vertex operators,  e.g. \eqref{eqn:isomPeculB} and
\eqref{eqn: finalDiso},  hence the allowance  for the different   powers of $z$ in the above definition.
\end{remark}
There are a variety of different vertex algebra like theories, each  designed to describe different sets of examples of collections of fields. The best known is the theory of super vertex algebras (see e.g. \cite{FLM}, \cite{Kac}, \cite{LiLep}, \cite{BorcVA}, \cite{FZvi}). The axioms of super vertex algebras are often given in terms of locality (see \cite{Kac}, \cite{FZvi}), as locality is a property that plays crucial importance for super vertex algebras (\cite{LiLocality}). On the other hand, there are vertex algebra like objects which do not satisfy the usual locality property, but rather a generalization. Twisted vertex algebras are among them, but there are also generalized vertex algebras, $\Gamma$-vertex algebras, deformed chiral algebras, quantum vertex algebras.
Twisted vertex algebras occupy intermediate step between super vertex algebras and  deformed chiral algebras of \cite{FR}). We chose here to define twisted vertex algebras with axioms closer to the deformed chiral algebra axioms,  including the analytic continuation axiom. But twisted vertex algebras are  closer to super vertex algebras in two aspects. First, they satisfy $N$-point locality (finitely many points of locality), see \cite{ACJ}, unlike the deformed chiral algebras which have lattices of points of locality. Second, as for super vertex algebras, the axioms requiring existence of analytic continuation of product of two vertex operators plus the  symmetry axiom do in fact enforce the property that analytic continuation of arbitrary product of fields exist; something that is not true for deformed chiral algebras (see \cite{FR}). This property, which we will not prove here (proof is given in  \cite{ACJ}), is why we chose to give the axioms for twisted vertex algebra in the form above. Further, we will derive formulas for the analytic continuations using the bicharacter construction which we will use to derive vacuum expectation values identities.
{\begin{prop} {\bf (Analytic continuation for arbitrary products of fields)}\\
Let $(V, W, \pi, Y)$ be a twisted vertex algebra.     There exists a  rational vector valued   function
\begin{equation*}
\label{eq:defX}
X_{z_1, z_2, \dots , z_k} : V^{\ten k}
\to W[[z_1, z_2, \dots , z_k]]\ten \mathbf{F}^N(z_1, z_2, \dots , z_k)^{+, z_k},
\end{equation*}
    such that
\begin{equation*}
Y(a_1, z_1)Y(a_2, z_2)\dots Y (a_k, z_k)1=i_{z_1,z_2,\dots,z_k} X_{z_1, z_2, \dots , z_k}(a_1\ten a_2\ten \dots \ten a_k)
\end{equation*}
for any $a_i \in V, i=1,\dots ,k$.
\end{prop}}
We do not discuss the axiomatics of twisted vertex algebras further in this paper due to length (for this see \cite{ACJ}). Instead we  show that the concept of a   twisted vertex algebra answers the question "what mathematical structures are the  boson-fermion correspondences of type B, C and D-A?". We claim that these boson-fermion correspondences  are isomorphisms of twisted vertex algebras.

Similarly to super vertex algebras,  twisted vertex algebras are often generated by a smaller number of fields. We will not prove theorems on what constitutes a generating set of fields (see \cite{ACJ}), instead we will take an alternative approach and use  the bicharacter construction which we will present in the next section. The bicharacter construction in some sense mimics the generation by a smaller set of generating fields, in that we start  with a smaller set of data which then "generates" the entire set of data for the twisted vertex algebra. The bicharacter construction  will thus replace the necessary theorems on generating sets of fields. We want to mention though, that if we have a set of generating fields, the full space of  fields is in turn  determined  by  requiring, as in a super vertex algebra,  that it be closed under OPEs (see modification above); that for any vertex operator  $a(z)$ the field $Da(z) =\partial_z a(z)$   has to be a vertex operator in the twisted vertex algebra. A new ingredient  in a twisted vertex algebra  is the  requirement  that the field $T_{\epsilon}a(z)=a(\epsilon z)$ is  a vertex operator in the same twisted vertex algebra as well. Note that this immediately violates the  stronger creation axiom for a classical vertex algebra, since:
\begin{equation}
\label{eqn:projection-creation}
(T_{\epsilon}a)(z)|0\rangle \arrowvert _{z=0}=a(\epsilon z)|0\rangle \arrowvert _{z=0}=a(z)|0\rangle \arrowvert _{z=0}=\pi(a).
\end{equation}
 Hence any such field $T_{\epsilon}a(z)$ cannot belong to a classical vertex algebra. This is the reason we require the  projection map $\pi$ to be a part of our data for a twisted vertex algebra, as well as we modify the  field-state correspondence with the \textbf{modified creation axiom}.
The projection map $\pi$ in the definition of a twisted vertex algebra $(V, W, \pi, Y)$ could be made more general, for example one can omit the requirement that $W\subset V$  and make $\pi$ a more general linear surjective projection map, but the current generality is sufficient for the examples we want to describe.

\subsection{Examples of twisted vertex algebras: the boson-fermion correspondence of type B}
\label{subsection:examplesB}

 We now proceed with the two examples of a twisted vertex algebra of order 2 which give the two sides of the boson-fermion correspondence of type B. This correspondence was first introduced in \cite{DJKM-4}, and was  interpreted as an isomorphism of twisted vertex algebras in  \cite{Ang-Varna2} (without proofs). Since this is a case of twisted vertex algebras of order two, the root of unity is $-1$, and we will write $T_{-1}$ instead of $T_{\epsilon}$. The proofs of all the statements and assertions in this section will be given in sections  \ref{section:freefermionB} and  \ref{section:freebosonB}.

 The fermionic side  is described via a single field $\phi ^B (z)=\sum _{n\in \mathbf{Z}} \phi^B_n z^{n}$, with OPE given by:
\begin{equation}
\label{equation:OPE-B}
\phi ^B(z)\phi ^B(w)\sim \frac{-2w}{z+w}, \quad \text{in \ modes:} \ [\phi^B_m,\phi^B_n]_{\dag}=2(-1)^m\delta _{m, -n}1.
\end{equation}
The  modes generate a Clifford algebra $\mathit{Cl_B}$, and the underlying space of states, denoted by $\mathit{F_B}$, of the twisted vertex algebra is a highest weight representation of $\mathit{Cl_B}$ with  the vacuum  vector $|0\rangle $, such that $\phi^B_n|0\rangle=0 \ \text{for} \  n<0$.   The space of fields, which is larger than the space of states,  is generated via the field $\phi ^B(z)$ together with its descendent $ T_{-1}\phi ^B(z)=\phi ^B(-z)$. We will prove in Section \ref{section:freefermionB} that there exists a twisted vertex algebra, denoted by $TVA (\mathit{F_B})$,  with a space of states $\mathit{F_B}$.
\begin{remark}\label{remark:shiftinB}
The  defining OPE, \eqref{equation:OPE-B},
shows why we adopted the modification of the completeness with respect to OPEs. The field defined by the residue at $z=-w$ in \eqref{equation:OPE-B} is
$c(w):=-2wId_W$. Note that $c(w)=-2wId_W$ is a field, but is not  a vertex operator in a super or twisted vertex algebra as  is. But a shift by $w^{-1}$ will produce the field  $w\cdot c(w)=-2Id_W$, which is the vertex operator $Y(-2|0\rangle, w)$ attached  to the state $-2|0\rangle$.
\end{remark}
\begin{remark} One often writes, especially in physics,  the OPE \eqref{equation:OPE-B} as
\begin{equation*}
\phi ^B(z)\phi ^B(w)\sim \frac{z-w}{z+w},
\end{equation*}
even though strictly speaking $\frac{z-w}{z+w}=\frac{-2w}{z+w} +1$ contains also a portion of the normal ordered product (the nonsingular part).
\end{remark}
\begin{lem}
\label{lem:VacExpFB}
The following formula for the vacuum expectation values on the fermionic side $\mathit{F_B}$ holds:
\begin{equation*}
\langle  0 |\phi ^B(z_1)\phi ^B(z_2)\dots \phi ^B(z_{2n})\mid 0 \rangle =i_{z}Pf \Big(\frac{z_i -z_j}{z_i +z_j}\Big)_{i =1}^{ 2n}.
\end{equation*}
Here $i_{z}$ stands for the expansion $i_{{z_1}, {z_1}, \dots , {z_{2n}}}$, and  $Pf$ denotes the Pfaffian of an antisymmetric matrix.
\end{lem}
This Pfaffian identity is well known, it was first derived in \cite{DJKM-4}; we will show if follows directly from  the bicharacter construction  in section \ref{section:freefermionB}.

The boson-fermion correspondence of type B is  determined once we write the image of the generating fields $\phi ^B (z)$ (and thus of $T_{-1}\phi ^B(z)=\phi(-z)$) under the correspondence. In order to do that, an essential ingredient is  the  so-called (see below) \textbf{twisted} Heisenberg field $h(z)$.
\begin{lem} (\cite{DJKM-4}, \cite{YouBKP}, \cite{Ang-Varna2}) Let
\begin{equation}
\label{eqn:normal-order-h-B}
h(z):=\frac{1}{4}(:\phi ^B(z)T_{-1}\phi ^B(z) :-1)= \frac{1}{4}(:\phi ^B(z)\phi ^B(-z): -1).
\end{equation}
The field $h(z)$ has only  odd-indexed modes,   $h(z)=\sum _{n\in \mathbb{Z}} h_{2n+1} z^{-2n-1}$, and  OPE:
\begin{equation}
\label{eqn:HeisOPEsB}
h(z)h(w)\sim \frac{zw(z^2 +w^2)}{2(z^2 -w^2)^2}.
\end{equation}
Its  modes, $h_n, \ n\in 2\mathbb{Z}+1$, generate a \textbf{twisted} Heisenberg algebra $\mathcal{H}_{\mathbb{Z}+1/2}$ with relations $[h_m,h_n]=\frac{m}{2}\delta _{m+n,0}1$, \ $m,n$--odd  integers; thus we call  the field $h(z)$  ``twisted Heisenberg field".
\end{lem}
 $\mathcal{H}_{\mathbb{Z}+1/2}$ has (up-to isomorphism) only one irreducible highest weight module $B_{1/2}\cong \mathbb{C}[x_1, x_3, \dots , x_{2n+1}, \dots ]$.
 \begin{lem} \label{lem:BosonicSpaceBB} (\cite{DJKM-4}, \cite{YouBKP}, \cite{Ang-Varna2})
The space of states $\mathit{F_B}$ can be decomposed as
\begin{equation}
\mathit{F_B}=B_{1/2} \oplus B_{1/2} \cong \mathbb{C}[\mathbb{Z}_2]\ten \mathbb{C}[x_1, x_3, \dots , x_{2n+1}, \dots ],
\end{equation}
where the  isomorphism is as  twisted Heisenberg modules for $\mathcal{H}_{\mathbb{Z}+1/2}$.  We  identify the group algebra $\mathbb{C}[\mathbb{Z}_2]$  with the polynomial algebra $\mathbb{C}[e_B^{\alpha}, e_B^{-\alpha}]$  with  the extra relation $e_B^{2\alpha}\equiv 1$. Denote $B_B:=\mathbb{C}[e_B^{\alpha}, e_B^{-\alpha}]\ten \mathbb{C}[x_1, x_3, \dots , x_{2n+1}, \dots ]$. In the isomorphism above the image of the vacuum vector $|0\rangle $ (a highest weight vector in $\mathit{F_B}$) is  $1\in B_B$. The image of the  second highest weight vector $\phi^B_0|0\rangle\in \mathit{F_B}$ is  $e_B^{\alpha}\in B_B$.
\end{lem}
  The (rather peculiar) group algebra notation above, as we will show and use later, is due to the Hopf algebra structure on $B_B$, which differs from the super Hopf algebra structure on $\mathit{F_B}$.  Moreover, we will show that there is a twisted vertex algebra structure, denoted by $TVA(B_B)$, such that $B_B$   is the  space of \textbf{states} of the bosonic side of the boson-fermion correspondence of type B.

The image of the  generating field $\phi ^B(z)$ (acting on $\mathit{F_B}$) is the field $ e_B^{\alpha}(z)$ (acting on $B_B$), which will determine the correspondence of type B:
\begin{lem} (\cite{Ang-Varna2}) The bosonization of type B is determined by
\begin{equation}
\label{eqn:imageB}
\phi ^B(z)\mapsto e_B^{\alpha}(z)=\exp \big(\sum _{k\ge 0}\frac{h_{-2k-1}}{k+1/2} z^{2k+1} \big)\exp \big(-\sum _{k\ge 0}\frac{h_{2k+1}}{k+1/2} z^{-2k-1} \big)e^{\alpha}.
\end{equation}
\end{lem}
The fields $e_B^{\alpha}(z)$ and  $e_B^{\alpha}(-z)=e_B^{-\alpha}(z)$ (observe the symmetry) will generate a resulting \textbf{twisted} vertex algebra, which we denote  by $TVA(B_B)$.
\begin{lem}
\label{lem:VacExpBB}
The following formula for the vacuum expectation values on the bosonic side $\mathit{B_B}$ holds:
\begin{equation*}
\langle  0 |e_B^{\alpha} (z_1)\dots e_B^{\alpha} (z_{2n}) | 0 \rangle =i_{z}\prod_{i<j}^{2n} \frac{ z_i-z_j}{z_i +z_j}.
\end{equation*}
Here $i_{z}$ stands for the expansion $i_{{z_1}, {z_1}, \dots , {z_{2n}}}$, and  $Pf$ denotes the Pfaffian of an antisymmetric matrix.
\end{lem}
Note that one Heisenberg  $\mathcal{H}_{\mathbb{Z}+1/2}$-module $B_{1/2}$ on its own can be realized as a \textbf{twisted module} for \ an ordinary super vertex algebra (see \cite{FLM}, \cite{BakKac}  for details), but the point is that we need \textbf{two} of them glued together for  the bosonic side of the correspondence. The two of them glued together as above no longer constitute a twisted module for an ordinary super vertex algebra.
\begin{thm}  (\cite{Ang-Varna2})
The boson-fermion correspondence of type B is an isomorphism between the  fermionic \textbf{twisted} vertex algebra $TVA(\mathit{F_B})$ and the bosonic \textbf{twisted} vertex algebra $TVA(B_B)$.
\end{thm}
\begin{cor} \label{lem:VacuumExpEqB} The Schur Pfaffian identity follows from the equality between the vacuum expectation values:
\begin{equation*}
AC \langle  0 |\phi ^B(z_1)\dots \phi ^B(z_{2n})| 0\rangle = P\!f \Big(\frac{z_i -z_j}{z_i +z_j}\Big)_{i, j =1}^{2n} =\prod_{i<j}^{2n} \frac{ z_i-z_j}{z_i +z_j} =AC \langle  0 |e_B^{\alpha} (z_1)\dots e_B^{\alpha} (z_{2n}) | 0 \rangle.
\end{equation*}
$AC$ stands for Analytic Continuation.
\end{cor}

\subsection{Examples of twisted vertex algebras: the boson-fermion correspondence of type D-A}
\label{subsection:examplesD}

Next are the two examples of a twisted vertex algebra of order 2 which give the two sides of the boson-fermion correspondence of type D-A.
The boson-fermion correspondence of type D-A is new, and the bosonisation of type D is one of the main results of this paper. This correspondence was discussed  during the work on both this  paper and on  \cite{Rehren}, where the  "multilocal fermionization" is discussed. The author thanks K. Rehren for the helpful discussions and a physics point of view on the subject. Here we  will show the bosonization of type D (including the split of the neutral fermion space into a direct sum of bosonic spaces) and prove that  the boson-fermion correspondence of type D-A is an  isomorphism of twisted vertex algebras.
The correspondence of type D-A  can be generalized to arbitrary order $N\in \mathbb{N}$ (see below, and also \cite{Rehren}), but we start with the case of twisted vertex algebras of order two. Again we write $T_{-1}$ instead of $T_{\epsilon}$.
The proofs of all the statements and assertions in this section are given in sections  \ref{section:freefermionD} and \ref{section:freebosonD}.

The fermionic side  is generated by a single field $\phi ^D(z)=\sum _{n\in \mathbf{Z+1/2}} \phi^D_n z^{-n-1/2}$ with OPE given by:
\begin{equation}
\label{equation:OPE-D}
\phi ^D(z)\phi ^D(w)\sim \frac{1}{z-w}, \quad \text{in \ modes:} \ [\phi^D_m,\phi^D_n]_{\dag}=\delta _{m, -n}1.
\end{equation}
The  modes generate a Clifford algebra $\mathit{Cl_D}$, with  underlying space of states denoted by $\mathit{F_D}$--- the  highest weight representation of $\mathit{Cl_D}$ with  the vacuum  vector $|0\rangle $, such that $\phi_n|0\rangle=0 \ \text{for} \  n<0$. Here it is recognized that on its own the field $\phi ^D(z)$ and its descendants $D^{(n)} \phi ^D(z)$ generate an (ordinary) super-vertex algebra (see e.g., \cite{Kac}, \cite{Wang}).
It is important to note that on its own this super-vertex algebra, called free neutral fermion vertex algebra, cannot be bosonized.
But, if we take not only the field $\phi ^D(z)$, but also its \textbf{twisted} descendant $ T_{-1}\phi ^D(z)=\phi ^D(-z)$, they together with all their  descendants will generate a \textbf{twisted} vertex algebra,  denoted  by $TVA(\mathit{F_D})$ which will be bosonized. We call  this twisted vertex algebra ``\textbf{free neutral fermion of type D-A}". The \textbf{space of fields} of the twisted vertex algebra $TVA(\mathit{F_D})$ strictly contains the super-vertex algebra $\mathit{F_D}$ as a subset, and we can think of it as an orbifolded double cover  of the super-vertex algebra.

The boson-fermion correspondence of type D-A is  determined once we write the images of the generating fields $\phi ^D (z)$ and  $T_{-1}\phi ^D(z)=\phi(-z)$ under the correspondence. In order to do that, an essential ingredient is once again the  Heisenberg field.
\begin{prop}\label{prop:HeisOPEsD} Let
\begin{equation}
\label{eqn:normal-order-h-D}
h^D(z):=\frac{1}{2}:\phi ^D(z)T_{-1}\phi ^D(z) := \frac{1}{2}:\phi ^D(z)\phi ^D(-z) :
\end{equation}
The field $h^D(z)$   has only  odd-indexed modes,   $h^D(z)=\sum _{n\in \mathbb{Z}} h^D_{n} z^{-2n-1}$ (note the  indexing), and  OPE:
\begin{equation}
\label{eqn:HeisOPEsD}
h^D(z)h^D(w)\sim \frac{zw}{(z^2-w^2)^2}.
\end{equation}
Its  modes, $h^D_n, \ n\in \mathbb{Z}$, generate an \textbf{untwisted}  Heisenberg algebra $\mathcal{H}_{\mathbb{Z}}$ with relations $[h^D_m,h^D_n]=m\delta _{m+n,0}1$, \ $m,n$--   integers; hence we call $h^D(z)$  a ``Heisenberg field".
\end{prop}
  Unlike the twisted Heisenberg algebra, the untwisted Heisenberg algebra  has infinitely many  irreducible highest weight modules, labeled by the action of $h^D_0$.
\begin{prop}\label{prop:decompD}
The space of states $\mathit{F_D}$  can be decomposed as
\begin{equation}
W=F_D \cong \oplus _{i\in \mathbb{Z}} B_i \cong  \mathbb{C}[e^{\alpha}, e^{-\alpha}] \ten \mathbb{C}[x_1, x_2, \dots , x_n, \dots ],
\end{equation}
$\mathbb{C}[e^{\alpha}, e^{-\alpha}]$ denotes the Laurent polynomials with one variable   $e^{\alpha}$.
The isomorphism above is as  Heisenberg modules, where $e^{n\alpha }$ denotes the highest weight vector for the irreducible Heisenberg module $B_n$, with highest weight $n$.
\end{prop}
We denote $B_D:=\mathbb{C}[e^{\alpha}, e^{-\alpha}] \ten \mathbb{C}[x_1, x_2, \dots , x_n, \dots ]$. We will show and use later  that $B_D$ has  a Hopf algebra structure that differs  from the super Hopf algebra structure on $\mathit{F_D}$.  Moreover, we will show that there is a twisted vertex algebra structure, denoted by $TVA(B_D)$, such that $B_D$   is the space of \textbf{states} of the bosonic side of the boson-fermion correspondence of type D-A.
\begin{prop}\label{prop:imageD}
The boson-fermion correspondence of type D-A is determined by the images of the  generating fields $\phi ^D(z)$ and $T_{-1}\phi ^D(z)=\phi(-z)$ (which act on $\mathit{F_D}$), as follows:
\begin{equation}
\label{eqn:imageD}
\phi ^D(z) \mapsto e_{\phi}^{-\alpha}(z) +e_{\phi}^{\alpha}(z), \quad (T\phi) ^D(z) = \phi(-z)\mapsto e_{\phi}^{-\alpha}(z) -e_{\phi}^{\alpha}(z),
\end{equation}
where $e_{\phi}^{-\alpha}(z)$ and  $e_{\phi}^{\alpha}(z)$ are fields on $B_D$ defined by the the formulas
\begin{align}
\label{eqn:ExponD-1}
e_{\phi}^{-\alpha}(z) & =\exp (-\sum _{n\ge 1}\frac{h_{-n}}{n} z^{2n})\exp (\sum _{n\ge 1}\frac{h_{n}}{n} z^{-2n})e^{-\alpha}z^{-2\partial_{\alpha}} =e^{-\alpha}_A(z^2),\\
\label{eqn:ExponD-2}
e_{\phi}^{\alpha}(z) & =\exp (\sum _{n\ge 1}\frac{h_{-n}}{n} z^{2n})\exp (-\sum _{n\ge 1}\frac{h_{n}}{n} z^{-2n})e^{\alpha}z^{2\partial_{\alpha}+1} =ze^{\alpha}_A(z^2),
\end{align}
\end{prop}
\begin{thm}
The boson-fermion correspondence of type D-A is the isomorphism between the  fermionic \textbf{twisted} vertex algebra $TVA(\mathit{F_D})$ and the bosonic \textbf{twisted} vertex algebra $TVA(B_D)$.
\end{thm}
\begin{remark}
The name "boson-fermion correspondence of type D-A" is given since the fermionic side is a double cover of the well know neutral fermion super vertex algebra that  gives  the basic representation of $d_{\infty}$, see e.g. \cite{WangKac}, \cite{Wang}. On the other hand
the operators $e^{\alpha}_A(z)$ and $e^{-\alpha}_A(z)$ in the right-hand side above are the vertex operators describing the boson-fermion correspondence of type A (\eqref{eqn:ExponA-1}, \eqref{eqn:ExponA-2}, as in  e.g. \cite{Kac}).
\end{remark}
\begin{cor} \label{lem:VacuumExpEqD} The following  Pfaffian identity follows from the equality between the vacuum expectation values:
\begin{align*}
& AC \langle  0 |\phi ^D(z_1)\dots \phi ^D(z_{2n})| 0\rangle   =P\!f \Big(\frac{1}{z_i -z_j}\Big)_{i, j =1}^{2n} =\frac{\sum_{i_1 <i_2\dots <i_n}^{2n} z_{i_1} z_{i_2}\cdots z_{i_n}  \prod_{k<l}^n (z_{i_k}^2 - z_{i_l}^2)\prod_{p < q}^n(z_{j_p}^2 - z_{j_q}^2)}{\prod_{k, p}^{n} (z_{i_k}^2 - z_{j_p}^2)}  \\
& \hspace{8cm}  =AC \langle  0 |(e_{\phi}^{-\alpha} (z_1)+e_{\phi}^{\alpha}(z_1))\dots (e_{\phi}^{-\alpha} (z_{2n})+e_{\phi}^{\alpha}(z_{2n})) | 0 \rangle
\end{align*}
$AC$ stands for Analytic Continuation, and $\{j_1, j_2, \dots j_n\}=\{1, 2, \dots 2n\}\setminus \{i_1, i_2, \dots i_n\}$.
\end{cor}

\subsection{Examples of twisted vertex algebras: the boson-fermion correspondence of type D-A of order $N$}
\label{subsection:examplesD-N}

The boson-fermion correspondence of type D-A from the previous section  can be generalized to  general order $N$ as follows. We again consider the free field $\phi ^D(z)=\sum _{n\in \mathbf{Z+1/2}} \phi^D_n z^{-n-1/2}$, with OPEs with itself given by
\eqref{equation:OPE-D}. Let $\epsilon$ be an  $N$-th order primitive root of unity  and consider
the following twisted vertex algebra descendants  $T_{\epsilon}^i\phi ^D (z)=\phi ^D (\epsilon ^i z)$, for any $0\le i\le N-1$, with OPEs
\begin{equation*}
T_{\epsilon}^i\phi ^D (z)T_{\epsilon}^j\phi ^D(w) \sim \frac{1}{\epsilon ^iz- \epsilon ^jw}.
\end{equation*}
Such  OPEs are not allowed in a super vertex algebra, but are allowed in a twisted vertex algebra.
These mean that on its own each of the fields $T_{\epsilon}^i\phi ^D (z)$, for any $0\le i\le N-1$ will generate a super vertex algebra, but the $N$ of them "glued"  together form a twisted vertex algebra. Again, this  resembles the gluing together of the $N$ sheets of the $N$-th root Riemann surface.
\begin{lem}
\label{lem:HeisOrderN}
The field $h(z)$ given by
\begin{equation}
\label{eqn:HeisOrderN}
h(z)=\frac{1}{N}\sum_{i=0}^{N-1} \epsilon^{i-1}  : T_{\epsilon}^{i-1}\phi^D (z) T_{\epsilon}^{i}\phi^D (z) : =\sum _{n\in \mathbb{Z}}h_n z^{-Nn-1}
\end{equation}
is a Heisenberg field with  OPE
\begin{equation}
\label{eqn:HeisOrderN-OPE}
h(z)h(w)\sim  \frac{z^{N-1}w^{N-1}}{(z^N -w^N)^2}.
\end{equation}
Thus  the commutation relations $[ h_m, h_n ] = m\delta _{m+n, 0} 1$ for the Heisenberg algebra $\mathcal{H}_{\mathbb{Z}}$  hold.
\end{lem}
\begin{lem} Let
\[
e_{\phi}^{\alpha} (w)=\frac{1}{N}(\sum_{i=0}^{N-1} \epsilon^{-i}T^{i}\phi ^D(w)), \quad e_{\phi}^{\epsilon^{k} \alpha}(w)=\frac{1}{N}(\sum_{i=0}^{N-1} \epsilon^{(k-1)i}T^{i}\phi ^D(w)).
\]
The boson-fermion correspondence of order $N$ is given by
\begin{equation}
e_{\phi}^{\epsilon^{k}  \alpha}(z)  \mapsto \exp (\epsilon^{-k} \sum _{n\ge 1}\frac{h_{-n}}{n} z^{Nn})\exp (\epsilon^{k} \sum _{n\ge 1}\frac{h_{n}}{n} z^{-Nn})e_{\phi}^{\epsilon^{k} \alpha}z^{1-k + N\partial_{\alpha}},
\end{equation}
 where $e_{\phi}^{\epsilon^{k} \alpha}$ is identified as  the  highest weight vectors of the Heisenberg submodule.
\end{lem}

\section{Bicharacter construction: constructing examples of twisted vertex algebras}

\subsection{Super bicharacters and free Leibnitz modules}
\label{subsection:super-bichLeibnitz}

   In this section we recall the components of the super-bicharacter construction (the super case was introduced in \cite{A2}, generalizing \cite{BorcQVA}).
\begin{notation} Henceforth we will assume that $M$ is a   supercommutative and supecocommutative bialgebra.
  To unclutter the language, we will just write commutative, cocommutative, omitting the term "super" as long as the parity is clear from the context.
\end{notation}
\begin{defn}\begin{bf}(Super-bicharacter)\end{bf}
\label{defn:bich}
Define a bicharacter  $r$ on $M$ to be
a linear map $r: M\ten M\to \mathbf{F}^N(z, w)$, such that
\begin{align}
\label{eq:iden}
r_{z_1,z_2}(1\ten a)&= \eta (a) = r_{z, w}(a\ten 1),\\
\label{eq:multleft}
r_{z, w}(ab\ten c)&=\sum \ (-1)^{\tilde{b} \tilde{c^{\prime}} } r_{z, w}(a\ten
c^{\prime})r_{z, w}(b\ten c^{\second}),\\
\label{eq:multright}
r_{z, w}(a\ten bc)&= \sum \ (-1)^{\tilde{a^{\second}} \tilde{b} } r_{z, w}(a^{\prime}\ten b)r_{z, w}(a^{\second}\ten c).
\end{align}
We say that a bicharacter $r$ is even if $r_{z, w}(a\ten b)=0$ whenever
$\tilde{a}\neq \tilde{b}$.
\end{defn}
From now on we will always work with \emph{even}  bicharacters. In most cases there  are no nontrivial bicharacters which are not even (see \cite{A2}). The identity bicharacter is given by $r(a\ten b)= \eta (a) \ten \eta
(b) $.
\begin{remark}
The notion of super bicharacter  is similar  to the notion of a twist induced by Laplace pairing (or the more general concept of a Drinfeld twist) as  described in \cite{Oeckl}.
\end{remark}
\begin{defn}\begin{bf}(Symmetric bicharacter)\end{bf}
The transpose bicharacter $r^\tau$ of a bicharacter  $r$ is defined by
\begin{equation}
r^\tau_{z, w}(a\otimes b)=r_{w, z}\circ \tau (a\ten b).
\end{equation}
A bicharacter $r$ is called \emph{symmetric} if $r=r^\tau$.
\end{defn}
\begin{defn}$(\mathbf{H^N_{T_{\epsilon}}\ten H^N_{T_{\epsilon}}}$\textbf{-covariant bicharacter)}
Let $M$ be a   Hopf supercommutative and supecocommutative superalgebra, $r$ be a bicharacter on $M$. Suppose in addition $M$ is an $H^N_{T_{\epsilon}}$-module algebra. We say that  the bicharacter $r$ is $H^N_{T_{\epsilon}}\ten H^N_{T_{\epsilon}}$-covariant if it  additionally satisfies :
 \begin{equation} r_{z, w}(h(a)\otimes g(b))=h_{z}g_{w}\cdot r_{z, w}(a\otimes b),
\end{equation} for all   $a,b\in M$, $h, g \in H^N_{T_{\epsilon}}$.
\end{defn}
We recall  the following result  from \cite{BorcQVA}, generalized to the super case:
\begin{lem/defn}
\label{lem/defn:hofm} \textbf{(Free} $\mathbf{H}-$\textbf{Leibnitz module)}
Suppose $M$ is a super commutative algebra and $H$ is an entirely even
cocommutative bialgebra. Then there is a universal
supercommutative algebra $H(M)$ such that there is a map $h\ten m \to
hm:=h(m)$ \  from $H\ten M$ to $H(M)$ such that $H(M)$ is a left module for $H$ and
\begin{equation}
\label{eq:modalgprop}
h(mn)=\sum h^{'}(m)h^{''}(n), \quad  h(1)=\eta (h),
\end{equation}
for any $ m,n \in M, \  h\in H$. We will call $H(M)$, defined as above, the "free $H$ Leibnitz module of $M$" (or universal $H$-Leibnitz module of $M$).
\end{lem/defn}
An $H$-module with the properties \eqref{eq:modalgprop}
   is by definition an  $H$-module algebra (see e.g. \cite{Kassel}), thus $H(M)$  is an $H$-module algebra. It is the  universal $H$-module algebra containing $M$ in the super-commutative category.
\begin{remark}
If $M$ is supercommutative and supercocommutative bialgebra (or
Hopf algebra), then so is $H(M)$. The extension of comultiplication and the counit  from $M$ to $H(M)$ is as follows:
If $a\in M, \ h\in H$ we have $ha\in H(M)$ and we define
\begin{equation}
\label{eq:freemodcoprd}
\del (ha)=\sum h^{\prime}a^{\prime}\ten h^{\second}a^{\second},\quad
\eta(ha)=\eta (h)\eta (a).
\end{equation}
It is easy to check that the comultiplication and  the counit  defined as above will turn $H(M)$ into a bialgebra.
\end{remark}
A source of examples of    $H^N_{T_{\epsilon}}\ten H^N_{T_{\epsilon}}$-covariant   bicharacters is as follows:
\begin{lem}\label{lem:ExtBich}(\cite{BorcQVA}) \begin{bf}(Extension of bicharacters)\end{bf}  Let $r: M\ten M \to \mathbf{F}^N(z, w)$ be a bicharacter on $M$. Then there exist an $H^N_{T_{\epsilon}}\otimes H^N_{T_{\epsilon}}$-covariant bicharacter on the free Leibnitz module $H^N_{T_{\epsilon}}(M)$, such  that its restriction to $M$ is the bicharacter $r$.
\end{lem}
We will denote the ``induced" bicharacter  also by $r$.\\
{\it Proof:}
To define the bicharacter on $H^N_{T_{\epsilon}}(M)$ we note that all elements in $H^N_{T_{\epsilon}}(M)$ are generated, as an algebra, from elements of the form $a=h\bar{a}, \ b=g\bar{b}$ for some $\bar{a}, \bar{b} \in M$, $g, h\in H$. Thus define a bicharacter
\begin{equation}
r: H^N_{T_{\epsilon}}(M)\ten H^N_{T_{\epsilon}}(M) \to \mathbf{F}^N(z, w)\quad \text{by}\quad r_{z, w}(a\ten b)=h_{z}g_w\cdot r_{z, w}(\bar{a}\ten \bar{b})
\end{equation}
and extend it by linearity and using multiplicativity (\eqref{eq:multleft} and \eqref{eq:multright}) of the bicharacter to the whole of $H^N_{T_{\epsilon}}(M)$. The extended bicharacter $r$ on $H^N_{T_{\epsilon}}(M)$ is by construction $H^N_{T_{\epsilon}}\otimes H^N_{T_{\epsilon}}$-covariant.$\square$

\subsection{Examples of free Leibnitz modules}
\label{subsection:freeLeModEx}

Most of the vector spaces  underlying our vertex algebras in this paper are going to be free Leibnitz modules.
The first two types of examples of free Leibnitz modules are entirely even, or \textbf{bosonic}.
\begin{example}\label{example:HeisLeibnitzMod}\begin{bf}(The free Leibnitz modules $H_D(\mathbb{C}[h])$ and $H^N_{T_{\epsilon}}(\mathbb{C}[h])$)\end{bf}\\
The free $H_D$-Leibnitz  module  over the  algebra $\mathbb{C}[h]$ (the polynomial algebra of a single variable, considered as a Hopf algebra with $h$  a primitive element)  is  isomorphic to the polynomial algebra $\mathbb{C}[x_{1},x_{2},\dots, x_n, \dots]$.
We can identify $x_1=h$ and  $D^{(n)} h =D^{(n)}x_1 =x_{n+1}$.
From equation \eqref{eq:D^n}, we have that  $x_n = D^{(n-1)} h$ are  primitive: according to \eqref{eq:freemodcoprd}
\begin{equation}
\del(D^{(l)}h)=(\sum_{p+q=l} D^{(p)} \ten D^{(q)})(h\ten 1 +1\ten h) = D^{(l)}h\ten 1 +1\ten D^{(l)}h,
\end{equation}
one similarly checks the counit and the anitpode.
These variables commute and   generate $H_D(\mathbb{C}[h])$. Thus
 $\mathbb{C}[x_{1},x_{2},\dots, x_n, \dots]$ is  isomorphic to the free $H_D$ module-algebra over $\mathbf{C}[h]$.

The free Leibnitz module $H^N_{T_{\epsilon}}(\mathbb{C}[h])$ is  isomorphic as a Hopf algebra to the polynomial algebra with $k$ groups of variables:
\mbox{$\mathbb{C}[x^0_{1},x^0_{2},\dots, x^0_n, \dots, x^{N-1}_{1},x^{N-1}_{2},\dots, x^{N-1}_n, \dots, ]$},  by identifying
\begin{equation}
x^k_l =T^kD^{(l)}h, \quad k=0, \dots, N-1, \quad  n=0, 1, \dots, l, \dots.
\end{equation}
\end{example}
\begin{example}\label{example: LeibntizFreeAb} \begin{bf}(The free Leibnitz modules over free abelian group algebras)\end{bf}\\
Let $L_1=\mathbb{C}[\mathbb{Z}\alpha]$ be the group algebra of the rank-one free abelian group $\mathbb{Z}\alpha $. The group algebra is generated by $e^{m\alpha}, \ m\in \mathbb{Z}$, \ with relations \mbox{$e^{m\alpha}e^{n\alpha }=e^{(m+n)\alpha}$}, \  $e^{0}=1$. Note that as an algebra $L_1=\mathbb{C}[e^{\alpha}, e^{-\alpha}]$, with  the relation \mbox{$R: \ e^{\alpha}e^{-\alpha}=1$}. The  Hopf algebra structure is determined by  $e^{\alpha}$ and $e^{-\alpha}$ being grouplike.
\begin{lem}
The free $H_D$ Leibnitz module $H_D(L_1)$ is isomorphic as an algebra to \mbox{$L_1\ten  \mathbb{C}[h]$.}
\end{lem}
{\it Proof:}
Since $H_D(L_1)$ is a free Leibnitz module, we can define an element  $h=(De^\alpha)e^{-\alpha}$.  It follows that $h$ is primitive, we have
\begin{equation*}
\Delta(h)=\Delta(De^\alpha)\Delta (e^{-\alpha}) =(De^{\alpha}\ten e^{\alpha} +(e^{\alpha}\ten De^{\alpha}))(e^{-\alpha}\ten e^{-\alpha})=
h\otimes 1+1\otimes h.
\end{equation*}
Similarly one checks that $\epsilon(h)=0$. Thus, since $H_D(L_1)$ is a free Leibnitz module, it has the subalgebra $H_D(\mathbf{C}[h])=\mathbf{C}[x_{1},x_{2},\dots, x_n, \dots ]$. Thus, $L_1\ten H_D(\mathbf{C}[h])$
is (isomorphic to) a subalgebra in $H_D(L_1)$. Conversely, for any $m\in \mathbf{Z}$, since
$De^{\alpha}=h\cdot e^{\alpha}$
we can write
\begin{equation*}
De^{m\alpha}=D(e^{\alpha})^m=m(De^{\alpha})(e^{\alpha})^{m-1}=m(he^{\alpha})(e^{\alpha})^{m-1}=mh\cdot e^{m\alpha}
\end{equation*}
 thus for any $m, n\in \mathbf{Z}$ the element $D^n(e^{m\alpha})$ is in $M\ten H_D(\mathbf{C}[h])$, and so $H_D(L_1)$ is (isomorphic to) a subalgebra in $M\ten H_D(\mathbf{C}[h])$.
Thus $H_D(L_1)$ is isomorphic to $L_1\ten \mathbb{C}[h] \simeq L_1\ten \mathbf{C}[x_{1},x_{2},\dots, x_n, \dots ], \ n\in \mathbf{N}$.
$\square$

Note that the primitive element $h$ is of particular importance for the boson-fermion correspondences, as it generates a Heisenberg subalgebra  for a variety of  choices for a bicharacter.

The free Leibnitz module $H^N_{T_{\epsilon}}(L_1)$ is isomorphic to $L_N\ten H^N_{T_{\epsilon}}(\mathbb{C}[h])$, where
$L_N$ is the group algebra $L_N=\mathbb{C}[\mathbb{Z}\alpha_1, \mathbb{Z}\alpha _2, \dots \mathbb{Z}\alpha _N]$ of the  free abelian group of rank N (one can identify $T^ke^{\alpha}$, which is grouplike, with $e^{\alpha _k}$).

One  proceeds similarly with the free Leibnitz modules over the  free abelian group of any rank.
\end{example}
The other examples we will use throughout the paper are \textbf{fermionic}, or super algebras.
\begin{example}\label{example:Cphi} \begin{bf}(The free Leibnitz modules $H_D(\mathbb{C}\{\phi \})$ and $H^N_{T_{\epsilon}}(\mathbb{C}\{\phi \})$)\end{bf} \\
Let $\mathbb{C}\{\phi \}$ be the Grassmann algebra  generated by one odd primitive element $\phi $, $\phi \cdot \phi =0 $.
Then the free Leibnitz module $H_D(\mathbb{C}\{\phi \})$ is the Grassmann algebra with odd anticommuting generators $\phi ^n =D^{(n)}\phi$, $\phi ^n\phi ^m + \phi ^m\phi ^n =0$,
which one checks to be primitive.

Similarly, the free Leibnitz module $H^N_{T_{\epsilon}}(\mathbb{C}\{\phi \})$ is the Grassmann algebra with odd anticommuting generators
\begin{equation}
\phi ^{n, k} =D^{(n)}T_{\epsilon}^k\phi, \quad k=0, \dots, N-1, \quad  n=0, 1, \dots, l, \dots.
\end{equation}
Note that  the ordering of the operators $D^{(n)}$ and $T_{\epsilon}^k$ matters; one should be careful to be consistent, as it may result in  rescaling of the basis: $\phi ^{n, k} =D^{(n)}T_{\epsilon}^k\phi =\epsilon ^{kn} T_{\epsilon}^kD^{(n)}\phi$.

 Of particular interest for the boson-fermion correspondences is going  the case of $N=2$: the free Leibnitz module $H^2_{T_{\epsilon}}(\mathbb{C}\{\phi \})$ is algebra-isomorphic to $H_D(\mathbb{C}\{\phi , T\phi \})$, where  write $T=T_{\epsilon} =T_{-1}$ ($\epsilon=-1$ in this case). Both the boson-fermion correspondences (the type B, and the type D-A) have $H_D(\mathbb{C}\{\phi , T\phi \})$ as underlying space of fields on its fermionic side.
Of particular interest is the element
\begin{equation}
h_{\phi}=\phi T\phi, \quad \text{with} \quad Th_{\phi}=-h_{\phi}.
\end{equation}
This element is even, and   although it is not primitive, we will see in the later sections that the element $h_{\phi}$ generates a Heisenberg subalgebra for particular choices of  bicharacter.
\end{example}
The last example of free Leibnitz modules in this paper is the following:
\begin{example}\label{example:Cphipsi}\begin{bf}(The free Leibnitz modules $H_D(\mathbb{C}\{\phi , \psi\})$ and $H^N_{T_{\epsilon}}(\mathbb{C}\{\phi ,\psi \})$)\end{bf} \\
Let $\mathbb{C}\{\phi , \psi \}$ be the Grassmann algebra  generated by two  odd primitive element $\phi , \psi$, such that $\phi \cdot \phi  =\psi \cdot \psi =\phi \psi +\psi \phi =0 $.
Then the free Leibnitz module $H_D(\mathbb{C}\{\phi , \psi \})$ is the Grassmann algebra with odd anticommuting primitive generators $\phi ^n =D^{(n)}\phi$, $\psi ^n =D^{(n)}\psi$. The $H_D(\mathbb{C}\{\phi , \psi \})$ is the underlying space on the fermionic side of the boson-fermion correspondence of type A.
Of particular interest is the element $h_{\phi, \psi}=\phi \psi$, which is again even, but not primitive, and has similar coproduct as the element $h_{\phi}$ above.
$h_{\phi, \psi}$  also generates a  Heisenberg subalgebra for particular choices of the bicharacter as we will see in the later sections (an example appeared in \cite{A2}).
Similarly $H^N_{T_{\epsilon}}(\mathbb{C}\{\phi ,\psi \})$ is isomorphic to $H_D(\mathbb{C}\{\phi , T_{\epsilon}\phi,  \dots , T_{\epsilon}^{N-1}\phi , \psi , T_{\epsilon}\psi,  \dots , T_{\epsilon}^{N-1}\psi \})$.
\end{example}

\subsection{Exponential map and its properties; Nonsingular twisted vertex algebras}

\begin{defn} \label{defn:HolomVA}\begin{bf}(Nonsingular vertex algebra)\end{bf} We call a  twisted vertex algebra   nonsingular if the analytic continuations $X_{z, w, 0}(a\ten b\ten c)$  have no poles for any $a, b, c \in V$.
\end{defn}
For super vertex algebras this definition coincides with the notion of a holomorphic super vertex algebra introduced in the previous literature (see for example \cite{Kac},  \cite{LiLep}).
 A holomorphic super-vertex algebra is in fact just a commutative associative unital differential algebra: if $V$ is a holomorphic super-vertex algebra, for any $a, b \in V$ we have (see for example \cite{Kac},  \cite{LiLep}):
\begin{equation*}
Y(a, z)b=(e^{zD}a)b, \quad \text{where} \quad e^{zD}=\sum_{n\ge 0}z^n D^{(n)}.
\end{equation*}
Thus the fields in a holomorphic super-vertex algebra are uniquely determined by the unique "exponential map" $e^{zD}$.
In the case of twisted vertex algebras the situation is not as simple, as  the "exponential map" is not unique. There are a variety of examples of "exponential maps" for twisted vertex algebras  that would satisfy the properties of a nonsingular twisted vertex algebra. Note that for twisted vertex algebras, we also have the concept of a projection map, thus we really  consider \textbf{pairs} of exponential and projection maps that satisfy the properties of a nonsingular twisted vertex algebra.
Hence  we will proceed to  define  a standard pair of projection and exponential maps on  given free $H^N_{T_{\epsilon}}$ Leibnitz modules.
In what follows let $M$ be a commutative cocommutative Hopf algebra,  $V$ be the free Leibnitz module $V=H^N_{T_{\epsilon}}(M)$.
Note that the free Leibnitz module $W=H_D(M)$ is a sub-Hopf algebra of $V$, and thus we can use the exponential map $e^{zD}$ on $W$. Moreover, each element  in $V$ can be written uniquely as a linear combination of elements of the form $a=\prod _{i=0}^{N-1} a_i$, where $a_i=T_{\epsilon}^i\bar{a}_i$, \ for some $\bar{a}_i\in W$.
\begin{defn}\label{defn:TprojMap}\begin{bf}(T-projection Map $\mathbf{\pi_T}$)\end{bf}
Let $M$ be a commutative cocommutative Hopf algebra, let $V$ be the free Leibnitz module $V=H^N_{T_{\epsilon}}(M)$, $W=H_D(M)$,  and let $a\in V$ is such that $a=\prod _{i=0}^{N-1} a_i$, where $a_i=T_{\epsilon}^i\bar{a}_i$  for some $\bar{a}_i\in W$. Define the projection map $\pi_T:V\to W$ to be the algebra homomorphism map defined by:
\begin{equation}
\pi_T(a_i)=\bar{a}_i, ,\ \ i=1, \dots, N-1, \quad \pi_T(a)= \prod _{i=0}^{N-1} \bar{a}_i.
\end{equation}
\end{defn}
\begin{defn} \label{defn:ExpMap}\begin{bf}(Exponential Map $\mathbf{\mathcal{E}_z}$)\end{bf}
Let as above $a_i=T_{\epsilon}^i\bar{a}_i, \ \bar{a}_i\in W$, $a=\prod _{i=0}^{N-1} a_i$.
Define the map $\mathcal{E}_z : V\to W[[z]]$ to be the algebra homomorphism map such that
\begin{align}
&\mathcal{E}_z (\bar{a}_i)=e^{zD}\bar{a}_i , \ \ \text{for any} \ \ \bar{a}_i\in W\\
&\mathcal{E}_z (a_i)=e^{\epsilon^i zD}\bar{a}_i,   \  \ i=0, \dots, N-1;\\
&\mathcal{E}_z (\prod _{i=0}^{N-1} a_i)=\prod _{i=0}^{N-1}e^{\epsilon^i zD}\bar{a}_i;
\end{align}
and extend $\mathcal{E}_z$ by  linearity to  $V$.
\end{defn}
\begin{example}
\label{example:EpsilonsAndExponential}
This example points out that one has to be careful with the specific order between $D^{(n)}$ and $T_{\epsilon}^i$ which we implicitly used in the definition of  the exponential map $\mathcal{E}_z$. Let as above $a_i=T_{\epsilon}^i\bar{a}_i, \ \bar{a}_i\in W$, and let $n\in \mathbb{N}$. To calculate $\mathcal{E}_z (D^{(n)}a_i)$, we need
\begin{equation}
\label{eq:D and exp}
D^{(n)}a_i=D^{(n)}T_{\epsilon}^i\bar{a}_i=\epsilon^{ni}T_{\epsilon}^iD^{(n)}\bar{a}_i;
\end{equation}
where now $D^{(n)}\bar{a}_i\in W$, and thus
\begin{equation}
\mathcal{E}_z (D^{(n)}a_i)= \epsilon^{ni}e^{\epsilon^i zD}(D^{(n)}\bar{a}_i)=\epsilon^{ni}D^{(n)}(e^{\epsilon^i zD}\bar{a}_i)=\epsilon^{ni}D^{(n)}\mathcal{E}_z (a_i).
\end{equation}
This equality  plays part in the "transfer of action" property of $\mathcal{E}_z$, and is  also used in proving the Modified Expansion property of the exponential map.
\end{example}
\begin{lem}
\label{lem:expProp}\begin{bf}(Properties of the Exponential Map $\mathbf{\mathcal{E}_z}$)\end{bf}
Let $M$ be a commutative cocommutative Hopf algebra, let $V$ be the free Leibnitz module $V=H^N_{T_{\epsilon}}(M)$,  $r$ is a $H^N_{T_{\epsilon}}\otimes H^N_{T_{\epsilon}}$-covariant bicharacter on $V$. Let $\pi_T: V\to W$ and \mbox{$\mathcal{E}_z: V\to W[[z]]$} be the pair projection-exponential  map defined above.
This pair of maps  satisfies the  following properties:
\begin{itemize}
\item Vacuum property: $\mathcal{E}_z(1)=1$, where $1$ is the unit in $V$;
\item Modified creation property: $\mathcal{E}_z(a)\arrowvert _{z=0}=\pi_T(a)$, for any $a\in V$;
\item Transfer of action: $\mathcal{E}_z(ha)=h_z \mathcal{E}_z(a)$, for any $h\in H^N_{T_{\epsilon}}$, $a\in V$;
\item Multiplicativity: $\mathcal{E}_z(ab)=\mathcal{E}_z(a)\mathcal{E}_z(b)$, for any $a, b\in V$;
\item Grouplike: $\del \mathcal{E}_z(a)=\mathcal{E}_z(a^{\prime})\ten \mathcal{E}_z(a^{\prime\prime})$;
\item Compatibility with bicharacters: $i_{z, w}r_{z, w}(a\ten b)=r_{z, 0}(a\ten \mathcal{E}_w(b))$, for any $a, b\in V$.
\item Modified expansion: $\mathcal{E}_z(a)=\sum _{n\ge 0} (z-\epsilon ^i w)^n \mathcal{E}_w (T^iD^{(n)}a),$
\end{itemize}
\end{lem}
{\it Proof:}
The proofs for most of the properties  are straightforward, and use the similar  properties  of the ordinary exponential map $e^{zD}$ and the definition of the map $\mathcal{E}_z$ via the projection map.
$\square$
\begin{remark}
Note that as we saw in the  example \ref{example:EpsilonsAndExponential} above $D^{(n)}$ and $\mathcal{E}_w$ do not commute. Thus even though $\sum _{n\ge 0} (z-\epsilon ^i w)^n \mathcal{E}_w (T^iD^{(n)}a) = \sum _{n\ge 0} \epsilon ^{-in}(z-\epsilon ^i w)^n \mathcal{E}_w (D^{(n)}Ta)$, and $\sum _{n\ge 0} \epsilon ^{-in}(z-\epsilon ^i w)^n D^{(n)}=e^{\epsilon ^{-i}(z-\epsilon ^{-i} w)D}$,  the modified expansion property can \textbf{not}  be rewritten using $e^{\epsilon ^{-i}(z-\epsilon ^{-i} w)D}$. That is the reason for the lack of a "modified associativity" property.
\end{remark}
\begin{lem} \label{lem:TwHolomVA}\begin{bf}(Nonsingular twisted vertex algebra)\end{bf}  Let $M$ be a commutative cocommutative Hopf algebra, let $V$ be the free Leibnitz module $V=H^N_{T_{\epsilon}}(M)$,  let $\pi_T: V\to W$ be the projection map  from definition \ref{defn:TprojMap}.
The map \mbox{$\mathcal{E}_z: V\to W[[z]]$} defines a structure of a nonsingular twisted vertex algebra $(V, W, \pi=\pi_T, Y)$ by:
\begin{equation}
Y(a, z)\pi_T(b)=\mathcal{E}_z (a)\cdot \pi_T(b)\quad \text{for any} \quad a, b\in V.
\end{equation}
\end{lem}

\subsection{Vertex operators, analytic continuations, OPEs and normal ordered products from a bicharacter}

In this subsection we combine together the bicharacter construction to produce fields and vertex algebras from a bicharacter.  A space of fields in the vertex algebra is given by the free Leibnitz module $V=H^N_{T_{\epsilon}}(M)$, a  space of states by the free Leibnitz module $W=H_D(M)\subset V$.  This is part of the data  needed for a twisted vertex algebra. Now we will define the fields and the field-state correspondence  via a bicharacter.
\begin{defn} \label{defn:DefSingmult}\begin{bf}(Two-variable fields from a bicharacter)\end{bf}
Let $M$ be a commutative cocommutative Hopf algebra, let $V$ be the free Leibnitz module $V=H^N_{T_{\epsilon}}(M)$,  $r$ a $H^N_{T_{\epsilon}}\otimes H^N_{T_{\epsilon}}$-covariant bicharacter on $V$ with values in $\mathbf{F}^N(z, w)^{+, w}$, $W=H_D(M)$ be the free $H_D$-Leibnitz sub-module-algebra of $V$. Let $\mathcal{E}_z$ be the exponential map $\mathcal{E}_z: V\to W[[z]]$ defined in  \ref{defn:ExpMap}.
Define  a  singular multiplication map
\begin{equation}
X_{z, w}\colon V^{\otimes 2}\to
W[[z, w]]\otimes\mathbf{F}^N(z, w),
\end{equation}
by
\begin{equation}
\label{eq:two-varX}
X_{z, w}(a\otimes b)=\sum (-1)^{\tilde{a^{\prime\prime}} \tilde{b^{\prime}} } (\mathcal{E}_za^\prime )(\mathcal{E}_wb^\prime )r_{z, w}(a^{\prime\prime}\otimes b^{\prime\prime}),
\end{equation}
where  $a, b$ are homogeneous elements of the super space $V$. The map $X_{z, w}$ is extended by linearity to $V$.
\end{defn}
\begin{defn}[\textbf{Vertex operators $Y(a, z)$ and field-state correspondence}]\label{def:vertexoperatorY}
 Let $V$, $W$, $\mathcal{E}_z$ be as above, $\pi_T: V\to W$ be the projection map defined in \ref{defn:TprojMap}. Define the vertex
  operator $Y(a,z)$ associated to a homogeneous element $a\in V$, by
\begin{equation}
  \label{eq:DefY}
  Y(a,z)\pi_T(b)=X_{z,0}(a\otimes b)=\sum (-1)^{\tilde{a^{\prime\prime}} \tilde{b^{\prime}} } (\mathcal{E}_za^\prime )\pi_T(b^\prime )r_{z, 0}(a^{\prime\prime}\otimes b^{\prime\prime}),
\end{equation}
  for any homogeneous element $b\in V$. Extend by  linearity to any elements $a, b\in V$. Then   $Y(a, z)$ is a field on $W$  and the map $Y: a\in V \to Y(a, z)$  is a field-state correspondence for  the twisted vertex algebra with space of fields $V$ and space of states $W$.
\end{defn}
Here we are implicitly  using the modified creation property of the exponential map (lemma \ref{lem:expProp}).
\begin{remark}
This definition is consistent, i.e., for each $\bar{b}\in W$ the vertex operator acting on $\bar{b}$ is independent from the choice of the $b\in V$ used in the definition, due to the following: If $\bar{b}=\pi_T(b_1)=\pi_T(b_2)$, then from the $H^N_{T_{\epsilon}}\otimes H^N_{T_{\epsilon}}$-covariance of the  bicharacter $r$ it follows that
$r_{z, 0}(a\otimes b_1)=r_{z, 0}(a\otimes b_2)=r_{z, 0}(a\otimes \bar{b})$. Also, since the map $\pi_T(b)$ is a surjection, this definition is sufficient for any $\bar{b}\in W$.
\end{remark}
\begin{remark} Note that \textbf{any} $H^N_{T_{\epsilon}}\otimes H^N_{T_{\epsilon}}$-covariant bicharacter $r$ will produce a (different) field-state correspondence, thus even with the same space of fields $V$ and space of states $W$ we can get a variety of examples of field-state correspondences by choosing different bicharacters on $V$.
\end{remark}
\begin{lem} \label{lem:n=2AnCont}$\mathbf{(n=2}$ \textbf{Analytic continuation)}
Let $V$, $W$, $\mathcal{E}_z$, $\pi_T: V\to W$ be as above. We have for any $a, b\in V$
\begin{align}
 \label{eq:YmodCr}
 & Y(a, z)1=\mathcal{E}_za,\\
  \label{eq:YvsX}
 & i_{z, w} X_{z,w}(a\otimes b)=Y(a, z)\mathcal{E}_wb =Y(a, z)Y(b, w)1.
\end{align}
\end{lem}
{\it Proof:}
\eqref{eq:YmodCr} follows from the vacuum property of the exponential map (see lemma \ref{lem:expProp}). Note that since $W=H_D(M)$ is a Hopf subalgebra of $V$, the unit 1 is in $W$. Thus we have for $a\in V$  a homogeneous element
\begin{equation*}
Y(a, z)1 =\sum (-1)^0(\mathcal{E}_za^\prime )1r_{z, 0}(a^{\prime\prime}\otimes 1)=\sum (\mathcal{E}_za^\prime )\eta(a^{\prime\prime})=\mathcal{E}_za.
\end{equation*}
The equation \eqref{eq:YvsX} follows from the "compatibility with bicharacters" and the "grouplike" properties of the exponential map (see lemma \ref{lem:expProp}):
\begin{align*}
i_{z, w} X_{z,w}(a\otimes b) &=\sum (-1)^{\tilde{a^{\prime\prime}} \tilde{b^{\prime}} } (\mathcal{E}_za^\prime )(\mathcal{E}_wb^\prime )i_{z, w}r_{z, w}(a^{\prime\prime}\otimes b^{\prime\prime})  =\sum (-1)^{\tilde{a^{\prime\prime}} \tilde{b^{\prime}} } (\mathcal{E}_za^\prime )(\mathcal{E}_wb^\prime )r_{z, 0}(a^{\prime\prime}\otimes \mathcal{E}_w b^{\prime\prime})  \\
& =\sum (-1)^{\tilde{a^{\prime\prime}} \tilde{b^{\prime}} } (\mathcal{E}_za^\prime )(\mathcal{E}_wb)^\prime r_{z, 0}(a^{\prime\prime}\otimes (\mathcal{E}_w b)^{\prime\prime})= Y(a, z)\mathcal{E}_wb,
\end{align*}
since $\pi_T(\mathcal{E}_wb)=\mathcal{E}_wb$ for any $b\in V$.
$\square$

Thus we have established a field-state correspondence $Y$, and in fact have also proved that it satisfies the analytic continuation property of twisted vertex algebras for $n=2$. It is immediate that this  field-state correspondence satisfies also the following required properties for a twisted vertex algebra:
\begin{lem} \label{lem:FirstPropTVA}
Let $V$, $W$, $\mathcal{E}_z$, \mbox{$\pi=\pi_T: V\to W$}, $Y: a\in V \to Y(a, z)$ be as above. This data satisfies the following properties:
\begin{itemize}
\item vacuum axiom: \ \ $Y(1, z)=Id_W$;
\item modified creation axiom: \ \ $Y(a, z)|0 \rangle \arrowvert _{z=0}=\pi(a)$, for any $a\in V$;
\item transfer of action:\ \  $Y(ha,z)=h_z\cdot Y(a, z)$ for any $h\in H^N_{T_{\epsilon}}$.
\end{itemize}
\end{lem}
{\it Proof:} The vacuum property and the modified creation property follow immediately from the corresponding properties of the exponential map. For the transfer of action property:
\begin{equation*}
Y(ha, z)\pi_T(b)=X_{z,0}(ha\otimes b)=\sum (-1)^{\tilde{a^{\prime\prime}} \tilde{b^{\prime}} } (\mathcal{E}_z(ha)^\prime )\pi_T(b^\prime )r_{z, 0}((ha)^{\prime\prime}\otimes b^{\prime\prime}),
\end{equation*}
which from the property of free Leibnitz modules equals
\begin{equation*}
\sum (-1)^{\tilde{a^{\prime\prime}} \tilde{b^{\prime}} } (\mathcal{E}_z(h^\prime a^\prime )\pi_T(b^\prime )r_{z, 0}(h^{\prime\prime}a^{\prime\prime}\otimes b^{\prime\prime}).
\end{equation*}
From the transfer of action property of the exponential map, and the covariance of the bicharacter this equals
\begin{align*}
\sum (-1)^{\tilde{a^{\prime\prime}} \tilde{b^{\prime}} } \big((h^\prime)_z\cdot \mathcal{E}_z(a^\prime )\big)\pi_T(b^\prime )&\big((h^{\prime\prime})_z \cdot r_{z, 0}(a^{\prime\prime}\otimes b^{\prime\prime})\big) \\
&=h_z\cdot \big(\sum (-1)^{\tilde{a^{\prime\prime}} \tilde{b^{\prime}} } (\mathcal{E}_z(a)^\prime )\pi_T(b^\prime )r_{z, 0}((a)^{\prime\prime}\otimes b^{\prime\prime})\big)=h_z \cdot \big(Y(a, z)\pi_T(b)\big). \square
\end{align*}
One of the main advantages  of the bicharacter construction is that there are explicit formulas for all the analytic continuation maps $X_{z_1,\dots,z_n}$ in terms of the bicharacter, similar to the formula \eqref{eq:two-varX}. We will start with  the formula for $X_{z_1, z_2, z_3}$  as it is needed also for the Operator Product Expansions.
\begin{defn} \label{defn:DefSingmult3}\begin{bf}(Three-variable fields from a bicharacter)\end{bf}
Let $V$, $W$, $\mathcal{E}_z$, \mbox{$\pi_T: V\to W$} be as above.  Let  $a, b, c$ be arbitrary homogeneous elements of the super space $V$.
Define  the three variable field
\begin{equation}
\label{eq:three-varX}
X_{z_1, z_2, z_3}\colon V^{\otimes 3}\to
W[[z_1, z_2, z_3]]\otimes\mathbf{F}^N(z_1, z_2, z_3),
\end{equation}
by
\begin{equation*}
 X_{z_1, z_2, z_3}(a\ten b\ten c)  = \sum (-1)^{f(\tilde{a}, \tilde{b},\tilde{c}) }\mathcal{E}_{z_{1}}a^{(1)}\mathcal{E}_{z_{2}}b^{(1)}\mathcal{E}_{z_{3}}c^{(1)} r_{z_1,z_2}(a^{(2)}\otimes b^{(2)})r_{z_1,z_3}(a^{(3)}\otimes c^{(2)})r_{z_2,z_3}(b^{(3)}\otimes c^{(3)}),
\end{equation*}
where $f(\tilde{a}, \tilde{b}, \tilde{c})=\tilde{b^{(3)}}(\tilde{c^{(1)}}+\tilde{c^{(2)}}) + (\tilde{a^{(2)}}+ \tilde{a^{(3)}})(\tilde{b^{(1)}}+\tilde{c^{(1)}})+\tilde{a^{(3)}}\tilde{b^{(2)}}+\tilde{b^{(2)}}\tilde{c^{(1)}}$.
Here as usual we denote $\Delta ^2(a)=a^{(1)}\ten a^{(2)}\ten a^{(3)}$ for any $a\in V$. The map $X_{z_1, z_2, z_3}$ is  extended  to  the whole of $V$ by linearity.
\end{defn}
\begin{lem}[$\mathbf{n=3}$\textbf{ Analytic continuation}]
\label{lem:analcont-3}
Let $V$, $W$, $\mathcal{E}_z$, $\pi_T: V\to W$ be as above. We have for any $a, b, c\in V$
\begin{equation*}
i_{z_1, z_2, z_3} X_{z_1, z_2, z_3}(a\otimes b\ten c)=Y(a, z_1)Y(b, z_2)\mathcal{E}_{z_{3}}c =Y(a, z_1)Y(b, z_2)Y(c, z_3)1.
\end{equation*}
\end{lem}
{\it Proof:}
Again from the "compatibility with bicharacters" and the "grouplike" property of the exponential map, and  Definition \ref{def:vertexoperatorY} we have
\begin{align*}
&i_{z_1, z_2, z_3} X_{z_1, z_2, z_3}(a\otimes b\ten c) \\
& = \sum (-1)^{f(\tilde{a}, \tilde{b},\tilde{c}) }\mathcal{E}_{z_{1}}a^{(1)}\mathcal{E}_{z_{2}}b^{(1)}\mathcal{E}_{z_{3}}c^{(1)} \cdot  i_{z_1, z_2}r_{z_1,z_2}(a^{(2)}\otimes b^{(2)})i_{z_1, z_3}r_{z_1,z_3}(a^{(3)}\otimes c^{(2)})i_{z_2, z_3}r_{z_2,z_3}(b^{(3)}\otimes c^{(3)}) \\
& =\sum (-1)^{f(\tilde{a}, \tilde{b},\tilde{c}) }\mathcal{E}_{z_{1}}a^{(1)}\mathcal{E}_{z_{2}}b^{(1)} \mathcal{E}_{z_{3}}c^{(1)}  \cdot r_{z_{1},0}(a^{(2)}\otimes \mathcal{E}_{z_{2}}b^{(2)})r_{z_{1},0}(a^{(3)}\otimes \mathcal{E}_{z_{3}}c^{(2)})r_{z_{2},0}(b^{(3)}\otimes \mathcal{E}_{z_{3}}c^{(3)} \\
&=\sum (-1)^{\tilde{(b^{\prime\prime})^{\second}} (\tilde{c^{\prime}}+\tilde{(c^{\second})^{\prime}})+(\tilde{(a^{\prime\prime})^{\prime}}+\tilde{(a^{\prime\prime})^{\second}}) (\tilde{b^{\prime}}+\tilde{c^{\prime}})+ \tilde{(b^{\second})^{\prime}}\tilde{c^{\prime}}+\tilde{(a^{\prime\prime})^{\second}} \tilde{(b^{\second})^{\prime}}}\mathcal{E}_{z_{1}}a^{\prime}\mathcal{E}_{z_{2}}b^{\prime}\mathcal{E}_{z_{3}}c^{\prime}\cdot \\ & \hspace{5cm} \cdot r_{z_{1},0}((a^{\second})^{\prime}\ten (\mathcal{E}_{z_{2}}(b^{\second})^{\prime}))r_{z_{1},0}((a^{\second})^{\second}\ten \mathcal{E}_{z_{3}}(c^{\second})^{\prime}))r_{z_{2},0}((b^{\second})^{\second}\ten \mathcal{E}_{z_{3}}(c^{\second})^{\second}\\
&=\sum (-1)^{\tilde{(b^{\prime\prime})^{\second}} (\tilde{c^{\prime}}+\tilde{(c^{\second})^{\prime}})+\tilde{a^{\prime\prime}} \tilde{b^{\prime}}+\tilde{a^{\prime\prime}} \tilde{c^{\prime}}+ \tilde{(b^{\second})^{\prime}}\tilde{c^{\prime}}}\mathcal{E}_{z_{1}}a^{\prime}\mathcal{E}_{z_{2}}b^{\prime}\mathcal{E}_{z_{3}}c^{\prime}\cdot r_{z_{1},0}(a^{\second}\ten (\mathcal{E}_{z_{2}}((b^{\second})^{\prime})(\mathcal{E}_{z_{3}}c^{\second})^{\prime}))r_{z_{2},0}((b^{\second})^{\second}\ten (\mathcal{E}_{z_{3}}c^{\second})^{\second}) \\
&=Y(a, z_1)Y(b, z_2)\mathcal{E}_{z_{3}}c.  \ \  \square
\end{align*}
Similar formulas can be derived for any $X_{z_1,\dots,z_n}$, $n\in \mathbb{N}$. Before we give those, as a corollary of the formula above we will derive a formula for  Operator Product  Expansions (OPEs).
Let $r$ be a $H^N_{T_{\epsilon}}\otimes H^N_{T_{\epsilon}}$-covariant bicharacter on $V$, with values in $\mathbf{F}^N(z, w)^{+, w}$.
For any $a, b\in V$ the bicharacter $r_{z,w}(a\ten b)$ is just a function of $z$ and $w$ in $\mathbf{F}^N(z, w)$ and  can be expanded as a Laurent series around $z=\epsilon ^i w$ for any $i=0, 1, \dots , N-1$: \mbox{$r_{z,w}(a\ten b)=\sum_{l=0}^{M_{a, b}-1}\frac{f^{i,l}_{a,b}}{(z- \epsilon ^i w)^{l+1}} +reg.$}
We denote by $M_{a, b}$  the order of the pole at $z= \epsilon ^i w$ and note that  $f^{i,l}_{a,b}=f^{i,l}_{a,b}(w)$ is a function only of $w$.

Recall we usually omit writing the indexing in $\del (a)=\sum_p a^{\prime}_p\ten a^{\second }_p$, and write it just as $\del (a)=\sum a^{\prime}\ten a^{\second }$ to unclutter notation, but this summation is always implicitly present.
\begin{thm}\begin{bf}(Bicharacter formula for the Residues)\end{bf}
\label{thm:comlH_{D, T}}
Let $V$, $W$, $\mathcal{E}_{z}$ and $\pi_T$ be as above.  Let $r$ be a $H^N_{T_{\epsilon}}\otimes H^N_{T_{\epsilon}}$-covariant bicharacter on $V$, with values in $\mathbf{F}^N(z, w)$, denote $M_{pq}=M_{a^{\second}_p, b^{\second}_q}$.
For any homogeneous $a,\ b \in V$ and any $0\le k\le M_{pq}-1$ we have
\begin{equation*}
\Res_{z= \epsilon ^i w}X_{z,w,0}(a\ten b\ten c)(z- \epsilon ^i w)^k dz= \sum_{p, q} \sum_{l=k}^{M_{pq}-1} (-1)^{\tilde{a^{\second}}\tilde{b^{\prime}}}f^{i,l}_{a^{\second}, b^{\second}} Y\big( (T_{\epsilon}^iD^{(l-k)}a^{\prime }).b^{\prime }, w\big) \pi_T(c).
\end{equation*}
\end{thm}
{\it Proof:}
By using coassociativity and cocommutativity we have from \ref{defn:DefSingmult3}
\begin{align*}
&X_{z,w,0}(a\ten b\ten c)=\sum_{p, q, r} (-1)^{\tilde{(b^{\second})^{\second}}(\tilde{c^{\prime}}+\tilde{(c^{\second})^{\prime}}) + (\tilde{(a^{\second})^{\prime}}+ \tilde{(a^{\second})^{\second}})(\tilde{b^{\prime}}+\tilde{c^{\prime}})+\tilde{(a^{\second})^{\second}}
\tilde{(b^{\second})^{\prime}}+\tilde{(b^{\second})^{\prime}}\tilde{c^{\prime}} }\cdot \\
& \hspace{5cm} \cdot (\mathcal{E}_za^{\prime})(\mathcal{E}_wb^{\prime})\pi_T(c^{\prime})r_{z,w}((a^{\second})^{\prime}\ten (b^{\second})^{\prime}))r_{z,0}((a^{\second})^{\second}\ten c^{\second})^{\prime})r_{w,0}((b^{\second})^{\second}\ten (c^{\second})^{\second}) \\
 &=\sum_{p, q, r} (-1)^{\tilde{(b^{\second})^{\prime}}(\tilde{c^{\prime}}+\tilde{(c^{\second})^{\prime}}) + (\tilde{(a^{\second})^{\second}}+ \tilde{(a^{\second})^{\prime}})(\tilde{b^{\prime}}+\tilde{c^{\prime}})+\tilde{(a^{\second})^{\prime}}
 \tilde{(b^{\second})^{\second}}+\tilde{(b^{\second})^{\second}}\tilde{c^{\prime}}+\tilde{(a^{\second})^{\prime}}
 \tilde{(a^{\second})^{\second}}+\tilde{(b^{\second})^{\prime}}\tilde{(b^{\second})^{\second}} }\cdot \\
 & \hspace{5cm}  \cdot (\mathcal{E}_za^{\prime})(\mathcal{E}_wb^{\prime})\pi_T(c^{\prime}) r_{z,w}((a^{\second})^{\second}\ten (b^{\second})^{\second}))r_{z,0}((a^{\second})^{\prime}\ten c^{\second})^{\prime})r_{w,0}((b^{\second})^{\prime}\ten (c^{\second})^{\second})\\
 &=\sum_{p, q, r} (-1)^{\tilde{(b^{\prime})^{\second}}(\tilde{c^{\prime}}+\tilde{(c^{\second})^{\prime}}) + (\tilde{a^{\second}}+ \tilde{(a^{\prime})^{\second}})(\tilde{(b^{\prime})^{\prime}}+\tilde{c^{\prime}})+\tilde{(a^{\prime})^{\second}}
 \tilde{b^{\second}}+\tilde{b^{\second}}\tilde{c^{\prime}}+\tilde{(a^{\prime})^{\second}}\tilde{a^{\second}}+
 \tilde{(b^{\prime})^{\second}}\tilde{b^{\second}}}\mathcal{E}_z(a^{\prime})^{\prime }\mathcal{E}_w(b^{\prime})^{\prime }\cdot \\
 & \hspace{5cm}  \cdot \pi_T(c^{\prime}) \cdot r_{z,w}(a^{\second}\ten b^{\second}))r_{z,0}((a^{\prime })^{\second }\ten (c^{\second})^{\prime})r_{w,0}((b^{\prime })^{\second}\ten (c^{\second})^{\second}).
\end{align*}
 Note that $r_{z,0}((a^{\prime })^{\second }\ten (c^{\second})^{\prime})$ is nonsingular at $z= \epsilon ^i w$, and therefore can be expanded in a power series in $(z- \epsilon ^i w)$:
\begin{align*}
r_{z,0}((a^{\prime })^{\second }  \ten  (c^{\second})^{\prime})&=\sum_{j\ge 0} \big( (\partial _z)^{(j)}r_{z,0}((a^{\prime })^{\second }\ten c^{\second})^{\prime}))\big) \arrowvert _{z= \epsilon ^i w}(z- \epsilon ^i w)^j =\sum_{j\ge 0}  r_{\epsilon ^i w,0}(D^{(j)}(a^{\prime })^{\second }\ten c^{\second})^{\prime}))(z- \epsilon ^i w)^j \\
& = \sum_{j\ge 0}  r_{w,0}(T_{\epsilon}^iD^{(j)}(a^{\prime })^{\second }\ten c^{\second})^{\prime}))(z- \epsilon ^i w)^j=\sum_{j\ge 0}  \epsilon^{-ij}r_{w,0}(D^{(j)}T_{\epsilon}^i(a^{\prime })^{\second }\ten c^{\second})^{\prime}))(z- \epsilon ^i w)^j.
\end{align*}
Next we use the modified expansion property of the exponential map, see lemma \ref{lem:expProp}, and the fact that $f^l_{a^{\second },b^{\second }}=0$ unless $\tilde{a^{\second }}=\tilde{b^{\second }}$, as the bicharacters are even.
\begin{align*}
&\Res_{z= \epsilon ^i w}X_{z,w,0}(a\ten b\ten c)(z- \epsilon ^i w)^k\\
 &=\sum_{p, q, r} (-1)^{f(\tilde{a}, \tilde{b}, \tilde{c})+\tilde{(a^{\prime})^{\second}}\tilde{a^{\second}}+\tilde{(b^{\prime})^{\second}}\tilde{b^{\second}}}r_{w,0}((b^{\prime })^{\second}\ten (c^{\second})^{\second})\cdot
\\ & \quad \cdot
Res_{z= \epsilon ^i w} \big((\mathcal{E}_z(a^{\prime})^{\prime })(\mathcal{E}_w(b^{\prime})^{\prime})\pi_T(c^{\prime})(\sum_j  \epsilon^{-ij} r_{w,0}(D^{(j)}T_{\epsilon}^i(a^{\prime })^{\second }\ten (c^{\second})^{\prime}))(z- \epsilon ^i w)^{j+k}) \cdot r_{z,w}(a^{\second}\ten b^{\second}))\big) \\
 &=\sum_{p, q, r} (-1)^{f(\tilde{a}, \tilde{b}, \tilde{c})+\tilde{(a^{\prime})^{\second}}\tilde{a^{\second}}+\tilde{(b^{\prime})^{\second}}\tilde{b^{\second}}}r_{w,0}((b^{\prime })^{\second}\ten (c^{\second})^{\second})\cdot \\
 &\quad \cdot Res_{z= \epsilon ^i w} \big(\sum _{n, j \ge 0} \epsilon ^{-i(n+j)}(z- \epsilon ^iw)^{n+j+k} \mathcal{E}_w (D^{(n)}Ta^{\prime})^{\prime})(\mathcal{E}_w(b^{\prime})^{\prime})  \pi_T(c^{\prime})  \cdot r_{w,0}(D^{(j)}(a^{\prime })^{\second }\ten (c^{\second})^{\prime})r_{z,w}(a^{\second}\ten
 b^{\second}))\big)\\
 &=\sum_{p, q, r} (-1)^{\tilde{(b^{\prime})^{\second}}(\tilde{c^{\prime}}+\tilde{(c^{\second})^{\prime}}) + (\tilde{a^{\second}}+ \tilde{(a^{\prime})^{\second}})(\tilde{(b^{\prime})^{\prime}}+\tilde{c^{\prime}})+\tilde{(a^{\prime})^{\second}}
 \tilde{b^{\second}}+\tilde{b^{\second}}\tilde{c^{\prime}}+\tilde{(a^{\prime})^{\second}}\tilde{a^{\second}}+
 \tilde{(b^{\prime})^{\second}}\tilde{b^{\second}}}\cdot\\
 &\quad \cdot \sum _{l=k}^{M_{pq}-1} \mathcal{E}_w((D^{(l-k)}Ta^{\prime})^{\prime }) (\mathcal{E}_w(b^{\prime})^{\prime })\pi_T(c^{\prime})r_{w,0}(D^{(l-k)}(Ta^{\prime })^{\second }\ten (c^{\second})^{\prime}))\cdot r_{w,0}((b^{\prime })^{\second }\ten c^{\second})^{\second})) )\big) f_{a^{\second }, b^{\second }}^k  \\
 &= \sum_{p, q, r} (-1)^{\tilde{(b^{\prime})^{\second}}\tilde{c^{\prime}} + \tilde{a^{\second}}(\tilde{(b^{\prime})^{\prime}}+\tilde{c^{\prime}})+ \tilde{(a^{\prime})^{\second}}\tilde{c^{\prime}}+\tilde{(a^{\prime})^{\second}}\tilde{b^{\second}}+\tilde{b^{\second}}\tilde{c^{\prime}}+\tilde{(a^{\prime})^{\second}}\tilde{a^{\second}}+\tilde{(b^{\prime})^{\second}}\tilde{b^{\second}}}.\\ &\hspace{0.5cm} .\Big( \sum_{l=k}^{M_{pq}-1} \big(\mathcal{E}_w ((D^{(l-k)}Ta^{\prime})(b^{\prime}))^{\prime } )\pi_T(c^{\prime}\big) r_{w,0}((D^{(l-k)}Ta^{\prime }b^{\prime })^{\second}\ten c^{\second})f^l_{a^{\second }, b^{\second }}\Big)\\
&= \sum_{p, q, r} \sum_{l=k}^{M_{pq}-1}(-1)^{(\tilde{(b^{\prime})^{\second}}+\tilde{(a^{\prime}})\tilde{c^{\prime}} + \tilde{a^{\second}}\tilde{b^{\prime}}} \mathcal{E}_w\big( ((D^{(l-k)}a^{\prime})(b^{\prime}))^{\prime }\pi_T (c^{\prime}\big) )\cdot r_{w,0}((D^{(l-k)}Ta^{\prime }b^{\prime })^{\second}\ten c^{\second})f^k_{a^{\second }, b^{\second }}\\
  & =\sum_{p, q} \sum_{l=k}^{M_{pq}-1} (-1)^{\tilde{a^{\second}}\tilde{b^{\prime}}}f^l_{a^{\second },b^{\second }}Y((D^{(l-k)}Ta^{\prime }).b^{\prime }, w)\pi_T(c). \square
\end{align*}
\begin{cor}
\label{cor:opes-bich}
Let $V, W, r$ be as above, let again $M_{pq}=M_{a^{\second}_p, b^{\second}_q}$.
For any $a,\ b \in V$ we have
\begin{equation*}
Y(a,z)Y(b,w)=i_{z,w}\sum _{p, q}\sum_{k=0}^{M_{pq}-1}\frac{\sum_{l=M_{p, q}-1-k}^{M_{p, q}-1} (-1)^{\tilde{a^{\second}}\tilde{b^{\prime}}}f^{i,l}_{a^{\second}, b^{\second}} Y\big( (T_{\epsilon}^iD^{(l-k)}a^{\prime }).b^{\prime }, w\big)}{(z- \epsilon ^i w)^{k+1}}  + Reg^{\epsilon^i}_{(z, w)}(a\ten b).
\end{equation*}
The term  $Reg^{\epsilon^i}_{(z, w)}(a\ten b)$ denotes the regular part in the Laurent expansion above,  it depends on $a, b\in V$, $z$ and $w$, and $i\in \{0, 1, \dots N-1\}$.
\end{cor}
Note that $Reg^{\epsilon^i}_{(z, w)}(a\ten b)$ is non-singular for $z=\epsilon ^i w$,  but it  can still be singular  at $z=\epsilon ^j w$ for $j\ne i$, and thus $Reg^{\epsilon^i}_{(z, w)}(a\ten b)$ can potentially differ from the normal ordered product $:a(z)b(w):$. On the other hand, as the normal ordered product \eqref{eqn:normordDef} is the regular part of an OPE of two fields, it is clear that if we have a single pole at $z=\epsilon ^i w$, then $Reg^{\epsilon^i}_{(z, w)}(a\ten b)=$ \ $:a(z)b(w):$. (For proof of that the normal ordered product is the regular part of the OPE of two fields in this more general context, see \cite{ACJ}). The above formula simplifies in the case of a simple pole:
\begin{cor}\begin{bf}(Bicharacter formula for OPEs for simple poles)\end{bf}
\label{cor:opes-simple}
Let $V, W, r$ be as above, and let  $a,\ b \in V$ are such that the bicharacters  $r_{z, w} (a^{\second }\ten b^{\second })$ have at most simple poles at each $a^{\second }, b^{\second }$. Then
\begin{equation*}
Y(a,z)Y(b,w)=i_{z,w}\sum_{p, q} \sum _{i}(-1)^{\tilde{a^{\second}}\tilde{b^{\prime}}}f^{i,0}_{a^{\second },b^{\second }} \frac{Y((T^{i} a^{\prime }).b^{\prime }, w)}{z- \epsilon ^i w} \ \ +:a(z)b(w):.
\end{equation*}
\end{cor}
 We derived formulas for the analytic continuations of products of two fields, as well as formulas for the OPEs and normal order products via the  bicharacter. These formulas always hold for any $H^N_{T_{\epsilon}}\otimes H^N_{T_{\epsilon}}$-covariant bicharacter on $V$, but if we want the vertex operators given by these formulas to satisfy all the axioms for a twisted vertex algebra the restriction that remains is  the shift restriction, see remark \ref{remark:shift}. Hence   we need to impose the following restriction on the bicharacter:
\begin{defn} \begin{bf}(Shift restricted bicharacter)\end{bf}
\label{defn:bich-shift-restr}
Let $r$ be a $H^N_{T_{\epsilon}}\otimes H^N_{T_{\epsilon}}$-covariant bicharacter on $V$, with values in $\mathbf{F}^N(z, w)$. Let as above $f^{i,l}_{a,b}$ stands for the coefficient in the expansion
\[
r_{z,w}(a\ten b)=\sum_{l=0}^{M-1}\frac{f^{i,l}_{a,b}(w)}{(z- \epsilon ^i w)^{l+1}} +\sum_{l=0}^{\infty }f^{i,-l-1}_{a,b}(w)\cdot (z- \epsilon ^i w)^{l}
\]
as a Laurent series in  $(z- \epsilon ^i w)$. We call the bicharacter $r$ \textbf{shift-restricted} if for any $a, b\in V$ \ $f^{i,l}_{a,b}=c^{i,l}_{a,b}\cdot w^{s^{i,l}_{a,b}}$, where $c^{i,l}_{a,b}$ is a constant, $c^{i,l}_{a,b}\in \mathbb{C}$, and $s^{i,l}_{a,b}\in \mathbb{Z}$ such that $|s^{i,l}_{a,b}|\leq (N-1)(l+1)$.
\end{defn}
Examples of shift-restricted bicharacters are given by functions in $\mathbf{F}^N(z, w)$ which have separately homogeneous numerators and denominators.

\begin{lem}\begin{bf}(Bicharacter formula for the normal order product for simple poles)\end{bf}
\label{lem:NormalOrderSimplePolesBich}
Let $V, W, r$ be as above, and let  $a,\ b \in V$ are such that the bicharacters  $r_{z, w} (a^{\second }_k\ten b^{\second }_l)$ have at most simple poles at each $a^{\second }_k, b^{\second }_l$.  Then the  normal order product $:Y(a,z)Y(b, z):$ of the fields $Y(a, z)$ and $Y(b, w)$ is given by
\begin{equation}
 :Y(a, z)Y(b, z):=\sum_{i} \sum_{k, l} c^{i,-1}_{a^{\second }_k, b^{\second }_l}(-1)^{\tilde{a^{\second}_k}\tilde{b^{\prime}_l}}Y( a^{\prime }_k.b^{\prime }_l, z)
\end{equation}
\end{lem}

We want to finish this section by giving the general formulas for multivariable fields and analytic continuation of  products of fields via bicharacter.
Recall the extended Sweedler notation for an element $a$ in a  commutative and cocommutative Hopf algebra,  $n\in \mathbb{N}, \ n\ge 2$:
\begin{equation}
\del ^{n-1}(a)=\sum_s \ a_s^{(1)}\ten a_s^{(2)}\ten \dots \ten a_s^{(n)},
\end{equation}
which again we will often write omitting the index $s$ to shorten the notation.
\begin{notation}\label{notation:CoprMatr}\begin{bf}{(Coproduct matrices)}\end{bf}\\
Let $a_1, a_2, \dots, a_n$ be $n$ elements of a commutative and cocommutative Hopf algebra. We can arrange the terms of the  $l$-coproducts of these elements as sets of $n$ by $(l+1)$ matrices  $M_{\Delta^{l}}^{\vec{k}} (a_1, a_2, \dots, a_n)=((a_{i}^{(j)})_{\vec{k}})_{j=1}^{l+1}$, where $\vec{k}=(k_1, k_2, \dots , k_n)$ is the coproduct index.
\end{notation}
Note that this is not one matrix, but a set of matrices, indexed by $\vec{k}$, with cardinality of the set dependent on the coproducts of  $a_1, a_2, \dots, a_n$.
Since we are going to encounter a lot of sign contributions, we introduce the following  notation:
\begin{notation}\label{notation:sign}
\begin{bf}(Sign notation)\end{bf}
Let $M=(m_{ij})_{i, j=1}^n$ be an $n$ by $n$ matrix with elements $m_{ij}\in V$.
Let $\textbf{sign}(M)= \textbf{sign}((m_{ij})_{i, j=1}^n)$ denote the following sign value:
\begin{equation*}
\textbf{sign}(M)= (-1)^{\sum_{i=2}^n\sum _{j=1}^{i-1}\sum _{k=2}^{n}\tilde{m}_{i1}\tilde{m}_{jk}}\cdot (-1)^{ \sum_{i=2}^n \sum _{j=2}^{n}(\sum _{k<i}\sum _{l\ge i+j-k}\tilde{m}_{ij}\tilde{m}_{kl})}.
\end{equation*}
\end{notation}
If $l=n-1$ in the coproduct matrices above, $M_{\Delta^{n-1}}^{\vec{k}} (a_1, a_2, \dots, a_n)$ are square matrices, and we can calculate $\textbf{sign}(M_{\Delta^{n-1}}^{\vec{k}} (a_1, a_2, \dots, a_n))$ for each one.
\begin{example}
Let  $M=\mathbb{C}\{\phi \}$, as in example \ref{example:Cphi}, $\phi$ is odd. We have $\Delta (\phi) =\phi \ten 1 +1\ten \phi$. Thus there are two matrices $M_{\Delta}^{\vec{k}}(\phi, 1)$, $\vec{k}\in \{(1, 1), \ (2, 1)\}$:\\
$M_{\Delta}^{(1, 1)} (\phi, 1)=\left(\begin{array}{cc}\phi & 1\\ 1 & 1\end{array}\right)$, \ \ \ $M_{\Delta}^{(2, 1)} (\phi, 1) =\left(\begin{array}{cc}1 &\phi \\ 1 & 1 \end{array}\right)$.
Both  have $\textbf{sign}=1$.

There are four matrices  $M_{\Delta}^{\vec{k}}(\phi, \phi)$, i.e., $\vec{k}\in \{(1, 1), \ (1, 2), \ (2, 1), \ (2, 2)\}$:\\
$M_{\Delta}^{(1, 1)} (\phi, \phi)=\left(\begin{array}{cc}\phi & 1\\ \phi & 1\end{array}\right)$, \ \ \ $M_{\Delta}^{(1, 2)} (\phi, \phi) =\left(\begin{array}{cc}\phi & 1\\ 1 & \phi \end{array}\right)$, \ \ \ $M_{\Delta}^{(2, 1)} (\phi, \phi) =\left(\begin{array}{cc}1 & \phi \\ \phi & 1\end{array}\right)$, \ \ \ $M_{\Delta}^{(2, 2)} (\phi, \phi) =\left(\begin{array}{cc}1 & \phi\\ 1& \phi \end{array}\right)$.\\
We have $\textbf{sign}(M_{\Delta}^{(1, 1)})=\textbf{sign}(M_{\Delta}^{(1, 2)})=\textbf{sign}(M_{\Delta}^{(2, 2)})=1$,  \mbox{$\textbf{sign}(M_{\Delta}^{(2, 1)})=-1$}.
\end{example}
\begin{defn}\label{defn:n-bich}
\begin{bf}($n$-characters)\end{bf}
Let $M$ be a commutative and cocommutative Hopf algebra, and let $r:M\ten M\to \mathbf{F}^N(z, w)$ be a super bicharacter on $M$. Let $a_1, a_2, \dots, a_n$ be $n$ elements of $M$.
Define an $n$-character $r_n:M^{\ten n} \to \mathbf{F}^N(z_1, z_2, \dots, z_n)$ by
\begin{align*}
&r_{z_1, z_2, \dots, z_n}(a_1\ten a_2\ten \dots \ten a_n)
=\sum_{\text{coproducts}} r_{z_1, z_2}(a_1^{(1)}\ten a_2^{(1)})r_{z_1, z_3}(a_1^{(2)}\ten a_3^{(1)})\dots r_{z_1, z_n}(a_1^{(n-1)}\ten a_n^{(1)})\cdot \\ &\hspace{5cm} \cdot  r_{z_2, z_3}(a_2^{(2)}\ten a_3^{(2)})\dots r_{z_2, z_n}(a_2^{(n-1)}\ten a_n^{(2)})\dots r_{z_{n-1}, z_n}(a_{n-1}^{(n-1)}\ten a_n^{(n-1)}).
\end{align*}
In particular, a tri-character $r_3:M\ten M\ten M \to \mathbf{F}^N(z_1, z_2, z_3)$ is given by
\begin{align*}
r_{z_1, z_2, z_3}(a_1\ten a_2 \ten a_3) &=\sum_{\text{coproducts}} r_{z_1, z_2}(a_1^{\prime} \ten a_2^{\prime})r_{z_1, z_3}(a_1^{\second}\ten a_3^{\prime}) r_{z_2, z_3}(a_2^{\second}\ten a_3^{\second}) \\
&=\sum_{k_1, k_2, k_3} r_{z_1, z_2}((a_1^{\prime})_{k_1} \ten (a_2^{\prime})_{k_2})r_{z_1, z_3}((a_1^{\second})_{k_1}\ten (a_3^{\prime})_{k_3}) r_{z_2, z_3}((a_2^{\second})_{k_2}\ten (a_3^{\second})_{k_3}).
\end{align*}
\end{defn}
\begin{defn} \label{defn:DefSingmult-n}\begin{bf}(Multivariable fields from a bicharacter)\end{bf}
Let $V$, $W$, $\mathcal{E}_z$, $\pi_T: V\to W$ be as above.  Let $a_1, a_2, \dots, a_n$ be any $n$   homogeneous elements of $V$.
Define  the $n$-variable  field
\begin{equation}
\label{eq:multivarX}
X_{z_1, z_2, \dots, z_n}\colon V^{\otimes n}\to
W[[z_1, z_2, \dots, z_n]]\otimes\mathbf{F}^N(z_1, z_2, \dots, z_n),
\end{equation}
by
\begin{equation*}
 X_{z_1, z_2, \dots, z_n}(a_1\ten a_2\ten \dots \ten a_n)= \sum_{\vec{k}} \textbf{sign}(M_{\Delta^{n-1}}^{\vec{k}} (a_1, a_2, \dots, a_n)) \mathcal{E}_{z_{1}}a_{1}^{\prime}\mathcal{E}_{z_{2}}a_{2}^{\prime}\cdot \cdot \cdot \mathcal{E}_{z_{n}}a_{n}^{\prime}  \cdot r_{z_1, z_2, \dots, z_n}(a_{1}^{\second}\ten a_{2}^{\second}\ten \dots \ten a_{n}^{\second})
\end{equation*}
\end{defn}
\begin{lem}(\textbf{ Analytic continuation})
\label{lem:analcont-n}
Let $V$, $W$, $\mathcal{E}_z$, $\pi_T: V\to W$ be as above. We have
\begin{equation*}
i_{z_1, z_2, \dots, z_n} X_{z_1, z_2, \dots , z_n}(a_1\ten a_2\ten \dots \ten a_n)=Y(a_1, z_1)Y(a_2, z_2)\dots \mathcal{E}_{z_{n}}a_n  =Y(a_1, z_1)Y(a_2, z_2)\dots Y(a_n, z_n)1
\end{equation*}
for any $a_1, a_2, \dots ,a_n \in V$.
\end{lem}
The proof is
very similar to the proof of lemma \ref{lem:analcont-3} and we omit it.
\begin{remark} \label{remark:sign-contr-z} (\textbf{Keeping track of sign contributions  with variables})\\
The use of the  $\textbf{sign}$ notation  can be restated as follows: In the definition \ref{defn:DefSingmult-n} one can think of the variables $z_1, z_2, \dots , z_n$ as "attached" to the arguments $a_1, a_2, \dots , a_n$ and correspondingly their coproducts. One then multiplies by a minus sign whenever an \textbf{odd} element with attached variable $z_j$ appears before ("transposes")  another \textbf{odd} element  with attached variable $z_i$, where $i<j$. There is no sign contribution unless both elements are odd.
\end{remark}
We  summarize  the entire construction  in the main bicharacter theorem:
\begin{thm}
\label{thm:Main}
Let $M$ be a commutative cocommutative Hopf algebra, let $V$ be the free Leibnitz module $V=H^N_{T_{\epsilon}}(M)$,  $r$ be a shift-restricted  $H^N_{T_{\epsilon}}\otimes H^N_{T_{\epsilon}}$-covariant symmetric bicharacter on $V$ with values in in $\mathbf{F}^N(z, w)^{+, w}$, $W=H_D(M)$ be the free $H_D$-Leibnitz sub-module-algebra of $V$. Let $\pi_T: V\to W$ be the projection map as in definition \ref{defn:TprojMap} and  $Y$ be the field-state correspondence defined by \eqref{eq:DefY}, via \eqref{eq:two-varX}. The set of data $(V, W, \pi_T, Y)$ constructed as above satisfies the definition of a twisted vertex algebra for \textbf{any} shift-restricted supercommutative $H^N_{T_{\epsilon}}\otimes H^N_{T_{\epsilon}}$-covariant bicharacter on $V$.
\end{thm}

\section{Examples of  twisted   vertex algebras based on a bicharacter}
\label{section:examplesexpl}

In most of the examples in the literature vertex operators are presented in terms of generating fields  and commutation relations. With the bicharacter construction as we saw in the previous section (Theorem \ref{thm:Main}) one starts instead with the commutative cocommutative Hopf algebra $M$ and its free Leibnitz module $H^N_{T_{\epsilon}}(M)$; the OPEs and thus the commutation relations are then dictated by the choice of the bicharacter $r$.  Moreover, for each commutative cocommutative Hopf algebra $M$ there are many choices of a symmetric bicharacter $r$, and so each such pair $(M, r)$ will give rise to a different twisted vertex algebra $(V, W, \pi_T, Y)$, even if the spaces $V$ and $W$ are the same as bialgebras--- since the field-state  correspondence $Y$  changes with the choice of a bicharacter.  This is the case in particular for the fermionic sides of the B and the D-A correspondences: there the space of fields $V$ and the space of states  $W$ coincide as free Leibnitz modules, but the generating  fields for each are different, which results in   highest weight modules for different Clifford algebras. Hence for the bicharacter construction examples are grouped based on  the Hopf algebra $M$, i.e. one starts by keeping $M$ the same, but changing the bicharacter $r$ on $M$. We want to stress  the fact that there is a  variety of examples even after we fix the  algebra $M$.  In the following  we will  list  the examples  grouped by the underlying Hopf algebra $M$.

\subsection{Twisted vertex algebras based on $\mathbb{C}\{\phi \}$ and a choice of a bicharacter}

Fix $M=\mathbb{C}\{\phi \}$ from example \ref{example:Cphi}. To define a  $H^N_{T_{\epsilon}}\ten H^N_{T_{\epsilon}}$-covariant  bicharacter on $H^N_{T_{\epsilon}}(M)$, one is only allowed to chose $r_{z, w}(\phi \ten \phi)$, as all the other values of the bicharacter on $H^N_{T_{\epsilon}}(\mathbb{C}\{\phi \})$ would be in turn determined by the covariance and the bicharacter properties (see section \ref{subsection:super-bichLeibnitz}). Thus a twisted vertex algebra $V$ based on $M=\mathbb{C}\{\phi \}$ will be determined entirely by a supersymmetric bicharacter value $r_{z, w}(\phi \ten \phi)$.

Let us specialize further, and consider the case of $N=2$--twisted vertex algebra ($\epsilon=-1$).
 From example \ref{example:Cphi}, the free  Leibnitz module $H^2_{T_{-1}}(\mathbb{C}\{\phi \})$ is isomorphic to $H_D(\mathbb{C}\{\phi, T\phi \})$. Before proceeding to specific examples (dependent on the value of $r_{z, w}(\phi \ten \phi)$), we want to present a formula for the vacuum expectation values valid for any choice of $r_{z, w}(\phi \ten \phi)$.

\subsection{Twisted vertex algebras based on $\mathbb{C}\{\phi \}$: Pfaffian vacuum expectation values}
\label{subsection:Pfaffian}

 Let $\langle
\ \mid\ \rangle: W\ten W \to \mathbb{C}$  be a  symmetric bilinear form on the space of states $W=H_D(\mathbb{C}\{\phi\})$. There is a very important concept  of an invariant bilinear form on a vertex algebra, for details see for example \cite{Li-bilinear} and \cite{Xu}. It is not our goal here to define a general  invariant bilinear form for a twisted vertex algebra, but for our bicharacter construction we will require that any bilinear form is such that the vacuum vector $1=|0\rangle$ is orthogonal to all other generators of the algebra $W$ and has norm 1, i.e.,
$\langle 1 \mid 1\rangle = \langle \langle 0| \mid |0\rangle \rangle =1$.
By abuse of notation we will just write $\langle  0 \mid 0 \rangle$ instead of $\langle \langle 0| \mid |0\rangle \rangle$. We can extend this  form to $W((z_1, z_2,\dots )) \ten W((z_1, z_2,\dots )) \to \mathbb{C}((z_1, z_2,\dots ))$ by bilinearity.
The values \mbox{$\langle 0 \mid Y(a, z_1) Y(a, z_2) \dots Y(a, z_n)|0\rangle$} of the bilinear form are usually called vacuum expectation values.
\begin{prop}
\label{prop:VacExpphi}
Let $V$ be a twisted vertex algebra based on $M=\mathbb{C}\{\phi \}$ and  a supersymmetric bicharacter $r$ (in particular, $V=H_D(\mathbb{C}\{\phi, T\phi \})$ and $W=H_D(\mathbb{C}\{\phi\})$).
Denote by $\phi(z)$ the field $Y(\phi, z)$ produced by  definition \ref{eq:DefY}, via \eqref{eq:two-varX}.
The following formula for the vacuum expectation values holds:
\begin{equation}
\langle 0 \mid \phi(z_1)\phi(z_2) \dots \phi(z_{2n})|0\rangle =i_{z} P\!f\big( r_{z_i, z_j}(\phi \ten \phi )\big)_{i, j=1}^{2n}.
\end{equation}
Here as usual $P\!f$ denotes the Pfaffian of an antisymmetric matrix and $i_{z}$ stands for the expansion $i_{{z_1}, {z_1}, \dots , {z_{2n}}}$.
\end{prop}
Note that the matrix on the right-hand side is antisymmetric since  the bicharacter $r$ is symmetric and $\phi$ is odd, i.e., $r_{z_i, z_j}(\phi \ten \phi )=-r_{z_j, z_i}(\phi \ten \phi )$.\\
{\it Proof:}
To calculate the vacuum expectation values we calculate instead the vacuum expectation values of the analytic continuation $X_{{z_1}, {z_2}, \dots , {z_{2n}}}(\phi \ten \phi \dots \phi)$ by use of Lemma \ref{lem:analcont-n}.
Since $\phi$ is a primitive element, we have
\begin{equation}
\del ^{2n}(\phi)=\phi \ten 1\ten \dots \ten 1 +1\ten \phi \ten \dots \ten 1 +1\ten 1\ten \phi \ten \dots \ten 1 +\dots 1\ten 1\ten \dots \ten \phi.
\end{equation}
We need three observations:
\begin{enumerate}
\item Since for the bilinear form  the vacuum vector $1=|0\rangle$ spans an  orthogonal subspace on its own (and in particular is orthogonal to $\phi$ and its descendants), the only contributions to the vacuum expectation values will come from the terms in the multivariable field where the coproducts have 1 as a first term; the other terms will not contribute. That forces us to work with the  $(2n)$-character $r_{z_1, z_2, \dots, z_{2n}}(\phi\ten \dots \ten \phi)$.
\item No sign contribution will come from the first $(-1)$ factor  of  $\textbf{sign}(M_{\Delta^{2n-1}}^{\vec{k}} (\phi, \phi, \dots , \phi))$, as the only contributing matrices are those with  the   first columns consisting  entirely  of 1s (1 is even).
\item Since $\phi $ is a primitive element  we have $r_{z, w}(\phi \ten 1)=r_{z, w}(1 \ten \phi)=0$ for  any  bicharacter. Thus the only contributions in the  $(2n)$-character $r_{z_1, z_2, \dots, z_{2n}}(\phi\ten \phi\ten \dots \ten \phi)$ will come from the following situation: Consider  a matrix $M_{\Delta^{2n-1}}^{\vec{k}} (\phi, \phi, \dots , \phi)$ with first column entirely consisting of 1s. A nonzero summand in the $(2n)$-character $r_{z_1, z_2, \dots, z_{2n}}(\phi\ten \phi\ten \dots \ten \phi)$ will be a product of nonzero bicharacter factors, and that happens  when  we have a sequence of  either $(1, 1)$  pairs (trivial, as $r_{z, w}(1 \ten 1)=1$), or $(\phi, \phi)$ pairs (nontrivial). If there is a  mixed pair $(1, \phi)$ or $(\phi, 1)$ as a factor in a summand,  that summand will be 0.
     So a nonzero summand will have exactly $n$ such nontrivial contributing pairs $(\phi, \phi)$ , and  each pair forms one bicharacter $r_{z_{i}, z_{j}}(\phi \ten \phi)$. The sign contribution will come only from the nontrivial pairs.
\end{enumerate}
Thus, we have
\begin{align*}
 X_{z_1, z_2, \dots, z_{2n}}(\phi&\ten \phi\ten \dots \ten \phi )\\
 &= \sum_{\text{coproducts}} \textbf{sign}(M_{\Delta^{2n-1}}^{\vec{k}} (\phi, \phi, \dots , \phi)) \mathcal{E}_{z_{1}}\phi^{\prime}\mathcal{E}_{z_{2}}\phi^{\prime}\cdot \cdot \cdot \mathcal{E}_{z_{2n}}\phi^{\prime} \cdot r_{z_1, \dots, z_{2n}}(\phi^{\second}\ten \phi^{\second}\ten \dots \ten \phi^{\second}) \\
& =\sum_{\substack{\text{contr.}\\
\text{coproducts}}}\textbf{sign}(M_{\Delta^{2n-1}}^{\vec{k}} (\phi, \phi, \dots , \phi)) 1\cdot r_{z_1, z_2, \dots, z_{2n}}(\phi\ten \phi\ten \dots \ten \phi) +\dots\\
& =\sum_P \epsilon (P) 1 \cdot r_{z_{i_{1}}, z_{i_{2}}}(\phi \ten \phi)r_{z_{i_{3}}, z_{i_{4}}}(\phi \ten \phi)\cdot \cdot \cdot r_{z_{i_{2n-1}}, z_{i_{2n}}}(\phi \ten \phi) +\text{other  terms}.
\end{align*}
The sum is over all permutations such that $i_1 <i_2, \ i_3 <i_4, \dots $, $i_{2n-1} <i_{2n}$, \ $i_1 <i_3 <\dots <i_{2n-1}$. The sign contribution from any \textbf{contributing} matrix $M_{\Delta^{2n-1}}^{\vec{k}} (\phi, \phi, \dots , \phi))$ is precisely the sign of the corresponding permutation, since $\phi$ is odd  (see remark \ref{remark:sign-contr-z} and observation 3 above). That produces precisely the Pfaffian
$P\!f\big( r_{z_i, z_j}(\phi \ten \phi )\big)_{i, j=1}^{2n}$.
$\square$

Next we will consider examples of twisted vertex algebras  arising from specific bicharacter values for $r_{z, w}(\phi \ten \phi)$ by use of Theorem \ref{thm:Main}.

\subsection{Twisted vertex algebras based on $\mathbb{C}\{\phi \}$: the neutral free fermion of type B}
\label{section:freefermionB}

We continue  working with space of fields $V=H^2_{T}(\mathbb{C}\{\phi \})\equiv H_D(\mathbb{C}\{\phi, T\phi \})$, and space of states $W=H_D(\mathbb{C}\{\phi\})$. The projection map (recall definition \ref{defn:TprojMap}) in this case is just the algebra homomorphism  defined by $\pi_T(T\phi)=\phi$.

Let the covariant bicharacter  $r^B:H_D(\mathbb{C}\{\phi, T\phi \}) \ten H_D(\mathbb{C}\{\phi, T\phi \}) \to \mathbf{F}^2(z, w)^{+, w}$ be defined by
\begin{equation}
r^B_{z, w}(\phi \ten \phi)=\frac{z-w}{z+w}.
\end{equation}
Note that the bicharacter $r^B$ is symmetric, as it is symmetric on the generator $\phi$, it is also shift-restricted,  and has a simple single pole.
Theorem \ref{thm:Main} asserts that  the pair $(\mathbb{C}\{\phi \}, r^B)$ will produce an example of a twisted vertex algebra of order 2. We claim that this N=2 twisted vertex algebra  is the free fermion of type B. To prove it, we need to show first that the field $\phi ^B(z)$ corresponding to the element $\phi$ via the field-state correspondence defined by \eqref{eq:DefY} coincides with  the free fermion field of type B introduced in section \ref{subsection:examplesB} (i.e., satisfies the OPE \eqref{equation:OPE-B}).
We  use corollary \ref{cor:opes-simple} to calculate the OPE  of $\phi ^B(z)\phi ^B(w)$. The only singular bicharacter from any of the coproducts  $\phi^{\second}$ and $\phi^{\second}$ is $r^B_{z, w}(\phi \ten \phi)$. Thus
\begin{equation*}
\phi ^B (z)\phi ^B(w) \sim i_{z,w}\sum(-1)^{\tilde{\phi ^{\second}}\tilde{\phi ^{\prime}}}f^{1,0}_{\phi ^{\second },\phi ^{\second }} \frac{Y((T \phi ^{\prime }).\phi ^{\prime }, w)}{(z +w)} \sim i_{z,w}(-1)^{\tilde{\phi}\tilde{1}}f^{1,0}_{\phi, \phi } \frac{Y((T1).1, w)}{(z +w)}\sim -\frac{2w\cdot 1}{z+w}. \quad \square
\end{equation*}
\begin{lem}
The normal ordered product field $h(z)$ from \eqref{eqn:normal-order-h-B} defined by $\frac{1}{4}(:\phi ^B (z)\phi ^B(-z):-1)$ corresponds to
\begin{equation*}
h(z)=\frac{1}{4}Y(\phi \cdot T\phi, z), \quad h=\frac{1}{4}h_{\phi}, \quad \text{where} \ h_{\phi}=\phi \cdot T\phi, \ \text{see \ Example}\ \ref{example:Cphi}.
\end{equation*}
\end{lem}
{\it Proof:}
From lemma \ref{lem:NormalOrderSimplePolesBich}
\begin{align*}
:\phi ^B (z)\phi ^B(-z): &= :\phi ^B (z)T\phi ^B(z): = \sum(-1)^{\tilde{\phi ^{\second}}\tilde{\phi ^{\prime}}}f^{1,-1}_{\phi ^{\second },\phi ^{\second }} Y(\phi ^{\prime }\cdot (T\phi )^{\prime }, z)\\
&= (-1)^{\tilde{\phi}\tilde{1}}f^{1,-1}_{\phi, \phi } Y(1\cdot 1, z) + (-1)^{\tilde{\phi}\tilde{1}}f^{1,-1}_{1, 1 } Y(\phi \cdot T\phi, z)=1_W +Y(\phi \cdot T\phi, z).\quad \square
\end{align*}
To calculate the OPEs  \eqref{eqn:HeisOPEsB} one uses theorem \ref{thm:comlH_{D, T}} to get
\begin{equation}
h_{\phi}(z)h_{\phi}(w)\sim +Y(1, w)r_{z, w}(h_{\phi}\ten h_{\phi}) \sim \frac{8zw(z^2 +w^2)}{(z^2 -w^2)^2}.
\end{equation}
Since the fermionic side of of the boson-fermion correspondence of type B is  known, see \cite{DJKM-4} and \cite{YouBKP}, we omit most of the calculations.

We have $Th=\frac{1}{4}T\phi \cdot \phi =-h$, and since $Y(Th, z)=Y(h, -z)$ from the transfer of action axiom of twisted vertex algebras, we have  $Y(h, -z)=-Y(h, z)$ in the twisted vertex algebra. Hence we have only \textbf{odd} powers of $z$ in
$Y(h, z)$, and the indexing in $h(z)=\sum _{n\in \mathbb{Z}}h_{2n+1} z^{-2n-1}$ is implied. The commutation relations $[ h_m, h_n ] = \frac{m}{2} \delta _{m+n, 0} 1$ for the twisted Heisenberg algebra $\mathcal{H}_{\mathbb{Z}+1/2}$ then follow from the OPE in a standard calculation.

The decomposition of the space of states $F_B =W=H_D(\mathbb{C}\{\phi\})$ into twisted Heisenberg modules, Lemma \ref{lem:BosonicSpaceBB}, was done first in  \cite{DJKM-4} and \cite{YouBKP}, and the calculations using the bicharacter formulas \eqref{eq:DefY} and \eqref{eq:two-varX} are also available.

 To calculate the image  of the  generating field $\phi ^B(z)\mapsto e^{\alpha}(z)$ we first calculate the OPE
$h(z)\phi ^B(w)$. As all the poles are simple here, from corollary \ref{cor:opes-simple} we get
\begin{align*}
h(z)\phi ^B(w)\sim \frac{1}{4}\big(\frac{2w}{z-w} +\frac{2w}{z+w}\big)Y(\phi , w) \sim \frac{zw}{z^2 -w^2}\phi ^B(w).
\end{align*}
 From this OPE and the exact description of the split of $W$ into irreducible Heisenberg submodules, by use of  the standard calculational lemmas (see for example \cite{Kac-Lie} and \cite{Wakimoto}) we  get that
the exponential boson formula \eqref{eqn:imageB} holds for the field $\phi ^B(w)$.

\subsection{Twisted vertex algebras based on $\mathbb{C}\{\phi \}$: the neutral free fermion of type D-A}
\label{section:freefermionD}

 As in the previous section the space of fields is $V=H^2_{T}(\mathbb{C}\{\phi \})$, and the space of states is $W=H_D(\mathbb{C}\{\phi\})$, with the same  projection map.
Let the bicharacter  $r^D:H_D(\mathbb{C}\{\phi, T\phi \}) \ten H_D(\mathbb{C}\{\phi, T\phi \}) \to \mathbf{F}^2(z, w)^{+, w}$ be given by
\begin{equation}
r^D_{z, w}(\phi \ten \phi)=\frac{1}{z-w}.
\end{equation}
The bicharacter $r^D$ is symmetric, it
 is  shift-restricted,  and has a simple single pole.
 By Theorem \ref{thm:Main}, the pair $(\mathbb{C}\{\phi \}, r^D)$ will produce an example  of a twisted vertex algebra of order 2. We claim that this twisted vertex algebra  is the free neutral fermion of type D-A (see Section \ref{subsection:examplesD}).
Since this is the new example of the boson-fermion correspondence of type D-A, we will go carefully over the details.

We use Corollary \ref{cor:opes-simple} to calculate the OPE  of $\phi ^D(z)\phi ^D(w)$. Since $\phi$  is primitive, $r^D_{z, w}(\phi \ten 1)=r^D_{z, w}(1\ten \phi )=0$, and the only nontrivial bicharacter from any of the coproducts  $\phi^{\second}$ and $\phi^{\second}$ is $r^D_{z, w}(\phi \ten \phi)$. Thus
\begin{equation*}
\phi ^D (z)\phi ^D(w) \sim i_{z,w}\sum(-1)^{\tilde{\phi ^{\second}}\tilde{\phi ^{\prime}}}f^{1,0}_{\phi ^{\second },\phi ^{\second }} \frac{Y((T \phi ^{\prime }).\phi ^{\prime }, w)}{(z -w)}\sim \frac{1}{z-w}.
\end{equation*}
This OPE coincides with \eqref{equation:OPE-D} and corresponds to the  anticommutation relations:
\begin{align*}
[\phi ^D (z), \phi ^D(w)]_{\dag}=i_{z, w}\frac{1}{z-w} +i_{w, z}\frac{1}{w-z}=(i_{z, w}-i_{w, z})\frac{ 1}{z-w}.
\end{align*}
Using the notation $\delta(z -w) =(i_{z, w} -i_{w, z})\frac{1}{z -w} =\sum_{j \in \mathbb{Z}} z^{-j-1}w^j$
we can write
\begin{equation*}
[\phi ^D (z), \phi ^D(w)]_{\dag}=\delta(z -w),
\end{equation*}
which if we index the field $\phi ^D(z)=\sum _{n\in \mathbf{Z+1/2}} \phi^D_n z^{-n-1/2}$ will give us the required anticommutation relations
$[\phi^D_m,\phi^D_n]_{\dag}=\delta _{m, -n}1$.
The field $\phi ^D (z)$ is well known as  the neutral free fermion (of type D). Since in the OPE of $\phi ^D (z)$ the only pole is at $z=w$, it is immediate that $\phi ^D (z)$ on its own will generate a super vertex algebra (see e.g. \cite{Kac}, \cite{WangKac}, \cite{WangDual}).  $\phi ^D (z)$ cannot be bosonized on its own, but as a new ingredient allowed in twisted vertex algebras we consider another descendant of $\phi ^D (z)$-- the field $T\phi ^D (z)=\phi ^D (-z)$.
Using the language of delta functions (see e.g. \cite{Kac}, \cite{Wakimoto}, \cite{ACJ}) we get
\begin{equation*}
[T\phi ^D (z), T\phi ^D(w)]_{\dag}=-\delta(z -w), \quad [T\phi^D_m, T\phi^D_n]_{\dag}=-\delta _{m, -n}1,
\end{equation*}
Thus on its own each of the fields $\phi ^D (z)$ and $T\phi ^D (z)$ (without the other) will generate a super vertex algebra, but the two  "glue together" to form a twisted vertex algebra. This  resembles the gluing together of the two sheets of the square root Riemann surface.

Since in a \textbf{twisted} vertex algebra  both the fields $\phi ^D (z)$ and $T\phi ^D (z)$  (and their descendants) are available,   we can form the  field
\begin{equation*}
h(z)=\frac{1}{2}:\phi ^D (z)\phi ^D(-z):=\frac{1}{2} :\phi ^D (z)T\phi ^D(z):.
\end{equation*}
$h(z)$ is the Heisenberg field from  Proposition  \ref{prop:HeisOPEsD}.
To prove this we  use lemma \ref{lem:NormalOrderSimplePolesBich}:
\begin{align*}
:\phi ^D (z)T\phi ^D(z): &= \sum(-1)^{\tilde{\phi ^{\second}}\tilde{\phi ^{\prime}}}f^{1,-1}_{\phi ^{\second },\phi ^{\second }} Y(\phi ^{\prime }\cdot (T\phi )^{\prime }, z)= \\
 & = (-1)^{\tilde{\phi}\tilde{1}}f^{1,-1}_{\phi, \phi } Y(1\cdot 1, z) + (-1)^{\tilde{\phi}\tilde{1}}f^{1,-1}_{1, 1 } Y(\phi \cdot T\phi, z)=0\cdot 1_W +Y(\phi \cdot T\phi, z).
\end{align*}
Thus the field $h(z)$ from \eqref{eqn:normal-order-h-D} is actually the vertex operator  $Y(\frac{1}{2}\phi \cdot T\phi, z)$ corresponding to the element $\frac{1}{2}h_{\phi}=\frac{1}{2}\phi \cdot T\phi$. To calculate the OPE of  \eqref{eqn:HeisOPEsD} we use theorem \ref{thm:comlH_{D, T}}. We  look at the coproduct :
\begin{equation*}
\del  h_{\phi} = h_{\phi}\ten 1 +1\ten h_{\phi} +\phi \ten T\phi - T\phi\ten \phi.
\end{equation*}
The first order poles in the OPE  come from
the $h^{\second}$ terms of  the coproduct which have first order poles in their bicharacter, namely from $r_{z, w}(\phi \ten T\phi)= -r_{z, w}(T\phi \ten \phi)=\frac{1}{z+w}$ and from  $r_{z, w}(\phi \ten \phi)=-r_{z, w}(T\phi \ten T\phi)=\frac{1}{z-w}$.
The OPE coefficients  in front of the first order poles then are:\\
for $\frac{1}{z-w}$ we get $Y(-\phi \cdot  \phi +T\phi  \cdot T\phi , w)$, which is zero as $\phi  \cdot \phi =0= T\phi  \cdot T\phi$,  \\
for $\frac{z-w}{z+w}$ we get $Y(-T\phi  \cdot T\phi +T^2\phi  \cdot \phi , w)$, which is zero as $T^2\phi =\phi$.
Thus there are no first order poles in the OPEs of $h(z)h(w)$.
The second order pole comes from
\begin{align*}
r_{z, w}(h_{\phi}\ten h_{\phi})& = r_{z, w}(\phi \cdot T\phi \ten \phi \cdot T\phi)= -r_{z, w}(\phi \ten \phi)r_{z, w}(T\phi \ten T\phi) +r_{z, w}(\phi \ten T\phi)r_{z, w}(T\phi \ten \phi)  \\
& =+\frac{1}{(z-w)^2} -\frac{1}{(z+w)^2} =\frac{4zw}{(z^2 -w^2)^2}.
\end{align*}
 Further, we have $Th=\frac{1}{2}T\phi \cdot \phi =-h$, and since $Y(Th, z)=Y(h, -z)$ from the transfer of action axiom of twisted vertex algebras, we have  $Y(h, -z)=-Y(h, z)$. Which  means that we have only \textbf{odd} powers of $z$ in $Y(h, z)$,  and we can  index it  $h(z)=\sum _{n\in \mathbb{Z}}h_{n} z^{-2n-1}$. With this indexing, we have from the OPE:
\begin{equation*}
[h(z), h(w)]=  (i_{z, w} -i_{w, z})\frac{zw}{(z^2 -w^2)^2} =\sum_{n\in \mathbb{Z}} n \frac{w^{2n-1}}{z^{2n+1}}=\frac{1}{4}\partial_w (\delta(z -w) +\delta(z +w)),
\end{equation*}
which gives us the required commutation relations for the Heisenberg algebra $\mathcal{H}_{\mathbb{Z}}$ and proves Proposition  \ref{prop:HeisOPEsD}.
\begin{remark} \label{remark:projofHeis}
The field $h(z)$ has a very special property (which was also true for the B case),  peculiar to  twisted vertex algebras, and not possible in a super vertex algebra. $h(z)$ corresponds to the nonzero element $h$ in the space of fields of the twisted vertex algebra $V$, i.e., $h(z)=Y(h, z)$. But the  projection of $h$ to the space of states  is zero, as
\begin{equation}
\pi_T(h)=\pi_T(\frac{1}{2}h_{\phi})= \pi_T(\frac{1}{2}\phi \cdot T\phi)=\frac{1}{2}\phi \cdot \phi =0.
\end{equation}
\end{remark}
We now prove Proposition \ref{prop:decompD}.
In order to decompose the space of states $F_D =W=H_D(\mathbb{C}\{\phi\})$ into Heisenberg submodules  we need to  identify the highest weight vectors in $W$. These are the elements $a_{\lambda}$ of $W$ such that $h(z)a_{\lambda}$ has only nonnegative powers of $z$.
Let $a^{\text{even}}_{n}=\phi D^{(2)}\phi \cdots D^{(2n)}\phi$, and denote as usual $\phi ^i=D^{(i)}\phi$. We claim $a^{\text{even}}_{n}$ is a highest weight vector for $\mathcal{H}_{\mathbb{Z}}$. To prove that, we will be using the bicharacter formula \eqref{eq:DefY} and \eqref{eq:two-varX}. We can easily calculate the coproduct of $a^{\text{even}}_{n}$:
\begin{equation*}
\del (a^{\text{even}}_{n}) =\sum _i (-1)^{i-1}\phi  \phi ^2\cdots \widehat{\phi ^{2i}} \cdots  \phi ^{2n}\ten \phi ^{2i} + \sum _{i, j} (-1)^{i+j}\phi \cdots \widehat{\phi ^{2i}} \cdots \widehat{\phi ^{2j}} \cdots \phi ^{2n}\ten \phi ^{2i}\phi ^{2j} +\dots
\end{equation*}
The only parts of the coproduct of consequence in this case  are  the parts with either single or quadratic terms in $(a^{\text{even}}_{n})^{\second}$, as the bicharacter with any term from $h^{\second}$ will be 0 otherwise.

We first consider the even $a^{\text{even}}_{0}=1$: We have
\begin{equation*}
h(z)1 = \mathcal{E}_{z}h = \frac{1}{2}e^{zD}\phi \cdot e^{-zD}\phi =\sum_{n\in \mathbb{Z}_{\ge 0}}z^{2n+1}\sum_{p+q=2n+1}(-1)^qD^{(p)}\phi D^{(q)}\phi= \sum_{n\in \mathbb{Z}_{\ge 0}}z^{2n+1}\sum_{p+q=2n+1}(-1)^q \phi ^p \phi ^q,
\end{equation*}
since we have $\sum_{p+q=\text{even}}(-1)^qD^{(p)}\phi D^{(q)}\phi =0$. Thus $a^{\text{even}}_{0}$ is a highest weight vector, with highest weight  0 (as $h_0 1=0$). Further, we have:
\begin{align*}
& r_{z, 0}(\phi T\phi \ten \phi ^i \phi ^j)=0 \quad  \text{if} \ i, j\ \text{both even, or } \ \text{if} \ i, j\ \text{both odd},\\
& r_{z, 0}(\phi T\phi \ten \phi ^i \phi ^j)=-\frac{2}{z^{i+j+2}}\quad  \text{if} \ i=\text{even}, \ j=\text{odd},\\
& r_{z, 0}(\phi T\phi \ten \phi ^i \phi ^j)=+\frac{2}{z^{i+j+2}}\quad  \text{if} \ i=\text{odd}, \ j=\text{even}.
\end{align*}
\begin{align*}
h(z) a^{\text{even}}_{n}&=\mathcal{E}_{z}h \cdot a^{\text{even}}_{n}   +\frac{1}{2}e^{zD}\phi \cdot  (\sum (-1)^{i}\phi \phi ^2 \phi ^4\cdots \widehat{\phi ^{2i}} \cdot \cdot \phi ^{2n}r_{z, 0}(T\phi \ten \phi ^{2i})) -\\
&\quad -\frac{1}{2}e^{-zD}\phi \cdot  (\sum (-1)^{i}\phi \phi ^2 \phi ^4\cdots \widehat{\phi ^{2i}} \cdot \cdot \phi ^{2n})r_{z, 0}(\phi \ten \phi ^{2i})+\\
&\quad  +\sum _{i, j} (-1)^{2i+2j}1\cdot \phi \phi ^2 \phi ^4\cdots \widehat{\phi ^{2i}} \cdots \widehat{\phi ^{2j}} \cdots \phi ^{2n}r_{z, 0}(\phi T\phi \ten \phi ^{2i} \phi ^{2j}).
\end{align*}
Hence
\begin{equation}
\label{eqn:provingHW}
h(z)a^{\text{even}}_{n} =\mathcal{E}_{z}h \cdot a^{\text{even}}_{n}   -\frac{1}{2}\sum_i (-1)^i(e^{zD}\phi + e^{-zD}\phi)\cdot  \Big( \phi \phi ^2 \phi ^4\cdots \widehat{\phi ^{2i}} \cdot \cdot \phi ^{2n}\frac{1}{z^{2i +1}}\Big).
\end{equation}
Thus, there will be no nonzero contribution to any power of z less than $-1$ in \eqref{eqn:provingHW}, as we are multiplying by one of the $\phi ^{2l}$ that is already in the product $\phi \phi ^2 \phi ^4\cdots \widehat{\phi ^{2i}} \cdot \cdot \phi ^{2n}$, and thus getting 0. On the other hand, the contribution to the coefficient in front of $z^{-1}$ is $-2na^{\text{even}}_{n}$, as we get a $z^{-1}$ precisely when  we are multiplying by the $\phi ^{2i}z^{2i}$ term from $\frac{1}{2}(e^{zD}\phi +(-1)^i e^{-zD}\phi)$ which exactly complements the $\phi \phi ^2 \phi ^4\cdots \widehat{\phi ^{2i}} \cdot \cdot \phi ^{2n}$. The minus sign in $-2na^{\text{even}}_{n}$ is due to the fact that when multiplying  $(-1)^i\phi ^{2i}\cdot \phi \phi ^2 \phi ^4\cdots \widehat{\phi ^{2i}} \cdot \cdot \phi ^{2n}$ we get $+a^{\text{even}}_{n}$. These considerations mean that $h_{n}$ annihilate $a^{\text{even}}_{n}$ for $n>0$, and $h_0a^{\text{even}}_{n}=-2na^{\text{even}}_{n}$. Hence $a^{\text{even}}_{n}$ is a highest weight vector with weight $-2n$.

Closer observation of the positive powers of $z$ in \eqref{eqn:provingHW} shows which elements of $W=F_D$ can be generated from the highest weight vector $a^{\text{even}}_{n}$: the elements with  $n+2m$ factors, $m\ge 0$, with exactly $m$ factors $\phi ^p$ where  $p$ is odd. We see that $W=F_D$ is bi-graded: first by the number $n$ of factors  in an element
$a=\phi ^{k_1}\cdot \phi ^{k_2} \cdot \cdot \cdot \phi ^{k_{n}}$, $k_1 <k_2 <\dots <k_{n}$, and second by the difference between how many of these $k_{i}$ are odd minus how many of them are even (we will call the second grading "derivative grading"). For example, the element $\phi D\phi$ is in the highest weight module generated by the highest vector $1=a^{\text{even}}_{0}$, as it has derivative grading 0 equal to the highest weight of $1$.

Now let $a^{\text{odd}}_{n}=\phi^1 \phi^3 \cdots \phi^{2n-1}$,  $a^{\text{odd}}_{1}=\phi^1=D\phi$. We claim $a^{\text{odd}}_{n}$ is a highest weight vector for $\mathcal{H}_{\mathbb{Z}}$. Similar calculations as for $a^{\text{even}}_{n}$ show that
\begin{equation*}
h(z)a^{\text{even}}_{n} =\mathcal{E}_{z}h \cdot a^{\text{odd}}_{n}   +\frac{1}{2}\sum_i (-1)^i(e^{zD}\phi - e^{-zD}\phi)\cdot  \Big( \phi^1 \phi ^3 \cdots \widehat{\phi ^{2i-1}} \cdot \cdot \phi. ^{2n-1}\frac{1}{z^{2i}}\Big)
\end{equation*}
Similar considerations as for $a^{\text{even}}_{n}$ hold for the $a^{\text{odd}}_{n}$, the difference is in the minus sign, i.e., $a^{\text{odd}}_{n}$ is a highest weight vector with highest weight $n$ for  the Heisenberg algebra $\mathcal{H}_{\mathbb{Z}}$. We also see that any element with derivative grading $n\in \mathbb{Z}$ will be in the highest module with highest weight $n$.

From the above consideration it follows that the following decomposition of the space of states $F_D=W$
holds:
\begin{equation}
W=F_D \cong \oplus _{i\in \mathbb{Z}} B_i .
\end{equation}
 Any highest weight module for $\mathcal{H}_{\mathbb{Z}}$ with highest weight $k$ is isomorphic to $\mathbb{C}[x_1, x_2, \dots , x_n, \dots ]$ via
\begin{equation*}
 \quad h_{n} =n\partial _{x_{n}} \quad \text{for} \ \ n>0; \quad
h_{-n} =x_{n}\cdot  \quad \text{for} \ \ n>0; \quad  h_{0} =k\cdot \ .
\end{equation*}
Thus we can  rewrite
\begin{equation}
\label{eqn:B_DcongF_D}
W=F_D \cong \oplus _{i\in \mathbb{Z}} B_i \cong  \mathbb{C}[e_{\phi}^{\alpha}, e_{\phi}^{-\alpha}] \ten \mathbb{C}[x_1, x_2, \dots , x_n, \dots ]=B_D,
\end{equation}
where   $e_{\phi}^{n\alpha}$, $e_{\phi}^{-n\alpha}$  are  for now just  labels for the highest weight vectors, but as we will see in the next section the notation is used because it is coming from the bosonic side, which is a twisted vertex algebra based on a Leibnitz module over a rank one abelian group (see Example \ref{example: LeibntizFreeAb}). Denote the right hand side of the isomorphism above by $B_D$, this is the bosonic space of states for the boson-fermion correspondence of type D-A. $\square$

We  modify the labeling of the highest weight vectors, as formulas \eqref{eqn:ExponD-1} and \eqref{eqn:ExponD-2} of Proposition \ref{prop:imageD} look simpler with such identification. Let
\begin{align}
& e_{\phi}^{n\alpha}\cong \phi^{2n-1}\cdot \phi^{2n-3}\cdots \phi^{3} \cdot \phi ^1 =(-1)^{n-1} a^{\text{odd}}_{n} \quad n>0\\
& e_{\phi}^{-n\alpha}\cong \phi^{2n}\cdot \phi^{2n-2}\cdots \phi^{2} \cdot \phi  =(-1)^{n} a^{\text{even}}_{n} \quad n\ge 0
\end{align}
 To prove Proposition  \ref{eqn:imageD}  we first calculate  the OPE
$h(z)\phi ^D(w)$. We use Corollary \ref{cor:opes-simple}, as all the poles are simple here.
The coefficients coming in front of the first order poles then are:
for $\frac{1}{z-w}$ we get $-Y(T\phi \cdot 1, w)=-Y(T \phi, w)$;  for $\frac{1}{z+w}$ we get $-Y(T \phi  \cdot 1, w)=-Y(T \phi, w)$. Thus
\begin{align*}
&h(z)\phi ^D(w)\sim \frac{1}{2}\big(-\frac{Y(T\phi , w)}{z-w} -\frac{Y(T\phi , w)}{z+w}\big) \sim \frac{-z}{z^2 -w^2}(T\phi) ^D(w),\\
&h(z)(T\phi) ^D(w)\sim \frac{1}{2}\big(-\frac{Y(\phi , w)}{z-w} -\frac{Y(\phi , w)}{z+w}\big) \sim \frac{-z}{z^2 -w^2}\phi ^D(w),
\end{align*}
and
\begin{align*}
&h(z)\frac{1}{2}(\phi ^D(w)+T\phi ^D(w)) \sim -\frac{z}{z^2 -w^2}\frac{1}{2}(\phi ^D(w)+T\phi ^D(w)),\\
&h(z)\frac{1}{2}(\phi ^D(w)-T\phi ^D(w))  \sim \frac{z}{z^2 -w^2}\frac{1}{2}(\phi ^D(w)-T\phi ^D(w)).
\end{align*}
Denote $e_{\phi}^{-\alpha}(w)=\frac{1}{2}(\phi ^D(w)+T\phi ^D(w))$, \ $e_{\phi}^{\alpha} (w)=\frac{1}{2}(\phi ^D(w)-T\phi ^D(w))$, we have
\begin{equation}
h(z) e_{\phi}^{-\alpha}(w) \sim -\frac{z}{z^2 -w^2}e_{\phi}^{-\alpha} (w),\quad
 h(z)e_{\phi}^{\alpha} (w)  \sim \frac{z}{z^2 -w^2}e_{\phi}^{\alpha}(w).
\end{equation}
These OPEs immediately imply the  commutation relations
\begin{align*}
&[h(z), e_{\phi}^{-\alpha} (w)] =-(i_{z, w} -i_{w, z})\frac{z}{z^2 -w^2}e_{\phi}^{-\alpha} (w)=-\frac{1}{2}(\delta (z-w) +\delta (z+w))e_{\phi}^{-\alpha} (w),\\
&[h(z), e_{\phi}^{\alpha} (w)] =(i_{z, w} -i_{w, z})\frac{z}{z^2 -w^2}e_{\phi}^{\alpha} (w)=\frac{1}{2}(\delta (z-w) +\delta (z+w))e_{\phi}^{\alpha} (w).
\end{align*}
 From these commutation relations  and the   exact description of the split of $W=F_D$ into Heisenberg submodules,  standard calculational lemmas (see for example \cite{Kac-Lie}, \cite{Wakimoto}) will give us
\begin{align}
\label{eqn:exponOperD-1}
\frac{1}{2}(\phi ^D(w)+\phi ^D(-w)) =e_{\phi}^{-\alpha}(z) & =\exp (-\sum _{n\ge 1}\frac{h_{-n}}{n} z^{2n})\exp (\sum _{n\ge 1}\frac{h_{n}}{n} z^{-2n})e_{\phi}^{-\alpha}z^{-2\partial_{\alpha}},\\
\label{eqn:exponOperD-2}
\frac{1}{2}(\phi ^D(w)-\phi ^D(-w)) =e_{\phi}^{\alpha}(z) & =\exp (\sum _{n\ge 1}\frac{h_{-n}}{n} z^{2n})\exp (-\sum _{n\ge 1}\frac{h_{n}}{n} z^{-2n})e_{\phi}^{\alpha}z^{2\partial_{\alpha}+1},
\end{align}
where the operators $e_{\phi}^{\alpha}$, $e_{\phi}^{-\alpha}$, $z^{\partial_{\alpha}}$ and $z^{-\partial_{\alpha}}$ act in an obvious way on the space $F_D\cong B_D$ as in \eqref{eqn:B_DcongF_D}.
This proves Proposition \ref{prop:imageD}.

Formulas \eqref{eqn:exponOperD-1} and \eqref{eqn:exponOperD-2} alone completely determine the twisted vertex algebra isomorphism between the two twisted vertex algebras: the fermionic  with space of states $F_D$ and the bosonic with space of states $B_D$.  Nevertheless,  in the  section \ref{subsection:latticesTVA} we will  present a bicharacter construction of the bosonic side of the boson-fermion correspondences.

Before moving to the bosonic sides, we want to give the bicharacter description of the fermionic side of the boson-fermion correspondence of type A. Even though it is a super vertex algebra (thus a twisted vertex algebra of order 1), the bicharacter construction gives us something new: a general formula for the vacuum expectation values that specializes to the determinant formula  \eqref{eqn:FermVEV-A} in the case of the charged free fermions of type A.

\subsection{Twisted vertex algebras based on $\mathbb{C}\{\phi , \psi \}$: determinant vacuum expectation values}
\label{subsection:Determinant}

 Recall that for the bicharacter construction examples are grouped based on  the Hopf algebra $M$, i.e. one keeps $M$ the same, but changes the bicharacter.  We dealt with two examples based on  $M=\mathbb{C}\{\phi \}$.
 Now we want to  work with  $M=\mathbb{C}\{\phi , \psi \}$, as in example \ref{example:Cphipsi}. A choice of  a super symmetric bicharacter on $M=\mathbb{C}\{\phi , \psi \}$ is determined  by the choice of three bicharacter values: $r_{z, w}(\phi \ten \phi)$, \ $r_{z, w}(\phi \ten \psi)$ and $r_{z, w}(\psi \ten \psi)$ (as $r_{z, w}(\psi \ten \phi) =-r_{w, z}(\phi \ten \psi)$ from the super-symmetry). We will restrict ourselves with the case when $r_{z, w}(\phi \ten \phi) =r_{z, w}(\psi \ten \psi) = 0$, thus we are only choosing $r_{z, w}(\phi \ten \psi)$. The pair $(\mathbb{C}\{\phi , \psi \}, r)$ for any covariant bicharacter will generate a  twisted vertex algebra with space of fields  $V=H^N_{T_{\epsilon}}(\mathbb{C}\{\phi , \psi \})$, and space of states $W=H_D(\mathbb{C}\{\phi , \psi \})$ as in theorem \ref{thm:Main}.

We want to derive a formula for the vacuum expectation values in any twisted vertex algebra based on the pair $(\mathbb{C}\{\phi , \psi \}, r)$, with the bicharacter $r$ chosen as above.
Recall $\langle
\ \mid\ \rangle: W\ten W \to \mathbb{C}$  is  a  symmetric bilinear form on the space of states $W$, such that the vacuum vector $1=|0\rangle$ is orthogonal to all other generators of the Hopf algebra $W$ and also has norm 1.
\begin{prop}
\label{prop:VacExpphipsi}
Let $V$ be a twisted vertex algebra based on $M=\mathbb{C}\{\phi, \psi \}$ and  a supersymmetric bicharacter $r$ (i.e., $V=H_D(\mathbb{C}\{\phi, T\phi, \psi , T\psi \})$ and $W=H_D(\mathbb{C}\{\phi , \psi\})$.
Denote by $\phi(z)$ and $\psi(z)$ the fields $Y(\phi, z)$ and $Y(\psi, z)$ produced by  \ref{eq:DefY}, via \eqref{eq:two-varX}.
The following formula for the vacuum expectation values holds:
\begin{equation}
\langle 0 \mid \phi(z_1)\phi(z_2) \dots \phi(z_{n}) \psi(w_1)\psi(w_2) \dots \psi(w_{n})|0\rangle = (-1)^{n(n-1)/2} i_{z, w} det \big( r_{z_i, w_j}(\phi \ten \psi )\big)_{i, j=1}^{n}.
\end{equation}
Here as usual $det$ denotes the determinant of an $n$ by $n$  square  matrix and $i_{z; w}$ stands for the expansion $i_{z_1, z_2, \dots , z_{n}, w_1, w_2, \dots , w_{n}}$.
\end{prop}
{\it Proof:}
To calculate the vacuum expectation values we calculate instead the vacuum expectation values of the analytic continuation
$ X_{z_1, z_2, \dots , z_{n}, w_1, w_2, \dots , w_{n}}(\phi \ten \phi \dots \phi\ten \psi \ten \psi \cdots \psi). $
We will use Lemma \ref{lem:analcont-n} which gives us a formula for the analytic continuation in terms of the bicharacter.
Both  $\phi$ and $\psi$ are  primitive elements, we have
\begin{align*}
& \del ^{2n}(\phi)=\phi \ten 1\ten \dots \ten 1 +1\ten \phi \ten \dots \ten 1 +1\ten 1\ten \phi \ten \dots \ten 1 +\dots 1\ten 1\ten \dots \ten \phi, \\
& \del ^{2n}(\psi)=\psi \ten 1\ten \dots \ten 1 +1\ten \psi \ten \dots \ten 1 +1\ten 1\ten \psi \ten \dots \ten 1 +\dots 1\ten 1\ten \dots \ten \psi.
\end{align*}
We need three observations:
\begin{enumerate}
\item Since for the bilinear form  the vacuum vector $1=|0\rangle$ spans an  orthogonal subspace on its own (and in particular is orthogonal to $\phi$ and $\psi$ and their descendants), the only contributions to the vacuum expectation values will come from the terms in the multivariable field where the coproducts have 1 as a first term; the other terms will not contribute. That forces us to work with the  $(2n)$-character \\ $r_{z_1, z_2, \dots , z_{n}, w_1, w_2, \dots , w_{n}}(\phi \ten \phi \dots \phi\ten \psi \ten \psi \cdots \psi).$
\item To continue the previous observation, no sign contribution will come from the first $(-1)$ factor  in the $\textbf{sign}(M_{\Delta^{2n-1}}^{\vec{k}} (\phi, \dots , \phi, \psi, \cdots , \psi))$ as the only contributing matrices are those with  the   first columns consisting  entirely  of 1s (as 1 is even).
\item Since $\phi $ and $\psi$ are  primitive elements  we have $r_{z, w}(\phi \ten 1)=r_{z, w}(1 \ten \phi)=0$ and
    $r_{z, w}(\psi \ten 1)=r_{z, w}(1 \ten \psi)=0$
    for  any  bicharacter. Thus the only contributions in the  $(2n)$-character $r_{z_1, z_2, \dots , z_{n}, w_1, w_2, \dots , w_{n}}(\phi \ten \phi \dots \phi\ten \psi \ten \psi \cdots \psi)$ will come from the following situation:  a nonzero summand in this $(2n)$-character  will be a product of nonzero bicharacter factors, and that happens  when in  the "permuted order" (see remark \ref{remark:sign-contr-z}) we have a sequence of  pairs: $(1, 1)$  pairs (trivial, as $r_{z, w}(1 \ten 1)=1$) or $(\phi, \psi)$ pairs (nontrivial). If there is a  mixed pair $(1, \phi)$ or $(\phi, 1)$ as a factor in a summand,  that summand will be 0. Note also that that the definition of a $(2n)$-character doesn't allow for pairs $(\psi, \phi)$, as it uses the "permuted order" in its definition.
     So a nonzero summand will have exactly $n$ such nontrivial contributing pairs $(\phi, \psi)$ , and  each pair forms one bicharacter $r_{z_{i}, w_{j}}(\phi \ten \psi)$. Which means that a nonzero summand  will consist of  the product $\prod_{k=1}^n r_{z_{k}, w_{j_k}}(\phi \ten \psi)$ times a sign factor, where $j_1, j_2, \dots, j_n$ is a permutation of $1, 2, \dots , n$.
\end{enumerate}
Thus, we have
\begin{align*}
& X_{z_1, z_2, \dots , z_{n}, w_1, w_2, \dots , w_{n}}(\phi \ten \phi \dots \phi\ten \psi \ten \psi \cdots \psi) \\ & = \sum_{\text{coproducts}} \textbf{sign}(M_{\Delta^{2n-1}}^{\vec{k}} (\phi, \dots , \phi, \psi, \dots, \psi)) \mathcal{E}_{z_{1}}\phi^{\prime}\cdot \cdot \cdot \mathcal{E}_{z_{n}}\phi^{\prime}\mathcal{E}_{w_{1}}\psi^{\prime} \cdots \mathcal{E}_{w_{n}}\psi^{\prime}\cdot \\
& \hspace{5cm} \cdot r_{z_1, z_2, \dots , z_{n}, w_1, w_2, \dots , w_{n}}(\phi^{\second}\ten \dots \phi^{\second}\ten \psi^{\second}\dots \ten \psi^{\second})\\
& =\sum_{\substack{\text{contr.} \\ \text{coproducts}}} \textbf{sign}(M_{\Delta^{2n-1}}^{\vec{k}} (\phi, \dots , \phi, \psi, \dots, \psi)) 1\cdot \\
& \hspace{5cm}  \cdot r_{z_1, z_2, \dots , z_{n}, w_1, w_2, \dots , w_{n}}(\phi \ten \phi \dots \phi\ten \psi \ten \psi \cdots \psi) +\text{other  terms} \\
& =(-1)^{\sfrac{n(n-1)}{2}} \sum_P \epsilon (P) 1 \cdot r_{z_{1}, w_{j_1}}(\phi \ten \psi)r_{z_{2}, w_{j_2}}(\phi \ten \psi)\cdots r_{z_{n}, w_{j_n}}(\phi \ten \psi) +\text{other  terms}.
\end{align*}
The sum is over all permutations $j_1, j_2, \dots, j_n$ of $1, 2, \dots , n$. The sign contribution from any \textbf{contributing} matrix $M_{\Delta^{2n-1}}^{\vec{k}} (\phi, \dots , \phi, \psi, \dots, \psi)) $ consist of two factors: the first factor is the sign due to all the  "$w$'s passing through the "$z$'s", which factor is exactly $(-1)^{\sfrac{n(n-1)}{2}}$. The second factor is precisely the sign of the corresponding permutation, since $\phi$ and $\psi $ are odd  (see remark \ref{remark:sign-contr-z} and observation 3 above, also permuting a pair across a pair contributes no minus sign). That produces precisely the determinant $(-1)^{\sfrac{n(n-1)}{2}} det \big( r_{z_i, w_j}(\phi \ten \psi )\big)_{i, j=1}^{n}$.
$\square$

\subsection{Super vertex algebra based on $\mathbb{C}\{\phi , \psi \}$: the charged free fermions of type A}
\label{section:freefermionA}

We start with the fermionic  side of the boson-fermion correspondence of type A, even though it is a super vertex algebra. We would like to recall that a super vertex algebra is in fact a twisted vertex algebra of order $N=1$. The bicharacter description of the fermionic side of the correspondence was given in \cite{A2}, we will recall it briefly for completeness, also the formula for the vacuum expectation values is new to this paper.
We choose
$V=H_D(\mathbb{C}\{\phi , \psi \})$, $W=V=H_D(\mathbb{C}\{\phi , \psi \})$, i.e., $\pi =Id_W$, and $r_{z, w}$ be the $H_D\ten H_D$-covariant bicharacter $r^{A_f}$ on $V=W$ generated by
\begin{equation}
r_{z, w}(\phi \ten \psi)=\frac{1}{z-w}, \quad r_{z, w}(\phi \ten \phi)=r_{z, w}(\phi \ten \psi) =0,
\end{equation}
which produces  a super-symmetric shift restricted bicharacter as in section \ref{subsection:Determinant}.
Let $Y$ be the field-state correspondence defined by \eqref{eq:DefY}, via \eqref{eq:two-varX}. The  set of data $(V, W, \pi=Id_W, Y)$ constructed as above satisfies the definition of a twisted vertex algebra, which is in fact a super vertex algebra, as the bicharacter has poles only at $z=w$.
Which vertex algebra it is is determined once we write out the OPEs for the Heisenberg relations.
We can define a Heisenberg element $h=\phi \cdot \psi$, which leads to $h(z)=:\phi (z)\psi (z)$ immediately from lemma \ref{lem:NormalOrderSimplePolesBich}. We have
from theorem \ref{thm:comlH_{D, T}}  that
\begin{equation}
h(z)h(w)\sim \frac{1}{(z-w)^2} \quad \text{and} \quad h(z) \phi (w) \sim \frac{1}{z-w}\phi (w),
\end{equation}
which in commutation relations is:
\begin{equation}
[h(z), \phi (w)] =(i_{z, w} -i_{w, z})\frac{1}{z-w}\cdot \phi (w)=\delta (z-w)\phi (w).
\end{equation}
From the last two equations it is a standard calculation that formulas \eqref{eqn:ExponA-1} and \eqref{eqn:ExponA-2} follow (these calculations are done for example in \cite{Kac} specifically for the boson-fermion correspondence of type A, as well as in \cite{Wakimoto}).

Thus we have shown that the pair $(\mathbb{C}\{\phi , \psi \}, r_{z, w}(\phi \ten \psi )=\frac{1}{z-w})$ generates and describes the super vertex algebra of the charged free fermions, which is the fermionic side of the boson-fermion correspondence of type A.

We can directly see that the determinant formula for the vacuum expectation values \eqref{eqn:FermVEV-A} is a special case of Proposition \ref{prop:VacExpphipsi}.

We won't give more examples here of a twisted vertex algebra based on $\mathbb{C}\{\phi , \psi \}$, but as was shown above such examples are easy to produce, as one just chooses a different bicharacter value $r_{z, w}(\phi \ten \psi )$.
Instead we move to the twisted vertex algebras that constitute  the bosonic sides of the boson-fermion correspondences.

\subsection{Twisted vertex algebras based on $\mathbb{C}[\mathbb{Z}\alpha]$ and a choice of a bicharacter}
\label{subsection:latticesTVA}

  In this section we fix   $M$ to  be the Hopf algebra  $L_1=\mathbb{C}[\mathbb{Z}\alpha]$,  the group algebra of the rank-one free abelian group $\mathbb{Z}\alpha $, as in Example  \ref{example: LeibntizFreeAb}.
 We  constructed the free Leibnitz module $\tilde{V} =H^N_{T_{\epsilon}}(L_1)$, and its sub-Hopf algebra $\tilde{W}=H_D(L_1)$. If we want to define a  $H^N_{T_{\epsilon}}\ten H^N_{T_{\epsilon}}$-covariant  bicharacter on $H^N_{T_{\epsilon}}(M)$ it is clear that we can only choose one bicharacter value, that of $r_{z, w}(e^{\alpha}\ten e^{\alpha})$. Recall (Example  \ref{example: LeibntizFreeAb}) that we  defined an element  $h=(De^\alpha)e^{-\alpha}$, which is primitive; note that $h\in \tilde{W}\subset \tilde{V}$.
\begin{lem}\label{lem:HeisFromGroup}
The following hold for any covariant bicharacter on $\tilde{V}$:
\begin{align}
\label{eqn:bichar-h-grouplike}
&r_{z, w}(h\ten e^{m\alpha})=m\partial _z \log r_{z, w}(e^{\alpha}\ten e^{\alpha});\\
\label{eqn:bichar-h-h-inlat}
&r_{z, w}(h\ten h)=\partial _z \partial _w\log r_{z, w}(e^{\alpha}\ten e^{\alpha});\\
\label{eqn:OPEgeneralHeis}
&h(z)h(w)\sim Id_W\cdot r_{z, w}(h\ten h);\\
\label{eqn:OPEgeneralExpon}
&h(z)e^{m\alpha}(w)\sim e^{m\alpha}(w)\cdot r_{z, w}(h\ten e^{m\alpha}).
\end{align}
\end{lem}

\subsection{Twisted vertex algebras based on $\mathbb{C}[\mathbb{Z}\alpha]$: product vacuum expectation values}
\label{subsection:Product}

Let $V$ be a twisted vertex algebra based on $L_1=\mathbb{C}[\mathbb{Z}\alpha]$ and  a supersymmetric bicharacter $r$ (by use of Theorem \ref{thm:Main}) with space of states  $W=H_D(L_1)$.
Denote by $e^{m\alpha}(z)$ the field $Y(e^{m\alpha}, z)$ produced by  definition \ref{eq:DefY}, via \eqref{eq:two-varX}. Let the projection map $\pi$ from the space of fields to the space of states satisfy $\pi (e^{m\alpha})\ne 0$ for any   $m\in \mathbb{Z}$  (which holds for $\pi=\pi_T$).
\begin{prop}
\label{prop:VacExpalpha}
The following formula for the vacuum expectation values holds:
\begin{equation*}
\langle 0 \mid e^{m_1\alpha}(z_1)e^{m_2\alpha}(z_2) \dots e^{m_n\alpha}(z_{n})|0\rangle =i_{z} \delta_{m_{1}+ m_{2}+ \dots + m_{n}, 0}\prod_{i<j=1}^n r_{z_i, z_j}(e^{m_i\alpha} \ten e^{m_j\alpha}).
\end{equation*}
Here  $i_{z}$ stands for the expansion $i_{{z_1}, {z_1}, \dots , {z_{n}}}$.
\end{prop}
{\it Proof:} From Lemma \ref{lem:analcont-n}, the fact that  the  parity is entirely even, and the elements $e^{m_k\alpha}$ are grouplike we have
\begin{equation*}
X_{z_{1}, z_{2}, \dots , z_{n}}(e^{m_1\alpha}\ten e^{m_2\alpha}\ten \dots \ten e^{m_n\alpha}) = \mathcal{E}_{z_{1}}e^{m_1\alpha}\cdot \mathcal{E}_{z_{2}}e^{m_2\alpha}\cdots \mathcal{E}_{z_{n}}e^{m_n\alpha}\cdot
r_{z_1, z_2, \dots, z_n}(e^{m_1\alpha}\ten e^{m_2\alpha}\ten \dots \ten e^{m_n\alpha}).
\end{equation*}
We have
\begin{equation*}
\langle 0 \mid \mathcal{E}_{z_{1}}e^{m_1\alpha}\cdot \mathcal{E}_{z_{2}}e^{m_2\alpha}\cdots \mathcal{E}_{z_{n}}e^{m_n\alpha}\rangle= \langle 0 \mid \pi_T(e^{(m_{1}+ m_{2}+ \dots + m_{n})\alpha }) +O(z) \rangle .
\end{equation*}
Since we required that the bilinear form is such that the vacuum vector $|0\rangle$ is orthogonal to all $e^{m\alpha}$, except for the $m=0$, then
\begin{equation*}
\langle 0 \mid e^{(m_{1}+ m_{2}+ \dots + m_{n})\alpha }\rangle =\delta_{m_{1}+ m_{2}+ \dots + m_{n}, 0}.
\end{equation*}
Note also that the $O(z)$
 terms contain non-vacuum descendants  of the $e^{m_k\alpha}$, and so do not contribute to the vacuum expectation value. Thus
\begin{equation*}
\langle 0 \mid X_{z_{1}, \dots , z_{n}}(e^{m_1\alpha}\ten \dots \ten e^{m_n\alpha})\rangle =\delta_{m_{1}+ \dots + m_{n}, 0} r_{z_1, z_2, \dots, z_n}(e^{m_1\alpha}\ten e^{m_2\alpha}\ten \dots \ten e^{m_n\alpha}).
\end{equation*}
Now since the elements $e^{m_k\alpha}$ are grouplike, the $n$-character $r_{z_1, z_2, \dots, z_n}(e^{m_1\alpha}\ten  \dots \ten e^{m_n\alpha})$ also has especially simple form:
\begin{equation*}
r_{z_1, z_2, \dots, z_n}(e^{m_1\alpha}\ten e^{m_2\alpha}\ten \dots \ten e^{m_n\alpha}) =\prod_{i<j=1}^n r_{z_i, z_j}(e^{m_i\alpha}\ten e^{m_j\alpha}).\quad \square
\end{equation*}

\subsection{Super vertex algebra based on $\mathbb{C}[\mathbb{Z}\alpha]$: the free boson of type A}
\label{section:freebosonA}

We start with the bosonic side of the boson-fermion correspondence of type A, even though it is a super vertex algebra. We would like to  recall that a super vertex algebra is in fact a twisted vertex algebra of order $N=1$. This section  gives the bicharacter description of the bosonic side of the correspondence.

We choose
$V=H_D(L_1)$, $W=V=H_D(L_1)$, i.e., $\pi =Id_W$, and $r^{A_b}_{z, w}$ be the $H_D\ten H_D$-covariant bicharacter  on $V=W$ generated by
\begin{equation}
r_{z, w}(e^{\alpha}\ten e^{\alpha})=z-w,
\end{equation}
which is a super-symmetric shift restricted bicharacter.
As was derived in the previous section for general bicharacter
\begin{equation}
r_{z, w}(h\ten e^{m\alpha})=m\frac{1}{z-w},\quad r_{z, w}(h\ten h)=\frac{1}{(z-w)^2}.
\end{equation}
Let $Y$ be the field-state correspondence defined by \eqref{eq:DefY}, via \eqref{eq:two-varX}. The  set of data $(V, W, \pi=Id_W, Y)$ constructed as above satisfies the definition of a twisted vertex algebra, which is in fact a super vertex algebra, as the bicharacter has poles only at $z=w$.
Which particular vertex algebra we get is determined once we write the OPEs for the Heisenberg relation \eqref{eqn:OPEgeneralHeis} and the exponential relation \eqref{eqn:OPEgeneralExpon} as commutation relations:
\begin{equation}
\label{eqn:OPEgeneralHeisA}
[h(z),h(w)]=(i_{z, w} -i_{w, z})\frac{1}{(z-w)^2}=\partial _w \delta (z-w),
\end{equation}
which is precisely \eqref{eqn:HeisOPEsA}; and
\begin{equation}
[h(z),e^{m\alpha}(w)] =(i_{z, w} -i_{w, z})\frac{m}{z-w}\cdot e^{m\alpha}(w)=m\delta (z-w)e^{m\alpha}(w).
\end{equation}
From the last two equations it is a standard calculation that formulas \eqref{eqn:ExponA-1} and \eqref{eqn:ExponA-2} follow (these calculations are done for example in \cite{Kac} specifically for the boson-fermion correspondence of type A, as well as in \cite{Wakimoto}).

Thus we have shown that the pair $(L_1, r_{z, w}(e^{\alpha}\ten e^{\alpha})=z-w)$ generates and describes the super vertex algebra of the rank one odd lattice, which is the bosonic side of the boson-fermion correspondence of type A.
To summarize all these considerations:
\begin{thm}
The boson-fermion correspondence of type A is the isomorphism between two super vertex algebras: the fermionic side, which is the vertex algebra based on the pair $(\mathbb{C}\{\phi , \psi \}, r^{A_f})$; and the bosonic side, which is the vertex algebra based on the pair  $(\mathbb{C}[\mathbb{Z}\alpha], r^{A_b})$.
\end{thm}
We can directly see that \eqref{eqn:BosVEV-A} is a special case of Proposition \ref{prop:VacExpalpha}, as
\begin{equation}
r_{z_i, w_j}(e^{\alpha} \ten e^{-\alpha}) =\frac{1}{z_i -w_j}, \quad r_{z_i, z_j}(e^{\alpha} \ten e^{\alpha}) =z_i -w_j \quad r_{w_i, w_j}(e^{-\alpha} \ten e^{-\alpha}) =w_i -w_j,
\end{equation}
thus
\begin{equation}
\langle  0 |e^{\alpha} (z_1)e^{\alpha} (z_2)\dots e^{\alpha} (z_n)e^{-\alpha} (w_1)e^{-\alpha} (w_2)\dots e^{-\alpha} (w_n) | 0 \rangle =i_{z, w}\frac{\prod_{i<j}^n ((z_i-z_j)(w_i-w_j))}{\prod_{i, j=1}^n (z_i-w_j)}.
\end{equation}
Here $i_{z; w}$  stands for the expansion $i_{z_1, z_2,\dots, z_n, w_1, \dots, w_n}$. We see that since the boson-fermion correspondence identifies the fields $e^{\alpha }(z)=\phi ^B(z)$, then Corollary  \ref{cor:VacuumExpEqA} follows directly.

\subsection{Twisted vertex algebra based on $\mathbb{C}[\mathbb{Z}\alpha]$: the free boson of type B}
\label{section:freebosonB}

We continue with the bosonic side of the boson-fermion correspondence of type B. This is the first example  where the spaces of states and fields are not free Leibnitz modules, but  quotients of a free Leibnitz module.

 Recall the free Leibnitz module $\tilde{V} =H^N_{T_{\epsilon}}(L_1)$, and its sub-Hopf algebra $\tilde{W}=H_D(L_1)$.  We again take $\epsilon =-1$ and write just $T$ instead of $T_{\epsilon}$. Let
 \begin{equation}
 V=\tilde{V}/ \{Te^{\alpha}=e^{-\alpha} \}.
 \end{equation}
Denote the quotient relations generated from $\{Te^{\alpha}=e^{-\alpha} \}$ by $\mathcal{R_B}$ .

 In order to define a  $H^2_{T_{\epsilon}}\ten H^2_{T}$-covariant  bicharacter on $V$  we can only choose the bicharacter value  $r_{z, w}(e^{\alpha}\ten e^{\alpha})$. Further,  in order to be able to restrict this bicharacter to a  bicharacter on $V=\tilde{V}/ \mathcal{R_B}$, it needs to be consistent with the relations $\mathcal{R_B}$, thus
\begin{equation}
\label{eqn:BichRelationsIn}
r_{-z, w}(e^{\alpha}\ten e^{\alpha})=r_{z, -w}(e^{\alpha}\ten e^{\alpha}) =\frac{1}{r_{z, w}(e^{\alpha}\ten e^{\alpha})}.
\end{equation}
Hence, we can  choose
\begin{equation}
\label{eqn:BichL1-B}
r_{z, w}(e^{\alpha}\ten e^{\alpha})=\frac{z-w}{z+w},
\end{equation}
and this bicharacter value will generate a bicharacter $r^{B_b}$ on $V$ by covariance.

Now we turn to the exact description of the space of fields $V$ and the space of states $W$ of the twisted vertex algebra generated by the pair $(L_1/\mathcal{R_B}, r^{B_b})$. From Example  \ref{example: LeibntizFreeAb},
the free Leibnitz module $H^2_{T}(L_1)$ is isomorphic to $L_2\ten H^2_{T}(\mathbb{C}[h])$, where $L_2$ is the group algebra $L_2=\mathbb{C}[\mathbb{Z}\alpha, \mathbb{Z}\alpha _1]$ of the  free abelian group of rank 2 (we identify $Te^{\alpha}$, which is grouplike, with $e^{\alpha _1}$). Denote by $h^B_{\alpha}$ the element $h^B_{\alpha}=\frac{1}{2}(De^{\alpha})Te^{\alpha}\in V$, which  coincides with  $\frac{1}{2}(De^{\alpha})e^{-\alpha}\in V$ under the relation $ \mathcal{R_B}$. It follows then that $TDe^{\alpha}=-DTe^{\alpha}=-De^{-\alpha}$. Therefore
\begin{equation}
Th^B_{\alpha} =h^B_{\alpha}.
\end{equation}
Thus under the imposed relations $\mathcal{R_B}$ in $V$ we have $H^2_{T}(\mathbb{C}[h])/ \mathcal{R_B} = H_D(\mathbb{C}[h^B_{\alpha}])$ and
\begin{equation}
V=L_1\ten H_D(\mathbb{C}[h^B_{\alpha}]).
\end{equation}
The space of states $W$ is defined via the
  projection map $\pi: V\to W$, and in order to apply Theorem \ref{thm:Main} we  use as projection map the map from definition \ref{defn:TprojMap} adapted to the relations $\mathcal{R_B}$. More precisely, define
$\pi: V\to W$ to be the linear map defined by
\begin{equation}
\pi(H^2_{T}(\mathbb{C}[h^B_{\alpha}])/ \mathcal{R_B})=Id, \quad \pi(Te^{n\alpha})=e^{n\alpha}, \  \ \pi(e^{n\alpha})=e^{n\alpha},  \ n\in \mathbb{Z}.
\end{equation}
Denote by $\bar{v}$ the element of $W$ that  is the projection of the element $v\in V$.
We have
\begin{equation*}
\overline{1}  = \overline{e^{\alpha}}\overline{e^{-\alpha}} =\pi(e^{\alpha}e^{-\alpha})= \pi(e^{\alpha}Te^{\alpha}) = \pi(e^{\alpha})\pi(Te^{\alpha}) =\pi (e^{2\alpha}) =\overline{e^{2\alpha}},
\end{equation*}
thus we have in $W$
\begin{equation}
\overline{e^{2\alpha}}=1, \quad \overline{e^{\alpha}} =\overline{e^{-\alpha}}.
\end{equation}
Thus $W=H_D(\mathbb{C}[h^B_{\alpha}]) \oplus e^{\alpha}H_D(\mathbb{C}[h^B_{\alpha}])$
 and  as expected,  $W=B_B$ as in Lemma \ref{lem:BosonicSpaceBB}.
Moreover, from Lemma \ref{lem:HeisFromGroup}  $h^B_{\alpha}$ is a Heisenberg element:
we use
\eqref{eqn:OPEgeneralHeis}, which in this case specializes from \eqref{eqn:bichar-h-h-inlat} to
\begin{equation*}
h^B_{\alpha}(z)h^B_{\alpha}(w)\sim 1\cdot  r_{z, w}(h^B_{\alpha}\ten h^B_{\alpha})\sim 1\cdot \frac{1}{4}\partial _w\partial _z \log \frac{z-w}{z+w}\sim 1\cdot \frac{z^2 +w^2}{2(z^2-w^2)^2}.
\end{equation*}
Now the unexpected twist here is that $Th^B_{\alpha} =h^B_{\alpha}$,
hence the field $h^B_{\alpha}(z)$ has only \textbf{even} powers of $z$, and we can write  $h^B_{\alpha}(z)=\sum _{n\in \mathbb{Z}}h_{2n+1} z^{-2n}$. This immediately leads to the required commutation relations
\begin{equation}
[h_m, h_n] =\frac{m}{2}\delta_{m +n, 0},  \quad m, n\ \ \text{odd \ integers}.
\end{equation}
Note that we can reindex this field
$h^B_{\alpha}(z)$ in the following form: $h^B_{\alpha}(z)=\sum _{n\in \mathbb{Z} +1/2}\tilde{h}_{(n)} z^{-2n-1}$ (i.e., $h_{2n}=\tilde{h}_{(n)}$ for $n\in \mathbb{Z} +1/2$),
which translates to
\begin{equation}
[\tilde{h}_{(m)}, \tilde{h}_{(n)}] =m\delta_{m +n, 0},  \quad m, n\in \mathbb{Z} +1/2,
\end{equation}
and explains the name $\mathcal{H}_{\mathbb{Z}+1/2}$ for this Heisenberg algebra.
\begin{remark}
Note that
\begin{equation}
\label{eqn:isomPeculB}
h^B_{\alpha}(z) =z\cdot h^B_{\phi}(z),
\end{equation}
 which is allowed in an isomorphism of  twisted vertex algebras: the doubly infinite sequences of the modes of the two fields $h^B_{\alpha}(z)$ and $h^B (z)$ are identical, except for  the \textbf{shift} in the indexing (the multiplication by $z$).
\end{remark}
For the OPEs of the fields $e^{m\alpha}(w)$ with $h^B_{\alpha}(z)$, from  \eqref{eqn:OPEgeneralExpon} we get
\begin{equation}
h^B_{\alpha}(z)e^{m\alpha}(w)\sim me^{m\alpha}(w)\cdot \frac{w}{z^2-w^2}.
\end{equation}
We see that if we identify $e^{\alpha }(z)=\phi ^B(z)$, this OPE   implies  the exponential operator formula  \eqref{eqn:imageB} of the boson-fermion correspondence.

Thus we have shown that the pair $(L_1/\mathcal{R_B}, r^{B_b})$ generates the twisted  vertex algebra  which constitutes the bosonic side of the boson-fermion correspondence of type B, and we have
\begin{thm}
The boson-fermion correspondence of type B is an isomorphism between two twisted vertex algebras: the fermionic side, which is the vertex algebra based on the pair $(\mathbb{C}\{\phi \}, r^{B_f})$; and the bosonic side, which is the twisted vertex algebra based on the pair  $(\mathbb{C}[\mathbb{Z}\alpha]/\mathcal{R_B}, r^{B_b})$.
\end{thm}
Lemma \ref{lem:VacuumExpEqB} follows directly:  Since the boson-fermion correspondence identifies the fields $e^{\alpha }(z)=\phi ^B(z)$, we can directly calculate the vacuum expectation values on each  side of the correspondence of type B and equate. As  a special case of Proposition \ref{prop:VacExpalpha} we have for the bosonic side:
\begin{equation}
\langle  0 |e^{\alpha} (z_1)e^{\alpha} (z_2)\dots e^{\alpha} (z_{2n}) | 0 \rangle =i_{z}\prod_{i<j}^{2n} \frac{z_i -z_j}{z_i +z_j}.
\end{equation}
Similarly, the vacuum expectation values on the fermionic side follow directly from Proposition \ref{prop:VacExpphi}.

\subsection{Twisted vertex algebra based on $\mathbb{C}[\mathbb{Z}\alpha]$: the free boson of type D-A}
\label{section:freebosonD}
\
We continue with the bosonic side of the boson-fermion correspondence of type D-A.

 We are again working with  the free Leibnitz module $\tilde{V} =H^N_{T_{\epsilon}}(L_1)$, and its sub-Hopf algebra $\tilde{W}=H_D(L_1)$, $T=T_{\epsilon}$.  For the bosonic space of type D-A we  let
 \begin{equation}
 V=\tilde{V}/ \{Te^{\alpha}=e^{\alpha} \},
 \end{equation}
i.e., $V$ is the quotient Leibnitz module modulo the relations $\mathcal{R_D}$ generated by $\{Te^{\alpha}=e^{\alpha} \}$.

 As we did in the previous section, if we want to define a  $H^2_{T}\ten H^2_{T}$-covariant  bicharacter on $V$, we need to choose  a bicharacter value  $r_{z, w}(e^{\alpha}\ten e^{\alpha})$ which is consistent with the relations $\mathcal{R_D}$. It follows that  $r_{z, w}(e^{\alpha}\ten e^{\alpha})$ needs to be even as a function of both  $z$ and $w$, as well as supersymmetric with exchange of $z$ and $w$.
Thus we can choose
\begin{equation}
\label{eqn:BichL1-D}
r_{z, w}(e^{\alpha}\ten e^{\alpha})=z^2 -w^2,
\end{equation}
which will generate a bicharacter $r^{D_b}$ on $V$ by covariance.

Now we turn to the exact description of the space of fields $V$ and the space of states $W$ of this twisted vertex algebra.  Denote by $h^D_{\alpha}$ the element  $\frac{1}{2}(De^{\alpha})e^{-\alpha}\in V$, which we know is a Heisenberg element. In $V$ due to the relations $\mathcal{R_D}$ we have $TDe^{\alpha}=-DTe^{\alpha}=-De^{\alpha}$, and  $Te^{-\alpha}=e^{-\alpha}$. Thus
\begin{equation*}
Th^D_{\alpha}=\frac{1}{2}(TDe^{\alpha})Te^{-\alpha} =-\frac{1}{2}DTe^{\alpha}e^{-\alpha}=-\frac{1}{2}De^{\alpha}e^{-\alpha}=-h^D_{\alpha}.
\end{equation*}
Hence
\begin{equation}
Th^D_{\alpha} =-h^D_{\alpha},
\end{equation}
 thus under the imposed relations $\mathcal{R_D}$ in $V$ we again  have $H^2_{T}(\mathbb{C}[h])/ \mathcal{R_D} = H_D(\mathbb{C}[h^D_{\alpha}])$, although now $h^D_{\alpha}$ is odd under $T$. Hence
\begin{equation}
V=L_1\ten H_D(\mathbb{C}[h^D_{\alpha}]).
\end{equation}
We define the space of states $W$  to be equal to $V$, i.e., we take as projection map $\pi: V\to W$  the identity map on $V$, which is consistent with the projection map $\pi_T$ modulo  $\mathcal{R_D}$. Hence we can again use Theorem \ref{thm:Main} to get a twisted vertex algebra.

From
\eqref{eqn:OPEgeneralHeis} and \eqref{eqn:bichar-h-h-inlat}  we have
\begin{equation}
\label{eqn:OPEgeneralHeisalphaD}
h^D_{\alpha}(z)h^D_{\alpha}(w)\sim 1\cdot \frac{zw}{(z^2-w^2)^2}.
\end{equation}
Here $Th^D_{\alpha} =-h^D_{\alpha}$,
hence the field $h^D_{\alpha}(z)$ has only \textbf{odd} powers of $z$, and we can write it as $h^D_{\alpha}(z)=\sum _{n\in \mathbb{Z}}h_n z^{-2n-1}$. Note that the field $h^D_{\alpha}(z)$ has the same OPE as the  Heisenberg field $h^D(z)$, \eqref{eqn:HeisOPEsD}, as required.

On the bosonic side  it is  easy to identify  the split into irreducible Heisenberg modules, as the highest weight vectors are precisely the elements  $e^{n\alpha}\in V\equiv W, \ n\in \mathbb{Z}$.  Hence as Heisenberg modules
\begin{align}
V\equiv W & \equiv    \mathbb{C}[e^{\alpha}, e^{-\alpha}] \ten \mathbb{C}[h^D_{\alpha}, Dh^D_{\alpha}, \dots , D^{(n)}h^D_{\alpha}, \dots ]\cong\\
& \cong    \mathbb{C}[e^{\alpha}, e^{-\alpha}] \ten \mathbb{C}[x_1, x_2, \dots , x_n, \dots ]\cong \oplus _{i\in \mathbb{Z}} B_i =B_D.
\end{align}
Now it remains to calculate the OPEs of the fields $e^{m\alpha}(w)$ with $h^D_{\alpha}(z)$, from \eqref{eqn:OPEgeneralExpon} we get
\begin{equation}
h^D_{\alpha}(z)e^{m\alpha}(w)\sim me^{m\alpha}(w)\cdot \frac{z}{z^2-w^2}.
\end{equation}
The  commutation relations for $e^{\alpha}(z)$ and $e^{-\alpha}(z)$ thus are:
\begin{equation*}
[h(z),e^{\pm \alpha}(w)] =\pm (i_{z, w} -i_{w, z}) \frac{z}{z^2-w^2} \cdot e^{\pm \alpha}(w)=\pm \frac{1}{2}(\delta(z-w) +\delta(z+w))e^{\pm \alpha}(w).
\end{equation*}
 From the standard vertex operator calculations this commutation relations immediately imply  the exponential operator formulas
\begin{align}
\label{eqn:exponOperD-B-1}
e^{-\alpha}_D (z)  =e^{-\alpha}(z) & =\exp (-\sum _{n\ge 1}\frac{h_{-n}}{n} z^{2n})\exp (\sum _{n\ge 1}\frac{h_{n}}{n} z^{-2n})e^{-\alpha}z^{-2\partial_{\alpha}},\\
\label{eqn:exponOperD-B-2}
e^{\alpha}_D (z) = e^{\alpha}(z) & =\exp (\sum _{n\ge 1}\frac{h_{-n}}{n} z^{2n})\exp (-\sum _{n\ge 1}\frac{h_{n}}{n} z^{-2n})e^{-\alpha}z^{2\partial_{\alpha}}.
\end{align}
Note that both of these are entirely even in the variable $z$  operators, i.e., $e^{-\alpha}_D(z)=e^{-\alpha}_D(-z)$ and $e^{\alpha}_D(z)=e^{\alpha}_D(-z)$, which is of course consistent with the relations $\mathcal{R_D}$  ($Te^{-\alpha}=e^{-\alpha}$ and $Te^{\alpha}=e^{\alpha}$). Note that
\begin{equation}
\label{eqn: D to A equiv}
e^{\alpha}_{D}(z) = e^{\alpha}_A (z^2), \quad e^{-\alpha}_{D} (z) = e^{-\alpha}_A (z^2),
\end{equation}
where the operators $e^{\alpha}_A (z)$ and $e^{-\alpha}_A (z)$ describe the boson-fermion correspondence of type A. This is a very interesting occurrence, and is discussed also  \cite{Rehren} from the point of view of the ``fermionization". One should note though that an isomorphism of the spaces of states as Heisenberg modules and the "change of variables" formula \eqref{eqn: D to A equiv}  do not in imply an  isomorphism as twisted vertex algebras---these two vertex algebras have a different set of singularities in the OPEs. The equivalence, as \cite{Rehren} notes, is as CAR algebras.

We see that the boson-fermion correspondence of type D-A identifies
\begin{equation}
\label{eqn: finalDiso}
\phi^D(z) = e^{-\alpha}_D(z) +ze^{\alpha}_D(z), \quad (T\phi) ^D(z) =e^{-\alpha}_D(z) -ze^{\alpha}_D(z).
\end{equation}
Here again we see the "shifts" allowed in an isomorphism of vertex algebras.

Thus we have shown that the pair $(L_1/\mathcal{R_D}, r_{z, w}(e^{\alpha}\ten e^{\alpha})=z^2 -w^2)$  describes the twisted  vertex algebra  which is the bosonic side of the boson-fermion correspondence of type D-A.
To summarize all these considerations we have:
\begin{thm}
The boson-fermion correspondence of type D-A is an isomorphism between two twisted vertex algebras: the fermionic side, which is the twisted vertex algebra based on the pair $(\mathbb{C}\{\phi \}, r^{D_f})$; and the bosonic side, which is the twisted vertex algebra based on the pair  $(\mathbb{C}[\mathbb{Z}\alpha]/\mathcal{R_D}, r^{D_b})$.
\end{thm}
In order to prove Lemma \ref{lem:VacuumExpEqD} we need to compare the vacuum expectation values on the bosonic side with those on the fermionic side.  We need to take into account the isomorphism formula \eqref{eqn: finalDiso} if we are to apply Proposition \ref{prop:VacExpalpha}.
 \begin{align*}
&\langle  0 |\phi^D (z_1)\phi^D (z_2)\dots \phi^D (z_{2n}) | 0 \rangle = \langle  0 |(e^{-\alpha}_D(z_1) +z_1e^{\alpha}_D(z_1)) (e^{-\alpha}_D(z_2) +z_2e^{\alpha}_D(z_2))\dots (e^{-\alpha}_D(z_{2n}) +z_{2n}e^{\alpha}_D(z_{2n}))| 0 \rangle  \\
& \quad \quad =\langle  0 |e^{-\alpha} (z_1)e^{-\alpha} (z_2)\dots e^{-\alpha} (z_{2n}) | 0 \rangle +\sum_{i=1}^{2n} z_i  \langle  0 |e^{-\alpha} (z_1)\dots e^{\alpha} (z_i)\dots e^{-\alpha} (z_{2n}) | 0 \rangle  \\
& \hspace{2cm} + \sum_{i<j}^{2n} z_i z_j  \langle  0 |e^{-\alpha} (z_1)\dots e^{\alpha} (z_i)\dots e^{\alpha} (z_j)\dots e^{-\alpha} (z_{2n}) | 0 \rangle +\dots \\
& \hspace{3cm} + \sum_{i_1 <i_2\dots <i_k}^{2n} z_{i_1} z_{i_2}\cdots z_{i_k}  \langle  0 |e^{-\alpha} (z_1)\dots e^{\alpha} (z_{i_1})\dots e^{\alpha} (z_{i_k}) \dots e^{-\alpha} (z_{2n}) | 0 \rangle +\dots
\end{align*}
The factor of $\delta_{m_{1}+ m_{2}+ \dots + m_{n}, 0}$ in the right-hand side of the Proposition \ref{prop:VacExpalpha} forces all the sums but one to vanish: the only remaining sum has the product  of exactly $n$ factors of $z_{i_k}$  in it, as it will have exactly $n$ $e^{\alpha}$'s  and as many $e^{-\alpha}$'s in it. Denote in this non-vanishing sum the indexes corresponding to $e^{-\alpha}$ by $z_{j_k}$. Hence we have a disjoint  split $\{1, 2, \dots 2n\} =\{i_1, i_2, \dots , i_n\} \sqcup \{j_1, j_2, \dots , j_n\}$ and we can write the non-vanishing sum as \begin{equation*}
\sum_{i_1 <i_2\dots <i_n}^{2n} z_{i_1} z_{i_2}\cdots z_{i_n}  \langle  0 |e^{-\alpha} (z_{j_1})e^{\alpha} (z_{i_1}) e^{-\alpha} (z_{j_2})e^{\alpha} (z_{i_2})\dots e^{-\alpha} (z_{j_n})e^{\alpha} (z_{i_n}) | 0 \rangle.
\end{equation*}
From Proposition \ref{prop:VacExpalpha} and
$r_{z_i, z_j}(e^{\alpha} \ten e^{-\alpha})= r_{z_i, z_j}(e^{-\alpha} \ten e^{\alpha})= \frac{1}{z_i^2 -z_j^2}$,
\begin{equation*}
\langle  0 | \phi^D (z_1)\phi^D (z_2)\dots \phi^D (z_{2n})  | 0 \rangle =i_{z}\frac{\sum_{i_1 <i_2\dots <i_n}^{2n} z_{i_1} z_{i_2}\cdots z_{i_n}  \prod_{k<l}^n (z_{i_k}^2 - z_{i_l}^2)\prod_{p < q}^n(z_{j_p}^2 - z_{j_q}^2)}{\prod_{k, p}^{n} z_{i_k}^2 - z_{j_p}^2}.
\end{equation*}
 Thus since the boson-fermion correspondence identifies the fields $\phi^D(z) = e^{-\alpha}_D(z) +ze^{\alpha}_D(z)$,  lemma \ref{lem:VacuumExpEqD} follows directly as the Pfaffian equality for the vacuum expectation values on the fermionic side is a special case of  Proposition \ref{prop:VacExpphi}). It is important that this  Pfaffian identity follows directly from the correspondence of type D, and is a representative "imprint" of the boson-fermion correspondence. $\square$

\subsection{Boson-fermion correspondence of type  D-A and order $N$ }
\label{section:orderN}

In this section we  briefly show how the boson-fermion correspondence of type D-A extends to order $N$ (and thereby give two examples of twisted vertex algebra of order $N$).
Similarly to section \ref{section:freefermionD} we are working with a space of fields $V=H^N_{T}(\mathbb{C}\{\phi \})$, and a space of states $W=H_D(\mathbb{C}\{\phi\})$, but now we allow  $N$ to be any natural number, $N\ge 2$. The projection map (recall definition \ref{defn:TprojMap})  is again the algebra homomorphism  defined by $\pi_T(T^i\phi)=\phi$, for any $0\le i\le N-1$.
As in Section \ref{section:freefermionD} the bicharacter  $r^D: V\ten V \to \mathbf{F}^N(z, w)^{+, w}$ is defined by
$r^D_{z, w}(\phi \ten \phi)=\frac{1}{z-w}$.
Theorem \ref{thm:Main} guarantees  that the pair $(\mathbb{C}\{\phi \}, r^D_{z, w})$ will generate  a twisted vertex algebra of order $N$.
In the twisted vertex  algebra we have the following  descendant fields  $T^i\phi ^D (z)=\phi ^D (\epsilon ^i z)$, for any $0\le i\le N-1$, with OPEs
\begin{equation*}
T^i\phi ^D (z)T^j\phi ^D(w) \sim \frac{1}{\epsilon ^iz- \epsilon ^jw}.
\end{equation*}
The boson-fermion correspondence is determined via the Heisenberg element
\begin{equation}
\label{eqn:HeisOrderN}
h=\frac{1}{N}\sum_{i=0}^{N-1} \epsilon^{i-1}  (T^{i-1}\phi ) (T^{i}\phi).
\end{equation}
We have $Th=\epsilon^{-1} h$,
hence we index the field $h(z):=Y(h, z)=\sum _{n\in \mathbb{Z}}h_n z^{-Nn-1}$. Lemma \ref{lem:HeisOrderN} then holds immediately.
Denote
\[
e_{\phi}^{\alpha} (w)=\frac{1}{N}(\sum_{i=0}^{N-1} \epsilon^{-i}T^{i}\phi ^D(w)), \quad
e_{\phi}^{\epsilon^{k} \alpha}(w)=\frac{1}{N}(\sum_{i=0}^{N-1} \epsilon^{(k-1)i}T^{i}\phi ^D(w)).
\]
We get
\begin{equation}
e_{\phi}^{\epsilon^{k}  \alpha}(z)  =\exp (\epsilon^{-k} \sum _{n\ge 1}\frac{h_{-n}}{n} z^{Nn})\exp (\epsilon^{k} \sum _{n\ge 1}\frac{h_{n}}{n} z^{-Nn})U_{\epsilon^{k} \alpha}(z),
\end{equation}
where $U_{\alpha}(z)$ acts as a constant on each Heisenberg submodule, $U_{\epsilon^{k} \alpha}(z) = e_{\phi}^{\epsilon^{k} \alpha}z^{1-k + N\partial_{\alpha}}$, and $e_{\phi}^{\epsilon^{k} \alpha}$ is the highest weight vector of the Heisenberg submodule. This establishes the boson-fermion correspondence of order $N$.

\subsection{Other examples of twisted  vertex algebras;  the correspondence of type C}
\label{section:misc}

 There are other important examples in the literature of boson-fermion (or boson-boson) correspondences which can be shown to be isomorphisms of  twisted vertex algebras: in particular  the CKP  correspondence (also called  correspondence of type C, \cite{DJKM6} and \cite{OrlovLeur}), and the  "super boson-fermion correspondence of type B" (\cite{KacLeur}),  are both isomorphisms of twisted vertex algebras.
\begin{lem} The correspondence of type C is an isomorphism of twisted vertex algebras, where one of the sides is a twisted vertex algebra  based on the pair $(\mathbb{C}[h], r^{C})$ via Theorem \ref{thm:Main}.\end{lem}
{\it Proof:}
The "left-hand side" of the CKP correspondence  has  a space of states $V=H^2_{T}(\mathbb{C}[h])$ (recall the free Leibnitz  module $H^2_{T}(\mathbb{C}[h])$ of example \ref{example:HeisLeibnitzMod}), and the twisted vertex algebra is generated by the specific bicharacter value
\begin{equation}
r^C_{z, w}(h\ten h)=\frac{1}{z+w}.
\end{equation}
The OPE for the corresponding  field $h_{\phi}(z)$ directly follows from the fact that $h_{\phi}$ is primitive:
\begin{equation}
h_{\phi}(z) h_{\phi}(w) \sim \frac{1}{z+w}\sim h_{\phi}(w) h_{\phi}(z).
\end{equation}
We use  $\phi_j$ as notation for the modes of $h_{\phi}(z)$ to follow \cite{OrlovLeur}. The field $h_{\phi}(z)$ is indexed as follows: $h_{\phi}(z) =\sum _{j\in \mathbb{Z} +1/2}\phi_j z^{j-1/2}$. Theorem \ref{thm:Main} guarantees  that the pair $(\mathbb{C}[h], r^C_{z, w})$ will generate  a twisted vertex algebra of order $2$.
$\square$

 Other  examples of twisted vertex algebras are supplied by the representation theory of affine Lie algebras and affine Lie super algebras, see for example \cite{Frenkel}, \cite{FeingoldFrenkel}, \cite{Triality} and \cite{WangKac}.

\section{Summary}

There are three main results of this paper: first,  we  derived  the new boson-fermion correspondence of type D-A.
 Second, we defined the new concept of a twisted vertex algebra of order $N$, which generalize  super-vertex algebras (in the sense that a super vertex algebra is a special case of twisted vertex algebras of order 1).
 This new concept of a twisted vertex algebra was designed to answer the following  question "What is a boson-fermion correspondence--- isomorphism of what mathematical structures?". As a technical set of tools
 we developed the bicharacter construction which provided us with a general way of producing examples of twisted vertex algebras. We proved formulas for the OPEs, analytic continuations, normal ordered products and vacuum expectation values using the underlying Hopf algebra structure and the bicharacter construction. Finally, we proved that the correspondences of types B, C and D-A  are isomorphisms of twisted vertex algebras.


\def\cprime{$'$}

\end{document}